\documentclass{aa}

\usepackage{graphicx}
\usepackage{txfonts}
\usepackage{siunitx}
\usepackage{hyperref}
\usepackage{upgreek}

\usepackage{tikz}

\usepackage{xcolor}
\usepackage{subfigure}

\usepackage{multirow}
\usepackage{multicol}

\usepackage{booktabs}
\usepackage{array}

\usepackage{makecell}
\usepackage{algorithm}
\usepackage{algpseudocode}

\newcommand{\C}{\texttt{COMPASS}\ }
\newcommand{\chem}{\texttt{Chempy}\ }

\begin{document} 

   \title{A \C to the Galaxy}
   \titlerunning{COMPASS}
   \subtitle{Joint nucleosynthetic yield set comparison and parameter inference in\\Galactic chemical evolution}%

   \author{
        Berkay Günes
          \inst{1,5}
          \and
          Giuseppe Viterbo
          \inst{1,2}
          \and
          Sven Buder\inst{3,4}
          \and
          Tobias Buck
          \inst{1,2}
          }

   \institute{Universität Heidelberg, Interdisziplinäres Zentrum für Wissenschaftliches Rechnen (IWR), Im Neuenheimer Feld 205, 69120 Heidelberg, Germany\\
   \email{tobias.buck@iwr.uni-heidelberg.de}
 \and
 Universität Heidelberg, Zentrum für Astronomie, Institut für Theoretische Astrophysik, Albert-Ueberle-Straße 2, D-69120 Heidelberg, Germany
 \and
 Research School of Astronomy and Astrophysics, Australian National University, Canberra, ACT 2611, Australia
 \and
 ARC Centre of Excellence for All Sky Astrophysics in 3 Dimensions (ASTRO 3D), Australia
 \and
 Max Planck Institute for Polymer Research, Ackermannweg 10, 55128 Mainz, Germany
             }

   \date{Received March 2026; accepted 04 November 2026}

 \abstract
   {Galactic Chemical Evolution (GCE) models connect stellar nucleosynthesis to observed abundance patterns, but their interpretation requires simultaneous parameter inference and comparison between competing stellar yield prescriptions. Yet, this task remains challenging due to intractable likelihoods and the growing complexity of stellar-abundance datasets.}
   {Using \C, a novel simulation-based inference framework, we perform, for the first time, amortized joint Bayesian model comparison and parameter inference for GCE models with intractable likelihoods within a unified architecture.}
   {\C employs a conditional score-based diffusion model with a transformer architecture to learn both the posterior and likelihood directly from complex forward simulations. This allows efficient, amortized inference and a likelihood-based ranking of competing nucleosynthetic yield sets.
   }
   {In controlled tests, \C identifies the generating yield set and recovers ground-truth parameters with accuracy comparable to existing simulation-based inference (SBI) methods, while reducing computational cost by orders of magnitude compared to traditional sampling approaches. Applied to Milky Way data, the framework favors a hybrid NuGrid + IllustrisTNG yield prescription, a preference confirmed by an independent classifier, and infers an initial mass function slope of $\alpha_{\rm IMF} = -2.65 \pm 0.02$ and a SN$_{\rm Ia}$ normalization of $\log(N_{\rm Ia}) = -2.81 \pm 0.02$, with a systematic spread of $\approx 0.3$ in $\alpha_{\rm IMF}$ between the best-ranked yield sets, consistent with independent constraints on solar-neighborhood enrichment. Crucially, the model’s attention mechanism confirms that the inferred parameters are driven by physically meaningful abundance signatures.}
   {By unifying amortized parameter inference, Bayesian model comparison, and interpretability within a single probabilistic framework, \C provides a scalable approach for precision chemical evolution studies with current and forthcoming spectroscopic surveys.}

\maketitle
\nolinenumbers
\section{Introduction}
\label{sec:intro}

Understanding the chemical enrichment of galaxies is central to reconstructing their formation and evolution. 
Stellar abundance patterns encode information about key physical ingredients of Galactic Chemical Evolution (GCE), including the high-mass slope of the stellar initial mass function (IMF), the normalization and delay-time distribution of Type Ia supernovae (SNe Ia), and the nucleosynthetic yields of stars. 
These parameters regulate the production of heavy elements \citep[e.g.,][]{2005A&A...430..491R, 2015MNRAS.449.1327V, 2015MNRAS.451.3693M}, stellar feedback \citep[e.g.][]{Agertz2013}, and the temporal build-up of chemical abundances \citep[e.g.][]{Kobayashi2020}, and are therefore essential inputs for hydrodynamical simulations of galaxy formation \citep[e.g.,][]{Sawala2016, Hopkins2018, Pillepich2018, Buck2020, Buck2020c, Font2020, Agertz2021} and for Galactic Archaeology studies of the Milky Way to inform inferences of the time evolution of the chemical abundances across the Milky Way disk \citep[e.g.][]{Ratcliffe2023,Ratcliffe2025,Wang2024}, of birth radii of stars in the Milky Way \citep[e.g.][]{Lu2022,Lu2024} or identification of accreted stars \citep[e.g.][]{Buder2024a,Buder2025,Buder2025b}.

Despite decades of work, fundamental uncertainties remain. 
The shape and potential non-universality of the IMF \citep{2010ARA&A..48..339B}, the relative contributions of different supernova channels and stellar winds \citep{1997ARA&A..35..503M}, the progenitors and delay-time distributions of SNe Ia \citep{2012PASA...29..447M},  and the choice of nucleosynthetic yield tables all introduce substantial degeneracies in GCE modeling. 
For instance, various studies have suggested that the high-mass slope of the IMF may be steeper than canonical values \citep[e.g.,][]{2015ApJ...806..198W, Rybizki2015, Chabrier2014}, and may itself vary as a function of metallicity, introducing further complexity \citep[e.g.,][]{2019MNRAS.482..118G, Martin2019}. 
Similarly, the normalization and delay-time distribution of SNe Ia, which are crucial for the enrichment of iron-peak elements, are still heavily debated \citep{2010ApJ...722.1879M, 2012MNRAS.426.3282M, 2015ApJ...810..137J, Buck2021}.

Furthermore, different stellar evolution calculations predict markedly different element production rates \citep[e.g.][]{2010A&A...522A..32R,Kobayashi2020}, leading to systematic variations in predicted abundance patterns. 
Recent studies have shown that no single commonly adopted yield set can simultaneously reproduce the observed chemical abundances across different galactic environments \citep[e.g.][]{Kobayashi2020,Buck2021,Liang2023}. Hence, the dominant source of systematic uncertainty in GCE modeling is not only parameter values, but the choice of nucleosynthetic prescriptions themselves. This underscores the need for principled methods to discriminate between competing physical prescriptions.

At the same time, large spectroscopic surveys such as APOGEE \citep{apogee17} and GALAH \citep{Buder2021, Buder2024} now provide high-precision elemental abundances for hundreds of thousands of stars. 
In principle, these data contain sufficient information to tightly constrain GCE parameters and nucleosynthetic models. 
In practice, however, inference in GCE is computationally challenging. 
Chemical evolution models are typically implemented as forward simulators that generate synthetic abundance patterns for a given parameter set, but do not provide a tractable analytic likelihood function. 
Traditional Bayesian approaches such as Markov Chain Monte Carlo (MCMC) or Hamiltonian Monte Carlo (HMC) therefore become computationally expensive and difficult to scale, particularly when likelihood evaluation must be repeated for large datasets. 
Moreover, principled model comparison between competing yield prescriptions requires evaluation of marginal likelihoods or likelihood ratios, which is especially challenging in likelihood-free settings.

Simulation-Based Inference \citep[SBI;][]{Cranmer_2020} provides a promising alternative by learning surrogate density estimators directly from simulations. 
In the context of GCE, recent work has demonstrated that SBI enables efficient parameter inference within a fixed physical model \citep[e.g.][]{buck2025inferringgalacticparameterschemical}. 
However, existing approaches focus exclusively on parameter estimation and do not provide a unified framework for jointly performing parameter inference and Bayesian model comparison across competing nucleosynthetic yield sets.

In this work, we employ \C \citep[Comparison Of Models using Probabilistic Assessment in Simulation-based Settings][]{Guenes2025}, a novel simulation-based inference framework that enables, for the first time, joint parameter inference and model comparison in Galactic Chemical Evolution. 
\C employs a conditional score-based diffusion model with a transformer architecture to learn both posterior and likelihood distributions from simulated data. 
This dual capability allows efficient amortized inference of galactic parameters while simultaneously enabling principled ranking of competing nucleosynthetic yield prescriptions.

We demonstrate the performance of \C using the one-zone GCE model \chem \citep{Rybizki_2017} and apply it to high-precision stellar abundance data from \citet{Nissen_2020}. 
By comparing 40 different combinations of asymptotic giant branch and core-collapse supernova yield tables, we identify the prescriptions most consistent with the chemical composition of the local Milky Way. 
Using these preferred models, we then infer constraints on the high-mass slope of the IMF and the normalization of SNe Ia.

This framework establishes a scalable and statistically principled approach to testing nucleosynthesis models and exploiting the full constraining power of next-generation spectroscopic surveys.

This paper is structured as follows. In Sec.~\ref{sec:methods} we start by briefly summarizing GCE models and describing the machine learning and statistics methods used in this work. We describe in detail how \C and its conditional score-based diffusion model with transformer architecture works and how we perform joint model comparison and inference. Sec.~\ref{sec:data} then discusses the simulated datasets used for model training and introduces the observed stellar abundances for the Milky Way which we use for inference. In Sec.~\ref{sec:results} we present our results, first for the model comparison part and then for the inference of IMF high mass slope and supernova Ia normalization. 
We discuss the implications and limitations of our work in Sec.~\ref{sec:disucssion}. 
Finally, Sec.~\ref{sec:conclusions} presents our conclusions and the Data availability section provides the link to our Github repository for reproducibility.

\section{Methods}
\label{sec:methods}

\subsection{Galactic chemical evolution}
\label{sec:GCE}

Galactic Chemical Evolution (GCE) describes the build-up of heavy elements as successive generations of stars form, evolve, and return nucleosynthetic products to the interstellar medium (ISM) \citep{Matteucci_2016, 2013ARA&A..51..457N, Kobayashi2020}. 
While Big Bang nucleosynthesis produced primarily hydrogen and helium, all heavier elements have been synthesized in stars and redistributed through stellar winds and supernova explosions. 
The resulting abundance patterns encode information about star formation histories, feedback processes, and stellar nucleosynthesis, forming the chemical foundation of Galactic Archaeology \citep{Freeman2002}.

The enrichment history of a galaxy is governed by a small set of key ingredients: the star formation rate (SFR), which controls how rapidly gas is converted into stars; the stellar IMF, which specifies the relative number of stars of different masses; the rates of different supernova channels, and the nucleosynthetic yields of stars as functions of mass and metallicity, which quantify the mass of each element ejected over a star’s lifetime. 
Massive stars ($M \gtrsim 8\,M_\odot$) explode as core-collapse supernovae (CC-SNe) on short timescales and dominate the production of $\alpha$-elements such as oxygen and magnesium. 
SNe~Ia, originating from thermonuclear explosions of white dwarfs in binary systems, enrich the ISM primarily in iron-peak elements on delayed timescales \citep{Maoz2014}. 
Low- and intermediate-mass stars contribute carbon, nitrogen, and $s$-process elements during their asymptotic giant branch (AGB) phase \citep{Karakas2014, Cristallo2015}. 
The distinct timescales of these channels produce characteristic abundance–ratio patterns, such as the [$\alpha$/Fe] plateau and its decline with increasing [Fe/H] \citep[e.g.][]{Hayden2015,Buder2021,Buder2024} and encode a fossil record of past star formation, gas accretion, and feedback processes \citep[e.g.][]{Spitoni2022, Vincenzo2019}.

Among the governing parameters, two play a particularly influential role in shaping these trends. 
The high-mass slope of the IMF ($\alpha_{\rm IMF}$) regulates the relative number of massive stars and thus the overall production of $\alpha$-elements \citep[e.g.][]{Dubay2024,denBrok2024}. 
The normalization of the SNe~Ia rate ($\log_{10}N_{\rm Ia}$) controls the integrated iron enrichment and the location of the [$\alpha$/Fe] knee \citep[e.g.][]{Dubay2024,Buck2021}. 
Variations in these parameters produce abundance signatures that are measurable with modern high-precision spectroscopic surveys and allow chemical tagging -- the identification of stellar populations that share common formation sites or accretion origins \citep[e.g.][]{Ness2018, Griffith2022, Griffith2025, Ting2022}.

However, GCE predictions depend not only on these global parameters but also critically on the adopted nucleosynthetic yield tables \citep{Cote2017,Weinberg2024,Buck2021,Liang2023,Philcox_2018} and of nucleosynthetic physics, including metallicity-dependent yields \citep{Kobayashi2020, Gronow2021}, the relative contributions of sub-Chandrasekhar versus Chandrasekhar-mass Type~Ia supernovae \citep{Alexander2025}, the type of SN~Ia delay-time distribution \citep{Dubay2024} or the slopes of the IMF \citep{denBrok2024}. 
Different stellar evolution calculations predict substantially different element production efficiencies across mass and metallicity, and these discrepancies propagate non-linearly into predicted abundance–ratio relations and can dominate the uncertainty budget of GCE inference. 
In many cases, systematic differences between yield sets rival or exceed the effects of varying $\alpha_{\rm IMF}$ or $N_{\rm Ia}$, making yield modeling one of the dominant uncertainties in chemical evolution studies.

In most existing GCE analyses, parameter inference is performed conditional on a fixed choice of yield prescriptions. 
Yet if the nucleosynthetic physics itself is uncertain, parameter constraints derived under a single yield model remain inherently conditional.
Consequently, inferring galactic parameters without simultaneously accounting for yield-set uncertainty risks biased conclusions. 
A statistically principled framework must therefore enable both parameter inference within a given GCE model and quantitative comparison between competing nucleosynthetic prescriptions. 
Addressing this challenge requires inference methods capable of operating in likelihood-free, simulator-based settings while remaining computationally scalable.

\begin{figure*}
    \centering
    \includegraphics[width=1\linewidth]{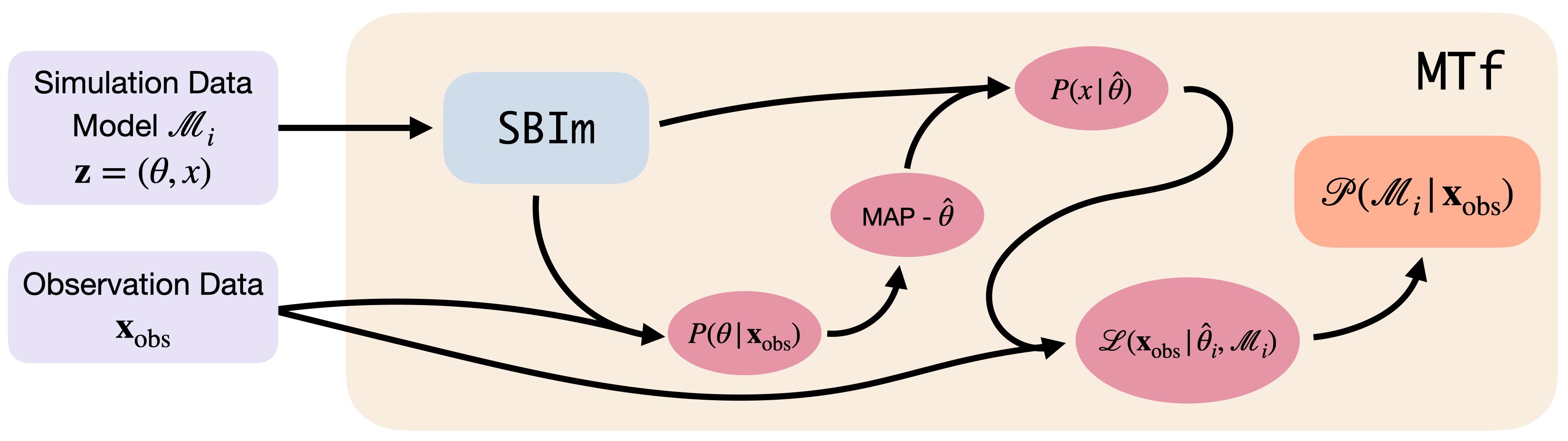}
    \caption{The \C model comparison workflow. 
Simulation data from model $\mathcal{M}_i$ train the \texttt{ScoreBasedInferenceModel} (\texttt{SBIm}). 
Then for some observations $\mathbf{x}_{\rm obs}$ \texttt{SBIm} estimates the posterior distribution $P(\theta|\mathbf{x}_{\rm obs})$ and then calculates its MAP $\hat \theta$.
\texttt{SBIm} then models the likelihood $P(x|\hat\theta)$ to get the maximized likelihood for each model $\mathcal{L}(\mathbf{x}_{\rm obs}|\hat\theta,\mathcal{M}_i)$, leading to the model posteriors $\mathcal{P}(\mathcal{M}_i|\mathbf{x}_{\rm obs})$. See \autoref{app:SBI} and following for more details.}
\label{fig:Workflow}
\end{figure*}

In this work, we employ the open-source one-zone GCE model \chem\ \citep{Rybizki_2017} as our forward simulator which provides a computationally efficient framework for exploring the impact of varying IMF slopes, SNe~Ia rates, and stellar yield sets.
By coupling \chem with our SBI and model comparison framework, \texttt{COMPASS} \citep{Guenes2025}, we jointly infer key chemical evolution parameters and assess the relative plausibility of competing enrichment models.

\subsection{Simulation-based inference}
\label{app:SBI}
Many astrophysical models are implemented as forward simulators that generate synthetic data $\mathbf{x}\!\sim\!p(\mathbf{x}|\theta)$ for a given parameter vector $\theta$, while the likelihood $p(\mathbf{x}|\theta)$ itself remains analytically intractable. 
This renders traditional likelihood-based Bayesian inference infeasible or computationally prohibitive. 
SBI provides a principled alternative by learning surrogate density estimators directly from simulated data \citep{Cranmer_2020}. In astronomy, SBI has recently been applied to a variety of problems ranging from galactic dynamics \citep{Viterbo2026,Viterbo2026b} over galactic Archaeology \citep{Viterbo2024,Sante2026}, gravitational-wave astronomy \citep{Dax_2025}, galaxy evolution \citep{Lovell2025}, binary star inference \citep{Knoell2026} up to model comparison in galaxy formation \citep{Zhou2025}. 

In practice, GCE analyses typically adopt a fixed set of nucleosynthetic yields and perform parameter inference conditional on this choice. However, if the assumed yields are themselves uncertain, parameter constraints become conditional statements rather than robust inferences. A principled statistical framework must therefore allow joint parameter inference and model comparison across competing yield prescriptions. 
In this work we employ \C --
Comparison Of Models using Probabilistic Assessment in Simulation-based Settings —
a Python framework\footnote{Publicly available at \url{https://github.com/bGuenes/COMPASS}.} 
designed for parameter inference and model comparison in simulator-based settings. 
Technical details are provided in \cite{Guenes2025}; here we summarize the methodological foundations with more details in appendix~\ref{sec:COMPASS.Core}.

\paragraph{Core principle.}
SBI methods train a neural density estimator on simulated pairs $(\theta_i, \mathbf{x}_i)$ drawn from the joint distribution $p(\theta, \mathbf{x})$. 
Depending on the formulation, the estimator approximates the posterior $p(\theta|\mathbf{x})$ (Neural Posterior Estimation; \citealt{Papamakarios2016}), the likelihood $p(\mathbf{x}|\theta)$ (Neural Likelihood Estimation; \citealt{99aad490fd134b849b5336473b4342c3}), or the likelihood ratio $p(\mathbf{x}|\theta)/p(\mathbf{x})$. 
Once trained, the surrogate enables fast amortized inference for new observations without rerunning the simulator.

Recent work has demonstrated that score-based diffusion models (Sec.~\ref{sec:COMPASS.Core.Diffusion}) can serve as powerful, flexible density estimators for SBI \citep{song2020, gloeckler2024allinonesimulationbasedinference}. 
These models learn the gradient of the data distribution’s log-density (the ``score’’) through a stochastic denoising process, enabling flexible modeling of complex, non-Gaussian distributions. 
In \C, the score is parameterized by a transformer architecture (Sec. \ref{sec:COMPASS.Core.Transformer}) that captures non-linear dependencies between observables and model parameters via multi-head self-attention that additionally allows for physical interpretability through the learned attention structure.

A key feature of \C is its dual-role density estimation. 
By applying an attention mask within the transformer, the same architecture can operate in:
(i) Neural Posterior Estimation (NPE, see~\ref{sec:COMPASS.Core.Transformer}) mode, learning $p(\theta|\mathbf{x})$, or 
(ii) Neural Likelihood Estimation (NLE) mode, learning $p(\mathbf{x}|\theta)$. 
This dual capability enables both amortized parameter inference (Sec. \ref{sec:COMPASS.Inference}) and likelihood-based model comparison. 
For each candidate model $\mathcal{M}_i$, the learned likelihood allows evaluation of
$
\ln P(\mathbf{x}|\hat{\theta}, \mathcal{M}_i),
$
 at the Maximum A Posteriori (MAP) parameters $\hat\theta$ 
facilitating a ranking of the models by relative weights $P(\mathcal{M}_i|\mathbf{x})$ (see Sec.~\ref{sec:COMPASS.Model-Comp}).

\subsection{Parameter inference with \C} \label{sec:COMPASS.Inference}

We now describe how \C performs parameter inference. 
We first consider single-observation inference and then show how information from multiple observations is combined.

\subsubsection{Single-observation posterior estimation} \label{sec:COMPASS.Inference.SingleObs}
For an individual observation $x_i$, \C estimates the posterior $P(\theta|x_i)$ using its conditional diffusion model in NPE mode (see Fig.~\ref{fig:Workflow} and Sec.~\ref{sec:COMPASS.Core.Transformer} in the appendix).
The observed data $x_i$ are specified via a condition mask $\mathcal{M}_C$: observed dimensions are set to $1$, while parameters $\theta$ are marked as latent ($0$). 
The reverse diffusion process is initialized at $t=1$ with $\theta$ drawn from the prior noise distribution. 
The reverse-time SDE (Eq.~\ref{eq:ReverseSDE}) is then solved numerically down to $t=0$ over $N_t$ timesteps (Sec.~\ref{sec:COMPASS.Core.Sampling}). 
At each timestep, the transformer-based score network 
$ s_{\phi}(\mathbf{z}_t, \mathcal{M}_C, t) $ 
provides the conditional score estimate. 
The resulting samples at $t=0$ constitute draws from $P(\theta|x_i)$.

From these posterior samples, point estimates such as the Maximum A Posteriori (MAP) parameters $\hat{\theta}$ are obtained by fitting a multivariate Gaussian Kernel Density Estimator (KDE) and minimizing the negative density using \texttt{scipy.optimize} \citep{2020SciPy-NMeth}. 
The architecture naturally accommodates missing data. If certain components of $x_i$ are unobserved, their mask entries are set to $0$, and the model jointly infers or marginalizes over them alongside $\theta$. 
This allows robust inference in incomplete or heterogeneous datasets and makes \C a versatile tool for real-world inference scenarios where datasets are often incomplete.

\subsubsection{Multi-observation posterior estimation} \label{sec:COMPASS.Inference.MultiObs}

For multiple observations $\mathbf{x} = \{x_1,\dots,x_N\}$ generated under common parameters $\theta$, \C is capable of combining information via posterior factorization to significantly increasing the certainty and precision of the inferred parameters $\theta$.
To derive the joint posterior distribution  $P(\theta|\mathbf{x})$, the individual posterior distributions obtained for each event are factorized by applying Bayes rule twice, as formulated by \citet{geffner2023compositionalscoremodelingsimulationbased}. Each single-observation posterior is learned from simulations drawn from the training proposal $\pi(\theta)$ (Sec.~\ref{sec:training_setup}), i.e. $\tilde{P}(\theta|x_i) \propto P(x_i|\theta)\,\pi(\theta)$. The joint posterior under the prior $P(\theta)$ is therefore
\begin{align}
P(\theta| \mathbf{x}) \propto P(\theta)P(\mathbf{x}|\theta)
    &= P(\theta) \prod_ {i=1}^{N} P(x_i|\theta) \\
    &\propto P(\theta) \prod_ {i=1}^{N} \frac{\tilde{P}(\theta|x_i)}{\pi(\theta)} \nonumber \\
    &\propto P(\theta) \prod_ {i=1}^{N} \tilde{P}(\theta|x_i), \label{eq:factorized-posterior}
\end{align}%
where the last step uses that $\pi(\theta)$ is uniform on its support. If the single-observation posteriors had instead been learned under $P(\theta)$ itself, the prior would enter as $P(\theta)^{1-N}$ \citep{geffner2023compositionalscoremodelingsimulationbased}.
This factorization allows the principled combination of individually estimated posteriors $P(\theta|x_i)$ with the prior $P(\theta)$ to construct the joint multi-observation posterior.

For analytical tractability, within \C each individual posterior $P(\theta|x_i)$ is approximated as a multivariate Gaussian with mean $\boldsymbol{\mu}_i$ and covariance $\boldsymbol{\Sigma}_i$, estimated from diffusion samples. 
The prior $P(\theta)$ on the global parameters $\alpha_{\rm IMF}$ and $\log_{10}N_{\rm Ia}$ is likewise modeled as Gaussian with mean $\boldsymbol{\mu}_{\text{Prior}}$ and covariance $\boldsymbol{\Sigma}_{\text{Prior}}$.
Under this approximation, the combined posterior remains Gaussian, with parameters:
\begin{align}
    \mathbf{\Sigma}_{\rm Posterior}^{-1} &= \mathbf{\Sigma}_{\rm Prior}^{-1} + \sum_i^N \mathbf{\Sigma}_i^{-1} \label{eq:post-cov} \\
\mathbf{\mu}_{\rm Posterior} &= \mathbf{\Sigma}_{\rm Posterior} \left[ \mathbf{\Sigma}_{\rm Prior}^{-1}\, \mathbf{\mu}_{\rm Prior} + \sum_i^N \mathbf{\Sigma}_i^{-1} \mathbf{\mu}_i \right] \label{eq:post-mean}
\end{align}

This formulation enables efficient aggregation of information from many stars, yielding increasingly precise parameter constraints without retraining the network.
Because the combined covariance shrinks as $1/N$, any systematic error shared by the single-star posteriors (e.g. a small bias of the network) does not average out, and its significance in units of the combined uncertainty grows with $N$. We quantify this on mock data in Sec.~\ref{sec:Results.Params.Chempy} (right panels of Figs.~\ref{fig:sbi-chempy-TNG}, \ref{fig:sbi-chempy-alt} and \ref{fig:sbi-chempy-TNG-sim}). We note that combination rules that are more robust to imperfect single-observation posteriors, such as compositional score modeling \citep{geffner2023compositionalscoremodelingsimulationbased}, are left for future work.

The assumption of Gaussianity for individual posteriors is a practical approximation.
While true posteriors can be complex, a Gaussian fit often adequately captures the central tendency and dispersion for well-behaved, unimodal posteriors, or when the primary goal is to combine information for parameters that are expected to be reasonably constrained by the data.
We test the resulting combined posteriors against known ground truth on mock data in Sec.~\ref{sec:Results.Params.Chempy} and Appendix~\ref{sec:model-misspecification}.
Future extensions of \C may explore more expressive distributional forms for $P(\theta|x_i)$ to relax this Gaussian assumption.

\subsubsection{Handling observational uncertainties}

A fundamental challenge in applying any inference framework to real-world data is the proper treatment of observational uncertainties.
Real observational data are subject to measurement noise and are therefore more accurately described by a probability distribution than by a single point estimate. 
A measurement $\mathbf{x}_{\rm obs}$ with uncertainty $\Sigma_{\rm obs}$ defines the measurement model $p(\mathbf{x}_{\rm obs} \mid x) = \mathcal{N}(\mathbf{x}_{\rm obs}; x, \Sigma_{\rm obs})$. The distribution of the true value given the measurement then follows from Bayes' theorem, $p(x \mid \mathbf{x}_{\rm obs}, \Sigma_{\rm obs}) \propto \mathcal{N}(\mathbf{x}_{\rm obs}; x, \Sigma_{\rm obs})\, p(x)$, where $p(x)$ is the distribution of abundances predicted by the simulator.

A statistically consistent parameter posterior must therefore marginalize over this measurement distribution rather than condition solely on the central value. 
The desired posterior is therefore given by the integral\footnote{This formulation assumes that the measurement distribution fully characterizes uncertainty in the observed quantity.
}:
\begin{align}
    p(\theta|\mathbf{x}_{\rm obs}, \sigma_{\rm obs}) = \int p(\theta|x) \cdot p(x | \mathbf{x}_{\rm obs}, \sigma_{\rm obs}) dx \label{eq:noise-marginalisation}
\end{align}

Although \C provides an accurate approximation to $p(\theta \mid x)$, direct evaluation of this integral would require nested inference within the diffusion process and is computationally impractical. 
Instead, \C employs a Monte Carlo approximation to perform the marginalization.

For each observation, $N$ realizations are drawn from the measurement distribution,
$ x_i \sim \mathcal{N}(\mathbf{x}_{\rm obs}, \Sigma_{\rm obs}) $,
and the posterior is evaluated for each realization,
$ \theta_i \sim p(\theta \mid x_i) $.
The aggregated set of samples $\{\theta_1, \dots, \theta_N\}$ constitutes a Monte Carlo approximation to the marginalized posterior
$ p(\theta \mid \mathbf{x}_{\rm obs}, \Sigma_{\rm obs}) $.
Drawing $x_i \sim \mathcal{N}(\mathbf{x}_{\rm obs}, \Sigma_{\rm obs})$ neglects the factor $p(x)$ in $p(x \mid \mathbf{x}_{\rm obs}, \Sigma_{\rm obs})$, which is only accurate when $p(x)$ is flat on the scale of the measurement errors. This holds for our data: the \citet{Nissen_2020} uncertainties (median $0.007$--$0.03$\,dex per element) are small compared with the spread of the simulated abundances ($0.22$--$0.73$\,dex), and evaluating $\mathrm{d}\ln p(x)/\mathrm{d}x$ from the simulations at each measurement, the implied shift of the noise draws, $\sigma_{\rm obs}^2\,\mathrm{d}\ln p/\mathrm{d}x$, is below $5\%$ of $\sigma_{\rm obs}$ for all stars and elements.
Note that the training simulations already include Gaussian noise of $0.05$\,dex (Sec.~\ref{sec:training_setup}), so the network itself conditions on noisy data. Eq.~\ref{eq:noise-marginalisation} propagates the star-specific errors on top of this, which enlarges the effective noise by a factor $\sqrt{1+\sigma_{\rm obs}^2/(0.05\,{\rm dex})^2} = 1.01$--$1.17$. This is a small and conservative double counting; conditioning the training on star-specific errors is left for future work.

By default, \C uses $N=1000$ realizations per observation.
This procedure propagates measurement uncertainties from data space into parameter space, so that inferred credible intervals reflect both model uncertainty and observational noise.

Finally, we assess the inference quality using established simulation-based validation methods. 
Simulation-Based Calibration (SBC; \citealt{talts2020validatingbayesianinferencealgorithms}) tests the self-consistency of inferred marginal posteriors by comparing the rank statistics of true parameters under repeated simulations. 
Complementarily, Tests of Accuracy with Random Points (TARP; \citealt{lemos2023samplingbasedaccuracytestingposterior}) evaluate the joint posterior coverage, ensuring that nominal credible intervals correspond to empirical frequencies. 
We use these diagnostics to assess the calibration of \C\unskip. For more details see appendix~\ref{sec:Results.Calibration}.

\subsection{Bayesian model comparison}
\label{sec:COMPASS.Model-Comp}
In many astrophysical applications, one must evaluate a set of competing models 
$\{\mathcal{M}_i\}$ that differ in their underlying physical assumptions -- for example, alternative nucleosynthetic yield prescriptions. 
The goal of Bayesian model comparison is to quantify the relative plausibility of these models given observed data $\mathbf{x}$.
A distinguishing capability of \C is that it provides a likelihood-based ranking of competing models in addition to parameter inference. The central question \C addresses in this context is: "Which model $\mathcal{M}_i$ is most strongly supported by the observed data $\mathbf{x}_{\text{obs}}$?"

\subsubsection{Bayesian evidence and practical approximation}
In the fully Bayesian framework, models are compared via their marginal likelihood or model evidence,
\begin{align}
p(\mathbf{x} \mid \mathcal{M}_i)
=
\int p(\mathbf{x} \mid \theta, \mathcal{M}_i)\,
p(\theta \mid \mathcal{M}_i)\, d\theta.
\label{eq:model-evidence}
\end{align}

This integral averages the likelihood over the parameter prior and automatically penalizes model complexity. 
However, in simulator-based settings the likelihood $p(\mathbf{x}\mid\theta,\mathcal{M}_i)$ is not available in analytic form, making direct evaluation of Eq.~\ref{eq:model-evidence} computationally prohibitive.

To enable practical model comparison, we adopt an approximation based on maximized likelihoods and information criteria.

\paragraph{Model ranking via maximized likelihood}

For each candidate model $\mathcal{M}_i$, we first perform parameter inference (Sec.~\ref{sec:COMPASS.Inference}) to obtain the Maximum A Posteriori (MAP) estimate $\hat{\theta}_i$. 
We then evaluate the likelihood at this parameter point,
\begin{align}
\mathcal{L}_i
=
p(\mathbf{x} \mid \hat{\theta}_i, \mathcal{M}_i).
\end{align}

To account for model complexity and approximate predictive performance, we use the Akaike Information Criterion (AIC) \citep{AKAIKE19781877,burnham2002model},
\begin{align}
\mathrm{AIC}_i
=
-2 \ln \mathcal{L}_i + 2 k_i,
\end{align}
where $k_i$ denotes the number of free parameters in model $\mathcal{M}_i$.
The AIC provides an asymptotic approximation to out-of-sample predictive performance under regularity conditions and large-sample limits. While it does not yield the full marginal likelihood, it offers a computationally efficient surrogate that is particularly attractive in simulation-based settings where exact evidence evaluation is infeasible. It allows us to estimate the relative quality of models based on their maximized log-likelihood and is suitable for non-nested models derived from simulators.

Differences in AIC values, $\Delta_i(\text{AIC}) = \text{AIC}_i - \text{AIC}_{\min}$, where $\text{AIC}_{\min}$ is the minimum AIC value among the models, can be converted into relative model weights,

\begin{align}
w_i
\propto
\exp\left(-\frac{1}{2}\Delta_i(\text{AIC})\right),
\end{align}

which approximate the probability that model $\mathcal{M}_i$ minimizes information loss within the candidate set. 
The normalized model posterior weights \citep{Wagenmakers2004, AKAIKE1979, Bozdogan_1987} are therefore
\begin{align}
\mathcal{P}(\mathcal{M}_i \mid \mathbf{x})
=
\frac{\exp\left(-\frac{1}{2}\Delta_i\right)}
{\sum_j \exp\left(-\frac{1}{2}\Delta_j\right)}.
\label{eq:model-posterior}
\end{align}

The resulting model weights therefore quantify relative predictive adequacy within the candidate set, rather than exact posterior model probabilities derived from full marginal likelihood integration. Given the computational constraints of simulator-based inference, this approximation provides a principled and scalable compromise between statistical rigor and feasibility.
For brevity we denote these normalized Akaike weights by $\mathcal{P}(\mathcal{M}_i \mid \mathbf{x})$; they do not account for the prior volume of each model and are in general more extreme than Bayes factors.

\paragraph{Equal-complexity simplification}

In the present study, all competing GCE models share identical parameter dimensionality ($k_i = k$ for all $i$). 
Under this condition, the complexity penalty cancels in the relative comparison, and Eq.~\ref{eq:model-posterior} reduces to a softmax over maximized log-likelihoods (see Appendix~\ref{sec:AIC}),
\begin{align}
\mathcal{P}(\mathcal{M}_i \mid \mathbf{x})
=
\frac{\exp\left(\ln \mathcal{L}_i\right)}
{\sum_j \exp\left(\ln \mathcal{L}_j\right)}.
\end{align}
Thus, model ranking is determined directly by relative predictive performance.
Because all candidate models share the same $k$ and are evaluated on the same $n$ observations, the Bayesian Information Criterion, ${\rm BIC}_i = -2 \ln \mathcal{L}_i + k \ln n$, differs from the AIC only by the model-independent constant $k(\ln n - 2)$. Both criteria therefore give identical rankings and identical weights and in this equal-$k$ setting the choice between them does not matter.

We emphasize that this simplification is specific to the equal-complexity case considered here. In future applications involving models with differing parameter dimensionality, the full information-criterion penalty term—or alternative approximations to the marginal likelihood—can be retained without modification of the \C inference architecture.

\subsubsection{Likelihood estimation with \C} \label{sec:COMPASS.Model-Comp.Likelihood}

The likelihood $\mathcal{L}_i = p(\mathbf{x}\mid\hat{\theta}_i,\mathcal{M}_i)$ is estimated using the dual posterior/likelihood capability of the conditional diffusion model (red parts in Fig.~\ref{fig:Workflow}).

After obtaining $\hat{\theta}_i$ via posterior inference (see Sec.~\ref{sec:COMPASS.Inference}), the diffusion model is operated in likelihood-estimation mode (by inverting the condition mask $\mathcal{M}_C$ relative to the NPE mode, see also Algorithm \ref{alg:Sampling} as detailed in the appendix) to generate samples $ \mathbf{x}_s \sim p(\mathbf{x}\mid\hat{\theta}_i,\mathcal{M}_i) $.
A multivariate Gaussian KDE is fitted to these generated samples, and the observed data are evaluated under this density to obtain $\ln \mathcal{L}_i$.

For multiple independent observations 
$\mathbf{x} = \{x_1,\dots,x_N\}$, 
the total log-likelihood is simply given by 
\begin{align}
\ln \mathcal{L}_i
=
\sum_{n=1}^{N}
\ln p(x_n \mid \hat{\theta}_i, \mathcal{M}_i).
\label{eq:combined-likelihood}
\end{align}

This summed log-likelihood enters directly into the model ranking framework above (Eq.~\ref{eq:model-posterior}).

Finally, we note that the procedure presented above approximates full Bayesian model comparison by evaluating predictive performance at the MAP parameter estimate rather than integrating over the full posterior.
While it does not compute the exact marginal likelihood (Eq.~\ref{eq:model-evidence}), it provides a scalable ranking criterion for simulator-based models. It does not include the Occam factor of the parameter prior, so the weights should be read as relative predictive support. In Appendix~\ref{sec:classifier} we provide an independent check of the resulting ranking.
Because all candidate models considered here have identical parameter dimensionality, the comparison isolates differences in predictive accuracy rather than model complexity.

\section{Data}
\label{sec:data}

Our analysis relies on two data sources: 
(i) simulated abundance data generated from a suite of GCE models for training and benchmarking \C, and 
(ii) high-precision stellar abundance observations used for scientific application.
We describe both, simulated and observational data, in detail below.

\begin{table*}
    \centering
    \caption{Free parameters of the \chem simulator.}\label{tab:GCEpriors}
    \resizebox{\textwidth}{!}{%
    \begin{tabular}{rl|c c c}
    \toprule
    Parameter & Description & $\overline{\theta}_\mathrm{prior}\pm\sigma_\mathrm{prior}$ & Prior from: & Training proposal \\
    \midrule
    \multicolumn{5}{c}{$\vec{\Lambda}$: Global stellar (SSP) parameters}\\
    \midrule
    $\alpha_\mathrm{IMF}$ & High-mass slope of the \citep{2003PASP..115..763C} IMF & $-2.3\pm0.3$ & \citep[Tab.\,1]{2003PASP..115..763C} & $\mathcal{U}(-3.8,\,-0.8)$ \\
    $\log_{10}\left(N_\mathrm{Ia}\right)$ & Number of SN\,Ia per $\mathrm{M}_\odot$ over 15\,Gyr & $-2.89\pm0.3$ & \citep[Tab.1\,]{2012PASA...29..447M} & $\mathcal{U}(-4.39,\,-1.39)$ \\
    \midrule
    \multicolumn{5}{c}{$\vec{\Theta}_i$: Local ISM parameters}\\
    \midrule
    $\log_{10}\left(\mathrm{SFE}\right)$ & Star formation efficiency governing gas infall & $-0.3\pm0.3$ & \citep{2008AJ....136.2846B} & $\mathcal{U}(-1.8,\,1.2)$ \\
    $\log_{10}\left(\mathrm{SFR}_\mathrm{peak}\right)$ & SFR peak in Gyr (scale of $k=2$ $\Gamma$-distribution) & $0.55\pm0.1$ & \citep[fig.\,4b]{2013ApJ...771L..35V} & $\mathcal{U}(0.05,\,1.05)$ \\
    x$_\mathrm{out}$ & Stellar feedback fraction & $\phantom{-}0.5\pm0.1$ & \citep[Tab.\,1]{Rybizki_2017} & $\mathcal{U}(0,\,1)$ \\
    \midrule
    \multicolumn{5}{c}{$T_i$: Timescale}\\
    \midrule
    $T_i$ & Time of stellar birth in Gyr & [$1$,$13.8$] & Observations & $\mathcal{U}(1,\,13.8)$ \\
    \bottomrule
    \end{tabular}}
    \tablefoot{Gaussian priors and uniform training proposal ($\pm5\,\sigma_\mathrm{prior}$ around the prior mean, Sec.~\ref{sec:training_setup}) of each parameter. The global parameters $(\alpha_\mathrm{IMF},\log_{10}\left(N_\mathrm{Ia}\right))$ are assumed to be shared across all stars in a given dataset. The local parameters and the stellar birth time $T$ are specific to each individual star's formation environment. For the \citet{Nissen_2020} stars, $T$ is observed via the stellar age and hence fixed.}
\end{table*}

\subsection{Nucleosynthetic yield sets for model comparison}
Following \citet{buck2025inferringgalacticparameterschemical}, we employ the one-zone GCE model \chem \citep{Rybizki_2017} to generate a large suite of chemical evolution simulations. 
For fixed galactic parameters $\theta$, \chem\ predicts abundance tracks as functions of time. 

Because this study focuses on model comparison, we construct competing GCE models by varying the adopted nucleosynthetic yield tables. 
Each candidate model $\mathcal{M}_i$ corresponds to a unique combination of AGB and CC-SN yield prescriptions. 
We consider four AGB yield sets and ten CC-SN yield sets, resulting in $4 \times 10 = 40$ distinct models. 
Type Ia supernova yields are fixed to \citet{Seitenzahl2013}. 
All yield configurations are listed in Table~\ref{table:yield-comparison-results}.

For the TNG configuration \citep{Pillepich2018}, we adopt the composite prescription described in their Table~2: CC-SN yields from \citet{2006ApJ...653.1145K,portinari} and AGB yields from \citet{2010MNRAS.403.1413K,2014MNRAS.437..195D,2014ApJ...797...44F}. 
The SN\,II yields are renormalized such that IMF-weighted yield ratios at fixed metallicity match those from \citet{2006ApJ...653.1145K}. 
We assume a maximum SN\,II progenitor mass of $100\,M_\odot$, adopt stellar lifetimes from \citet{portinari}, and exclude hypernova channels \citep[in contrast to][]{Philcox_2018}.

\paragraph{Chemical elements:}
Our simulations track nine elements: H, He, C, N, O, Ne, Mg, Si, Fe since these are the only elements traced by the TNG simulations. We principally compare the logarithmic abundances [X/Fe] and [Fe/H] defined by $[\mathrm{X}/\mathrm{Y}] = \log_{10}(N_\mathrm X/N_\mathrm Y)_\mathrm{star} - \log_{10}(N_\mathrm X/N_\mathrm Y)_\odot$  
where $N_\mathrm{X}$ denotes the number fraction of element X and solar abundances are taken from \citet{2009ARA&A..47..481A}. 
Using hydrogen for normalization leaves $n_{\rm el}=8$ independent abundance dimensions\footnote{In observational contexts, it may be more appropriate to compute abundances relative to Mg rather than Fe, as in \citet{2019ApJ...874..102W}, since Mg is only significantly produced by SN\,II and hence a simpler tracer of chemical enrichment.}. 
Although \chem\ tracks additional elements, our observational application (Sec.~\ref{sec:obs_data}) uses only the subset common to both simulations and data.

\subsection{Training and inference setup}
\label{sec:training_setup}

For each of the 40 competing yield configurations, a dedicated \C model is trained.
The six \chem parameters and their priors are listed in Table~\ref{tab:GCEpriors}: two global parameters shared by all stars, the high-mass IMF slope $\alpha_{\rm IMF}$ and the SN$_{\rm{Ia}}$ normalization $\log_{10}N_{\rm Ia}$; three local ISM parameters specific to each star's birth environment, the star formation efficiency $\log_{10}({\rm SFE})$, the peak of the star formation rate $\log_{10}({\rm SFR}_{\rm peak})$ and the outflow fraction $x_{\rm out}$; and the stellar birth time $T$.
Each training dataset consists of $10^6$ simulated pairs $(\theta,\mathbf{x})$, with an additional $10^5$ pairs reserved for validation.
Parameters are drawn from uniform priors spanning $\pm 5$ times the Gaussian prior widths reported in Table~1 of \citet{buck2025inferringgalacticparameterschemical}.
This broad sampling ensures coverage well beyond the prior mean and mitigates training bias.
Note, this uniform distribution is the training proposal; the Gaussian priors of Table~\ref{tab:GCEpriors} are the inference priors and enter only when single-star posteriors are combined (Eqs.~\ref{eq:factorized-posterior}--\ref{eq:post-mean}).

To increase robustness against measurement uncertainties, simulated abundances are perturbed with Gaussian noise of standard deviation $0.05$\,dex in every abundance. This exceeds the formal precision of the observational dataset (typically $0.01$--$0.03$\,dex, Sec.~\ref{sec:obs_data}), ensuring that the learned density estimator remains stable across a realistic error regime.

Hyperparameters of the neural architecture were optimized using an \texttt{optuna} study (Trial 275; Table~\ref{table:optuna-results}; Appendix~\ref{sec:optuna}). 
The inference settings for the reverse SDE solver were adapted to the specific task to balance accuracy and computational cost (see appendix~\ref{sec:Results.Calibration.Time} for more details).
For large-scale model comparison, 50 diffusion steps are used, yielding $\sim0.1$\,s per likelihood evaluation (in the controlled comparison of Sec.~\ref{sec:Results.Yields}, $50$ steps suffice to identify the generating model in all three cases).
For the final parameter inference on the \citet{Nissen_2020} dataset using the best-performing models, the number of diffusion steps was increased to $1000$. 
Given the relatively small sample size ($69$ stars), this maximized the accuracy of the final parameter constraints.
For all benchmark tests on mock data shown in Figures \ref{fig:sbi-chempy-TNG}, \ref{fig:sbi-chempy-alt} and \ref{fig:sbi-chempy-TNG-sim}, the default configuration of $500$ diffusion steps was used to ensure a consistent and fair comparison with the default settings of the baseline methods. 
All inference employs the first-order DPM-Solver unless otherwise stated.

\subsection{Observational data}
\label{sec:obs_data}

For the scientific application, we use the high-precision stellar abundance dataset of \citet{Nissen_2020}. 
The sample comprises 69 nearby solar-type stars\footnote{Of the 72 stars in the catalog we exclude one with a negative age and two without an [O/H] measurement.} observed with the HARPS spectrograph, providing some of the most precise abundance measurements available for Milky Way disk stars.

The stars have solar-like atmospheric parameters ($T_{\rm eff}=5777\pm180$\,K, $\log g > 4.15$) and metallicities in the range $-0.3 \lesssim \mathrm{[Fe/H]} \lesssim 0.3$. 
They span ages from 0 to 11.3\,Gyr with typical uncertainties of $0.5$–$1.3$\,Gyr (approximately $8$–$10\%$). 
These stars are well-sampled from both the thin and thick (or young and low-$\alpha$ as well as old and high-$\alpha$) stellar disk of the Milky Way in terms of age and metallicity.

The defining strength of this dataset is its abundance precision. 
For elements such as C, O, Mg, Si, and Fe, quoted $1\sigma$ uncertainties are typically $\sim0.01$\,dex and always below $0.03$\,dex. 
This precision enables sensitivity to subtle abundance differences induced by variations in nucleosynthetic yields or global galactic parameters, making the sample well-suited for discriminating between competing enrichment models.

\paragraph{Matching observables to model outputs}

\begin{figure*}
    \centering
    \includegraphics[width=.33\linewidth]{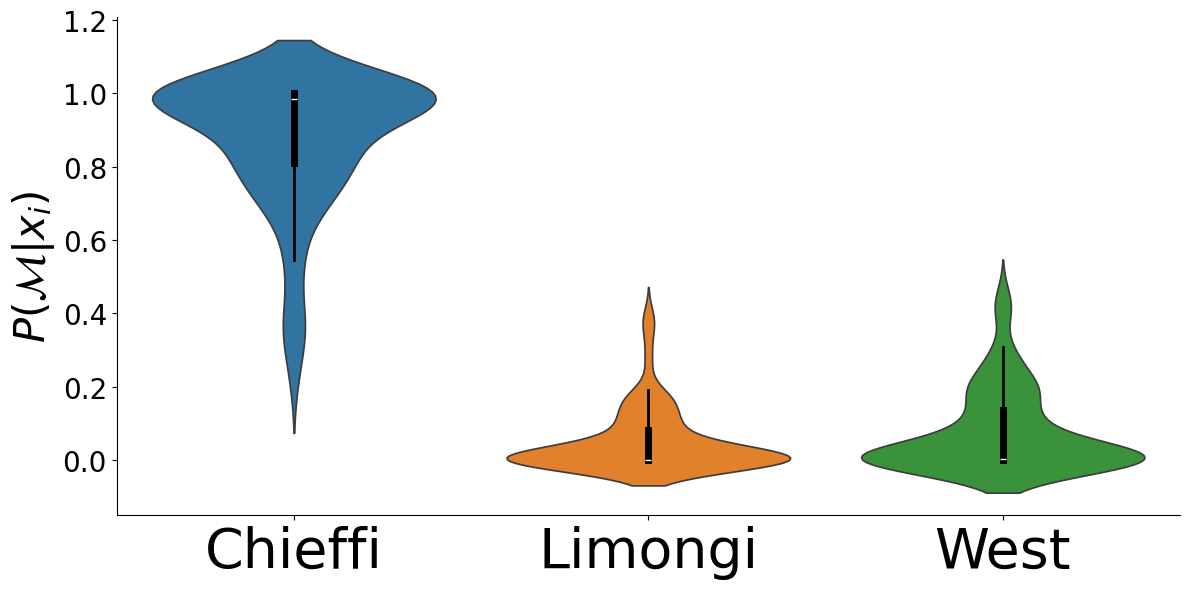}
    \includegraphics[width=.33\linewidth]{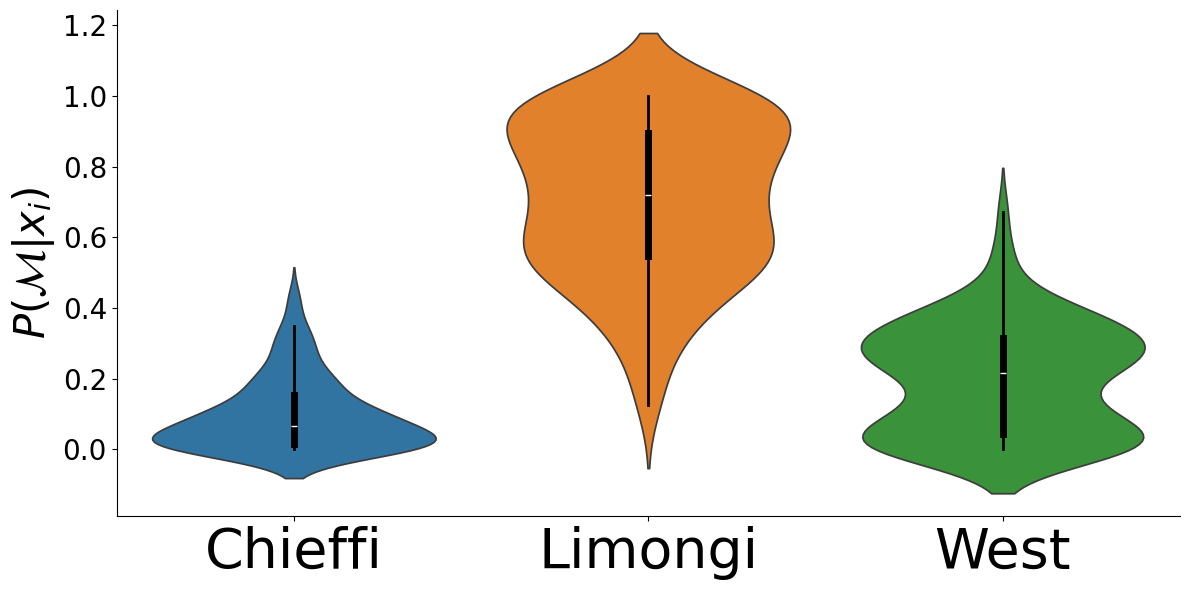}
    \includegraphics[width=.33\linewidth]{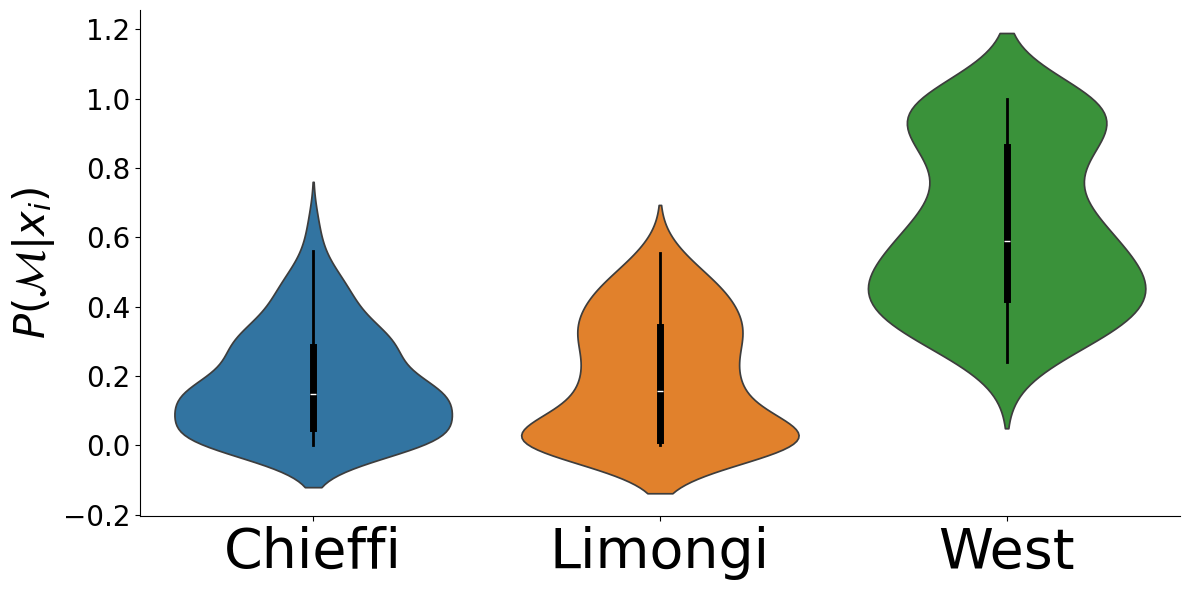}
    \vspace{-1em}
\caption{Violin plots of model weights.
Violin plots show the distribution of single-observation relative model weights (Eq.~\ref{eq:model-posterior}) $P(\mathcal{M}_k|x_i)$ for three competing CC-SN yield models (Chieffi, Limongi and 
West; see text for more details), when tested on mock \chem data.
Each panel corresponds to a different ground-truth model used to generate the mock observations: (Left) Data generated with \citet{Chieffi04} yields; (Middle) \citet{Limongi_2018} yields and (Right) \citet{West17} yields.
The framework correctly identifies the true generating model with high probability in most individual observations, even with only $50$ diffusion timesteps (see Sec.~\ref{sec:Results.Calibration.Time} for more details).
} \label{fig:MTf-tests}
\end{figure*}

Although \chem tracks nine elements, not all are measured in \citet{Nissen_2020}. 
We therefore restrict the analysis to the subset common to both simulations and observations: C, O, Mg, Si, and Fe. 
Elements tracked in simulations but absent in the observational dataset are treated as latent variables within the \C framework. 
The transformer-based architecture naturally handles such missing dimensions via its attention masking mechanism (Sec.~\ref{sec:COMPASS.Inference}), enabling consistent inference without retraining for each observational subset. 
In addition to abundances, stellar ages are incorporated as conditioning variables, i.e. the stellar age fixes each star's birth time $T$, which would otherwise be an additional nuisance parameter.

The \citet{Nissen_2020} sample serves as a high-quality testbed to demonstrate joint parameter inference and model comparison in a realistic astrophysical setting. 
While the sample size of 69 stars does not permit definitive constraints on global Milky Way parameters, its exceptional abundance precision and well-characterized uncertainties provide a stringent proof-of-concept benchmark. 
In the next Section~\ref{sec:results}, we first rank the 40 competing yield combinations using Bayesian model comparison. 
The best-performing prescriptions are then used to infer the high-mass IMF slope and Type Ia supernova normalization.

\section{Results}
\label{sec:results}

Our results are structured in three main parts progressing from model validation and astrophysical inference to an investigation of the model's internal reasoning process.
First, we validate the ability of \C to discriminate between competing nucleosynthetic yield prescriptions (Sec.~\ref{sec:Results.Yields}) using controlled mock experiments. 
Second, we apply the framework to high-precision stellar abundance data from \citet{Nissen_2020} to identify the yield sets most consistent with Milky Way disk stars. 
These preferred prescriptions are then fixed for subsequent inference of global galactic parameters (Sec.~\ref{sec:Results.Params}). 
Finally, we examine the internal attention patterns of the model to assess interpretability (Sec.~\ref{sec:Results.Interpretability}). This final section details the patterns of information flow learned by the model, providing the basis for a deeper physical interpretation.

\subsection{Bayesian Comparison of Nucleosynthetic Yield Sets} \label{sec:Results.Yields}

As demonstrated by e.g. \citet{Buck2021}, the choice of nucleosynthetic yield sets has a significant impact on the chemical abundance patterns of simulated galaxies, making it a critical axis of uncertainty in GCE modeling.
Because different prescriptions imprint distinct abundance patterns, yield comparison provides a stringent test of simulation-based model selection.
We first assess the robustness of the comparison pipeline on mock data, where the ground truth is known, before applying it to real stellar abundances.

\subsubsection{\chem mock data}

To validate model identification, we construct a controlled experiment using three CC-SN yield prescriptions:
\citet{Chieffi04} (Chieffi), \citet{Limongi_2018} (Limongi), and \citet{West17} (West), as these sets exhibit significant variations in their predicted yields.
SN,Ia and AGB yields are fixed to \citet{Seitenzahl2013} and \citet{Karakas2016}, respectively.
Each CC-SN prescription is used in turn to generate an independent mock dataset.
\C is then tasked with identifying the true generating model among the three competitors.

\paragraph{Single-observation posteriors}
Figure~\ref{fig:MTf-tests} shows the distribution of single-star model posteriors $P(\mathcal{M}_k \mid x_i)$ for individual mock observations $x_i$, where $\mathcal{M}_k$ represents one of the three competing CC-SN yield models.
The left panel shows the relative model weight $P(\mathcal{M}|x_i)$ for data created with the \citet{Chieffi04} CC-SN yields.
The middle panel shows $P(\mathcal{M}|x_i)$ for data created with the \citet{Limongi_2018} CC-SN yields.
And the right panel shows $P(\mathcal{M}|x_i)$ for data created with the \citet{West17} CC-SN yields.
For all three ground-truth cases, the correct model is assigned the highest weight for the vast majority of mock observations.
Importantly, these probabilities are normalized across the candidate model set via Eq.~\ref{eq:model-posterior} and therefore lie in $[0,1]$.

\begin{figure}
    \centering
    \includegraphics[width=1\linewidth]{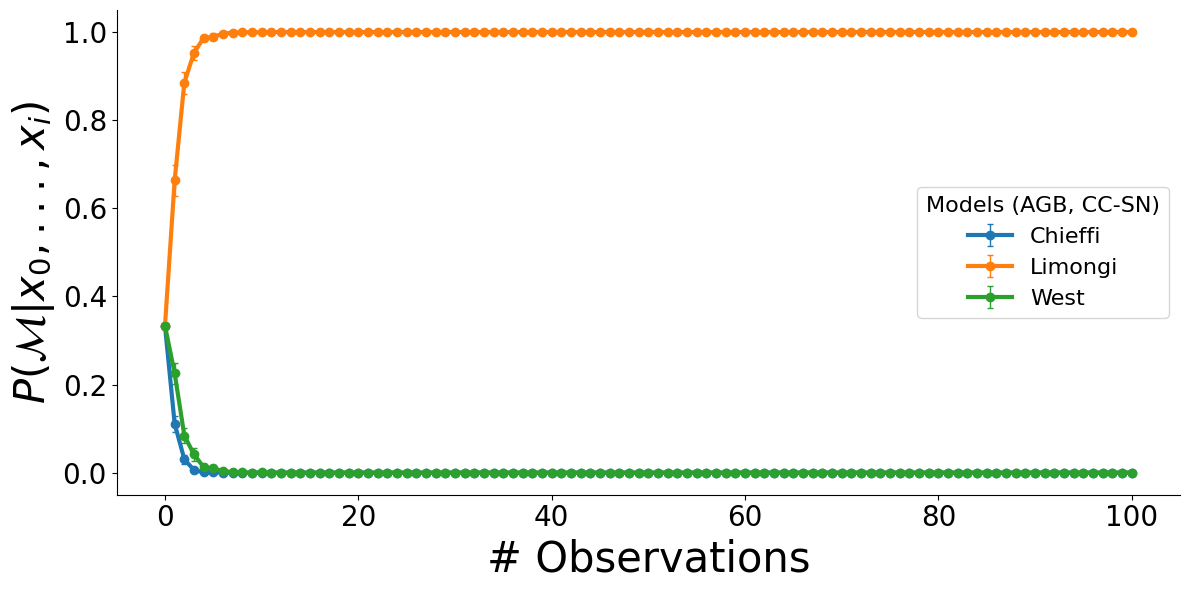}
    \vspace{-1em}
\caption{Cumulative relative model weight $P(\mathcal{M}_k|x_0,\dots,x_i)$ as a function of the number of combined observations $i$.
This example illustrates the case where the mock data was generated using the \citet{Limongi_2018} CC-SN yields corresponding to the middle panel of Fig. \ref{fig:MTf-tests}. 
The probability rapidly converges to $100\%$ for the correct model (Limongi) after only $10$ observations, demonstrating the increased confidence gained by combining evidence from multiple data points.}
\label{fig:MTf-test-updated}
\end{figure}

\paragraph{Combined evidence}
While individual observations already favor the correct prescription, combining evidence across stars (see Sec. \ref{sec:COMPASS.Model-Comp}) rapidly increases confidence.
Figure~\ref{fig:MTf-test-updated} shows the cumulated relative model posterior after combining $i$ observations $ P(\mathcal{M}_k \mid x_1,\dots,x_i) $, obtained by multiplying individual likelihood contributions (Eq.~\ref{eq:combined-likelihood}).

Figure~\ref{fig:MTf-test-updated} shows an example for the inference results from the middle panel of Fig.~\ref{fig:MTf-tests}, where the model posterior for data generated with CC-SN yields from \citet{Limongi_2018} are inferred.
For this mock example, the correct model reaches a near-unity weight after only $\sim 10$ observations.
This experiment demonstrates that
(i) in this controlled setting the likelihood estimator is accurate enough to identify the generating model, and (ii)
model identification is data-efficient.

\subsubsection{Nissen Milky Way Data} \label{sec:Results.Yields.Nissen}

\begin{figure}
    \centering
    \includegraphics[width=1\linewidth]{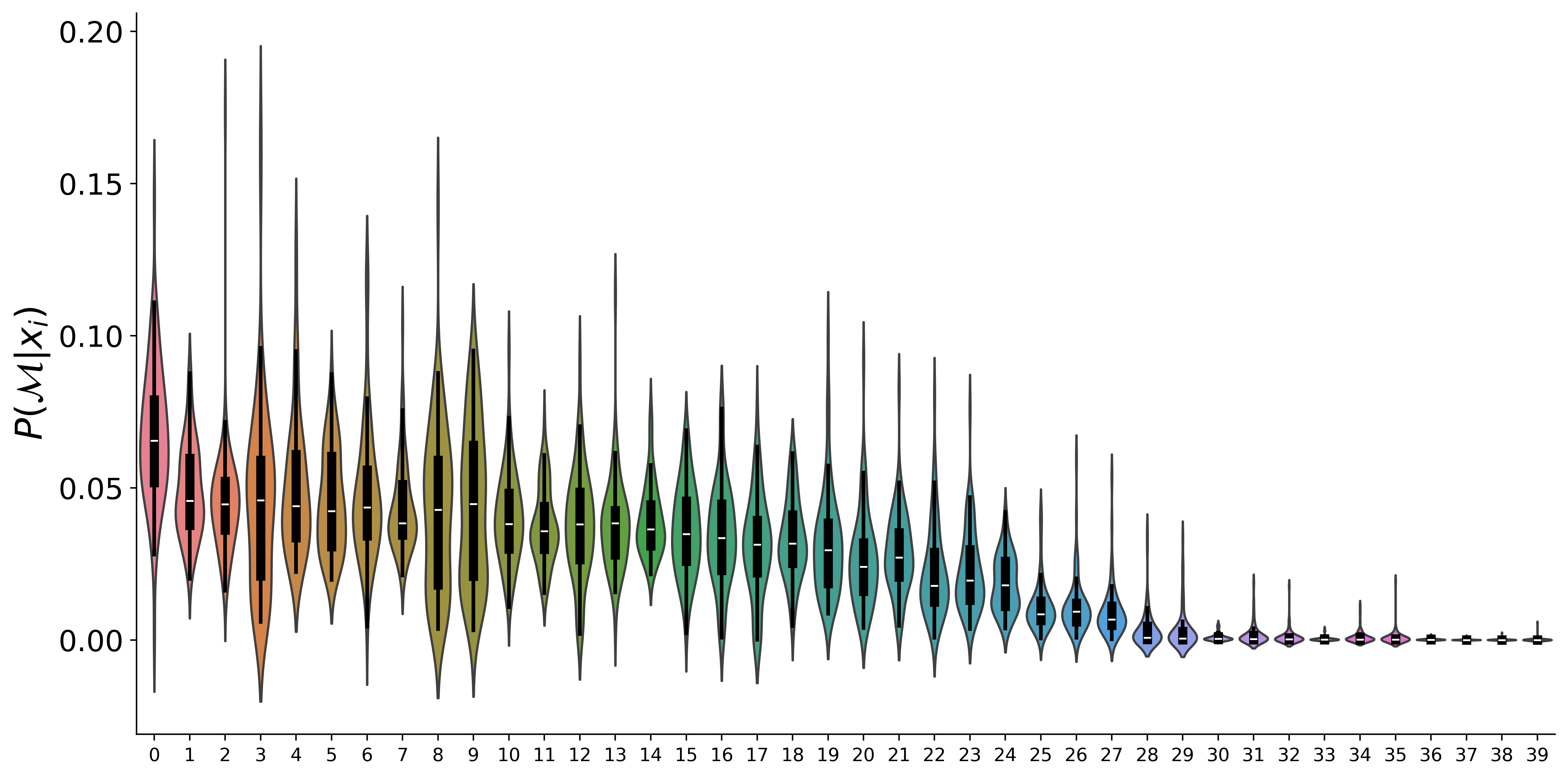}
    \includegraphics[width=1\linewidth]{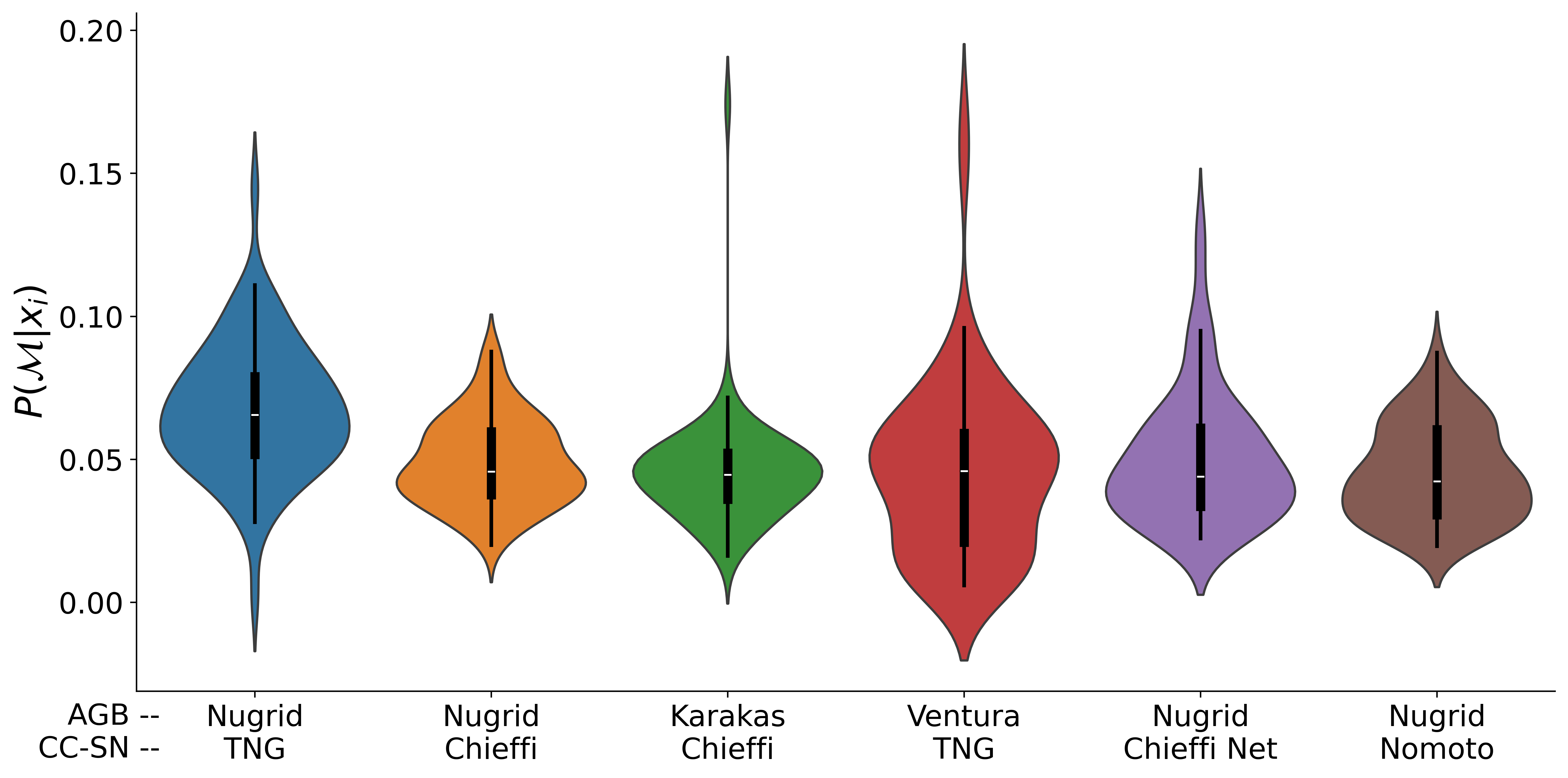}
    \includegraphics[width=1\linewidth]{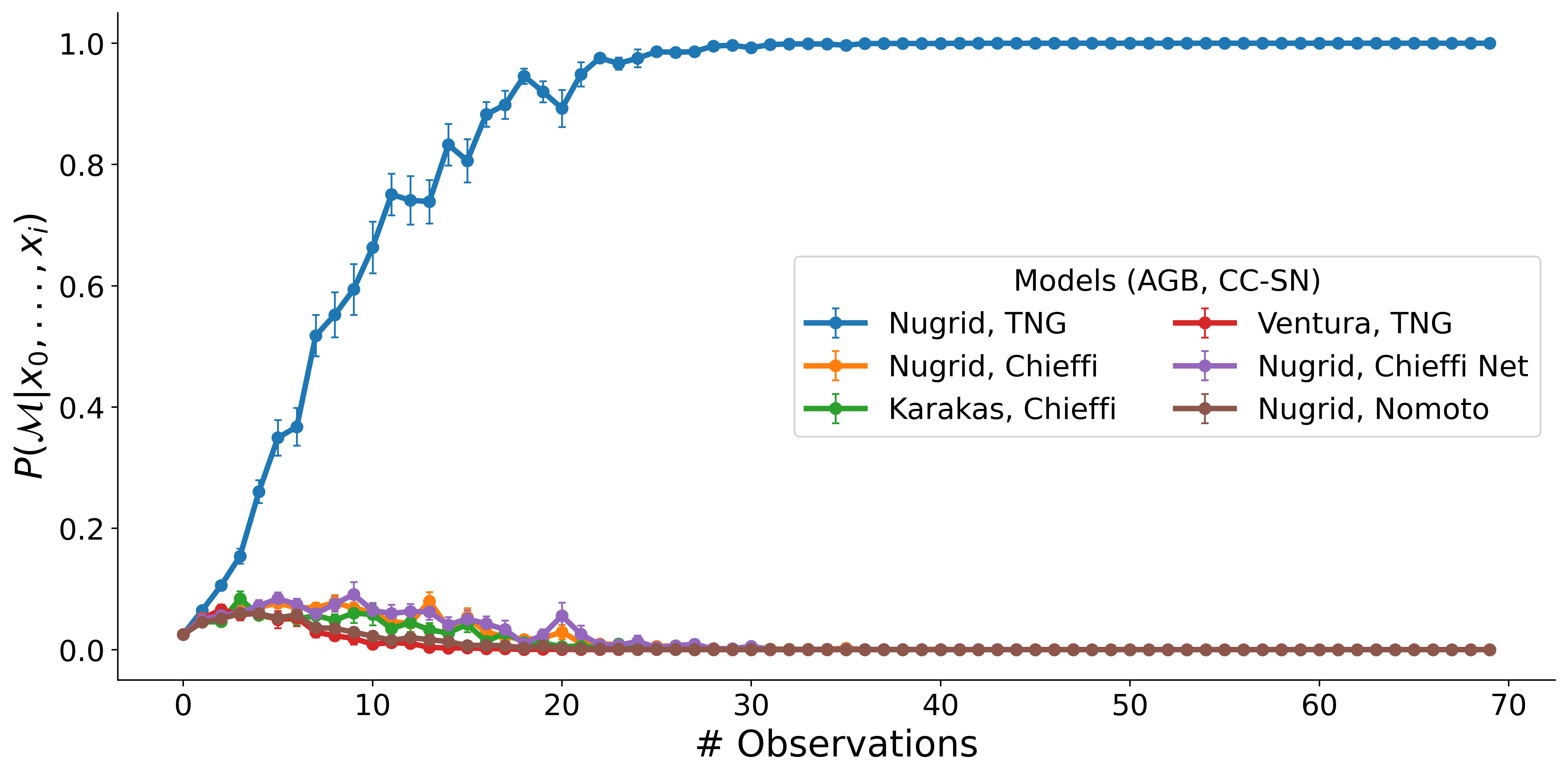}
    \vspace{-1em}
\caption{Observational yield-set inference.
Bayesian comparison of $40$ nucleosynthetic yield set combinations using observational data from \citet{Nissen_2020}. 
(Top) Single-star relative model weights for all $40$ tested combinations.
(Middle) Top six single-star model weights $P(\mathcal{M}_k|x_i)$.
(Bottom) Cumulative model weight $P(\mathcal{M}_k|x_0,\dots,x_i)$.}
\label{fig:MTf-nissen}
\end{figure}

\begin{table*}
    \centering
    \caption{Ranking of the nucleosynthetic yield-set combinations for the \citet{Nissen_2020} data.} \label{table:yield-comparison-results}
    \resizebox{\textwidth}{!}{%
    \begin{tabular}{l c c c c c c c c c c}
        \toprule
        AGB Yields & \multicolumn{10}{c}{Core-Collapse Supernovae (CC-SN) Yields} \\
        \cmidrule(lr){2-11}
        & Chieffi\tablefootmark{a} & Nomoto\tablefootmark{b} & Portinari\tablefootmark{c} & Chieffi Net & Nomoto Net & NuGrid\tablefootmark{d} & West\tablefootmark{e} & TNG\tablefootmark{f} & CL18\tablefootmark{g} & Frischknecht\tablefootmark{h} \\
        \midrule
        NuGrid\tablefootmark{d} & 1 & 6  & 29 & 4  & 5  & 22 & 19 & 0 & 30 & 36 \\
        Karakas\tablefootmark{i} & 2 & 18 & 31 & 14 & 21 & 26 & 23 & 9  & 35 & 39 \\
        Ventura\tablefootmark{j} & 10 & 16 & 34 & 13 & 17 & 27 & 20 & 3  & 32 & 38 \\
        TNG\tablefootmark{f} & 7  & 15 & 28 & 11 & 12 & 25 & 24 & 8  & 33 & 37 \\
        \bottomrule
    \end{tabular}}
    \tablefoot{Results of the Bayesian model comparison (Sec.~\ref{sec:Results.Yields.Nissen}). Lower ranks indicate higher relative model weight and correspond to the numbers in Fig.~\ref{fig:MTf-nissen}.
    \tablefoottext{a}{\citet{Chieffi04}}; \tablefoottext{b}{\citet{2013ARA&A..51..457N}}; \tablefoottext{c}{\citet{portinari1997galacticchemicalenrichmentnew}}; \tablefoottext{d}{\citet{Ritter_2018}}; \tablefoottext{e}{\citet{West17}}; \tablefoottext{f}{\citet{Pillepich2018}}; \tablefoottext{g}{\citet{Limongi_2018}}; \tablefoottext{h}{\citet{Frischknecht_2015}}; \tablefoottext{i}{\citet{Karakas2016}}; \tablefoottext{j}{\citet{Ventura13}}.
    }
\end{table*}

Having validated the methodology, we apply \C to the $69$ high-precision solar-type stars from \citet{Nissen_2020}. The objective is to determine which combination(s) of theoretical yields best reproduce the observed abundance patterns in the \citet{Nissen_2020} sample, according to the relative model posterior $P(\mathcal{M}_k|x)$.
All $40$ combinations of AGB and CC-SN yield prescriptions are evaluated.

\paragraph{Star-by-star preferences}
The top panel of Figure~\ref{fig:MTf-nissen} displays single-star posteriors for all $40$ yield combinations, ranked by their median posterior probability (see Tab.~\ref{table:yield-comparison-results} for all combinations and their numbering).
The middle panel highlights the six most competitive models. 
Here we find that individual stars exhibit moderate discrimination power: several yield combinations receive non-negligible posterior support, reflecting intrinsic degeneracies in abundance space.

\paragraph{Combined constraints}
The bottom panel of Fig.~\ref{fig:MTf-nissen} shows the cumulative model posterior as stars are sequentially combined. Note, the individual stars are not ordered in any particular way. We randomly combine the model posteriors according to Equ.~\ref{eq:factorized-posterior}. Hence, the probability is not increasing monotonically but can also go down if we add in a star that shows an abundance pattern that is less likely to originate from this yield set combination.
As expected from Bayesian updating, evidence strengthens with increasing sample size.

From Fig.~\ref{fig:MTf-nissen} a clear preference emerges:
The combination of NuGrid AGB yields \citep{Ritter_2018} and the TNG CC-SN prescription \citep{Pillepich2018,Kobayashi_2006, portinari1997galacticchemicalenrichmentnew} attains the highest cumulative posterior probability, approaching unity when all $69$ stars are included.
Other combinations that perform competitively on individual stars are strongly disfavored once the full dataset is considered.
Table~\ref{table:yield-comparison-results} summarizes the ranking of all $40$ prescriptions.
As an independent check that does not use the \C likelihood, we trained a neural classifier to identify the generating yield set from the same observables (Appendix~\ref{sec:classifier}). It also prefers the NuGrid + TNG combination by a large margin, and its ranking correlates with Table~\ref{table:yield-comparison-results} (Spearman $\rho = 0.89$; 7 of the top 10 in common). The ordering below the top few models is therefore less robust than the identification of the preferred yield set. In particular, the classifier ranks the TNG CC-SN combinations higher than Table~\ref{table:yield-comparison-results} does.

\paragraph{Interpretation}

Our findings highlight three things:
(i) High-precision abundance data contain sufficient information to discriminate between nucleosynthetic yield models. (ii) The ranking obtained with simulation-based likelihood estimation identifies a preferred yield set that is confirmed by an independent classifier (Appendix~\ref{sec:classifier}).
(iii) The \citet{Nissen_2020} sample strongly favors a specific combination of AGB and CC-SN yields that is consistent with the chemical makeup of the observed stellar populations.
The preferred yield sets identified here will be used for the subsequent parameter inference in Section~\ref{sec:Results.Params}.

\begin{figure}
    \centering
    \includegraphics[width=0.49\linewidth]{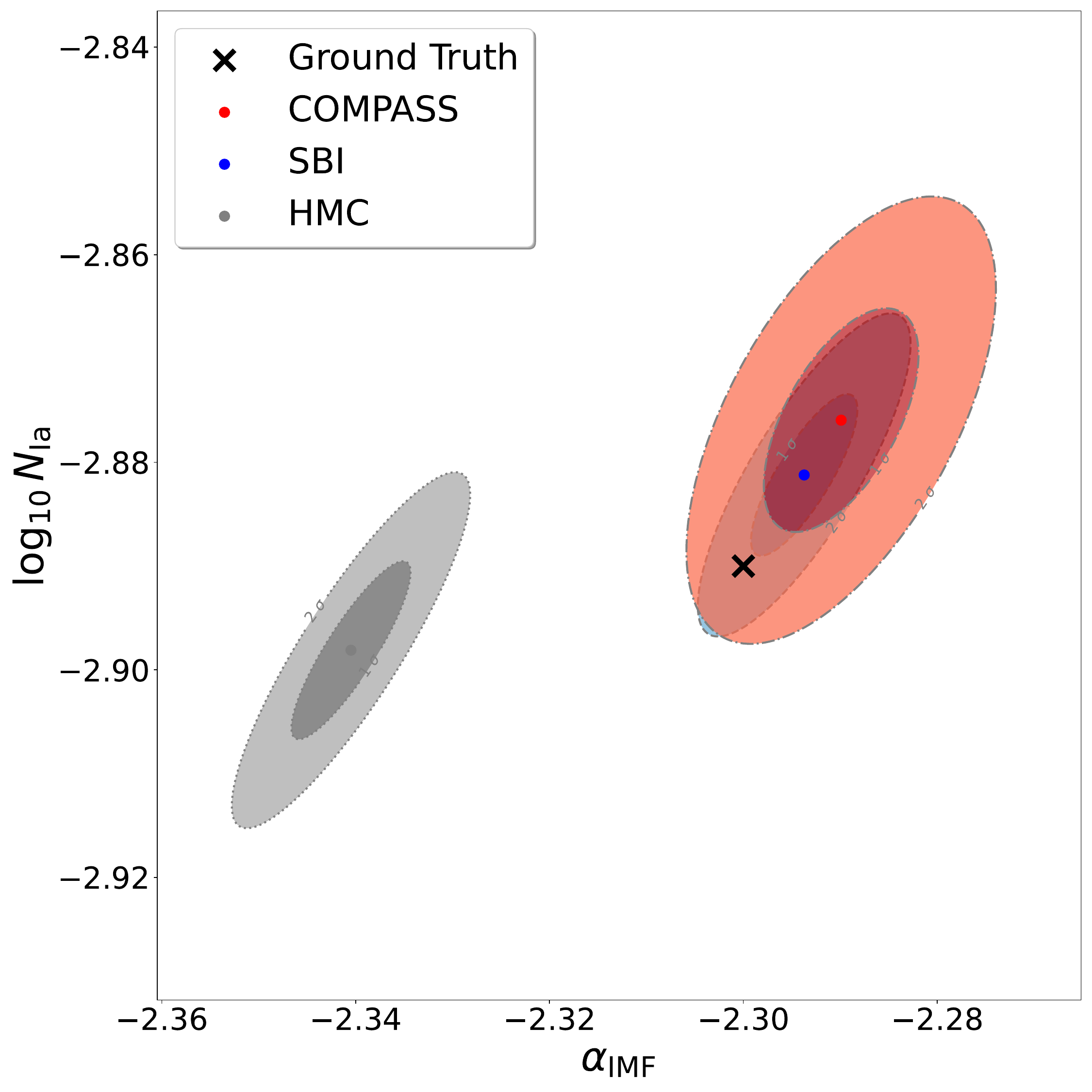}
    \includegraphics[width=0.49\linewidth]{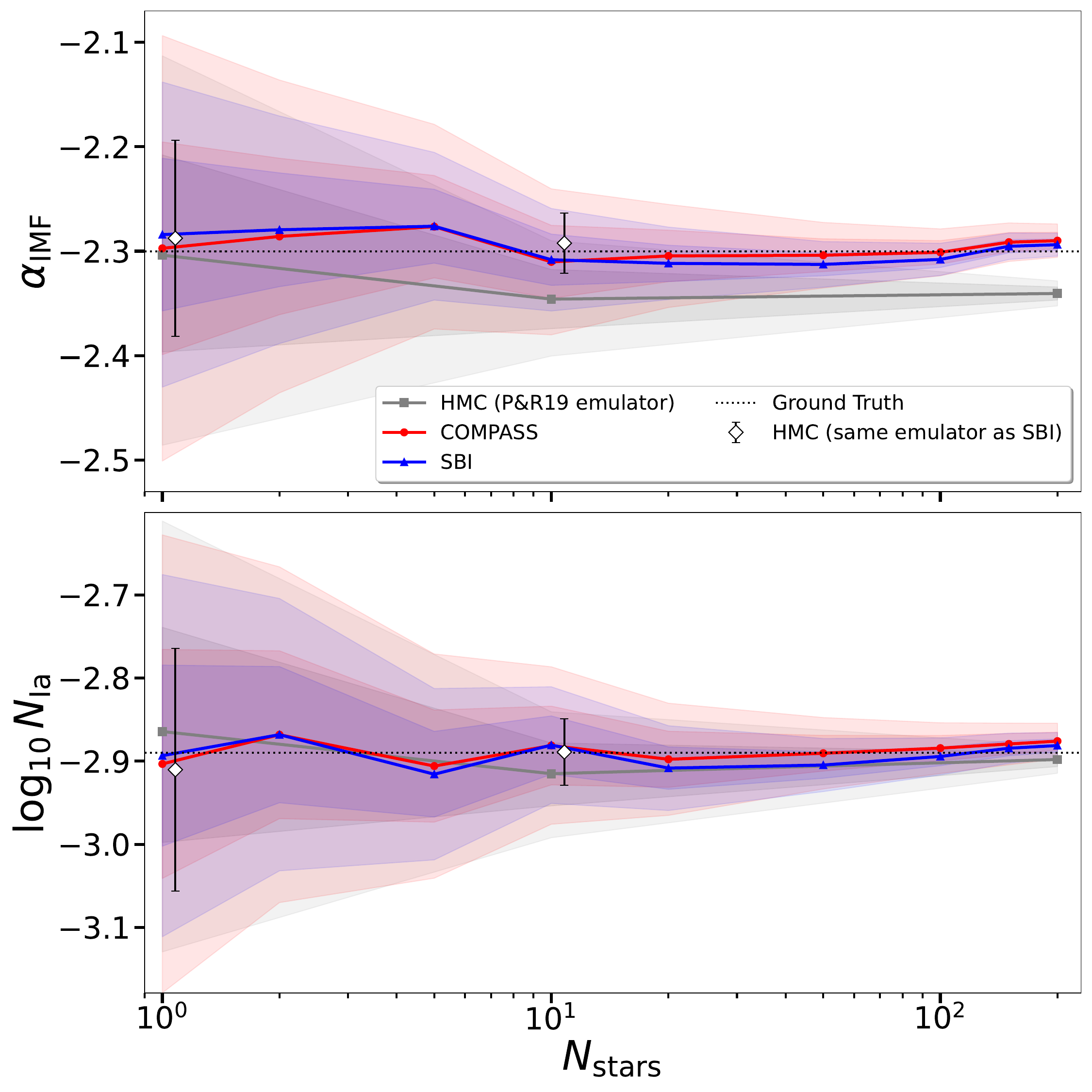}
\caption{Parameter inference with matched TNG yields.
Comparison of inference results for the global galactic parameters -- the initial mass function's high-mass slope $\alpha_{\mathrm{IMF}}$ and the Type Ia supernova normalization $\log_{10}N_{\mathrm{Ia}}$ -- using \C (red), SBI (blue) and HMC (gray) on mock data generated with TNG yields.
(Left) Joint posterior distribution inferred from $200$ stars with contours showing $1\sigma$ and $2\sigma$ ellipses. For \C and SBI we show combined single-star posteriors (Eqs.~\ref{eq:factorized-posterior}--\ref{eq:post-mean}). For HMC we show results of the hierarchical HMC of \citet{Philcox_2019} on the same stars, using their original emulator, which was trained on a different TNG yield compilation (Sec.~\ref{sec:Results.Params.Chempy}).
(Right) Convergence of the inferred parameters as a function of $N_{\rm stars}$, with $1\sigma$ and $2\sigma$ bands. Open diamonds show HMC results with the emulator of \citet{buck2025inferringgalacticparameterschemical}, which is trained on the same yields as SBI and \C\unskip, and are only shown where its chains converged.
}
\label{fig:sbi-chempy-TNG}
\end{figure}

\subsection{Inferring Galactic Parameters} \label{sec:Results.Params}
Having identified the yield prescriptions most consistent with the data (Sec.~\ref{sec:Results.Yields}) and assessed the calibration of the score model (Appendix~\ref{sec:Results.Calibration}), we now turn to the primary scientific objective: inferring the global galactic parameters governing chemical enrichment.
We focus on two key quantities:
(i) the high-mass slope of the IMF, $\alpha_{\mathrm{IMF}}$, and 
(ii) the normalization of the SN\,Ia rate, $\log_{10}N_{\mathrm{Ia}}$.

The following subsections we first benchmark \C against established inference pipelines from our previous work  \citep{buck2025inferringgalacticparameterschemical} and traditional Hamiltonian Monte Carlo (HMC) methods from \cite{Philcox_2019} using controlled \chem mock data. This validation includes tests with matched yield sets and evaluations under conditions of model misspecification, such as using alternative yield sets or data from the more complex IllustrisTNG cosmological simulation \citep{Pillepich2018} and tests how accurately \C recovers known ground-truth parameters.
We then apply the validated \C framework to the \citet{Nissen_2020} dataset employing the optimal yield set combinations identified in Section \ref{sec:Results.Yields.Nissen}, to derive constraints (conditional on the yield set) on $\alpha_{\mathrm{IMF}}$ and $\log_{10}N_{\mathrm{Ia}}$ for the local Milky Way.

\subsubsection{Validation on \chem mock data} \label{sec:Results.Params.Chempy}
\paragraph{Matched yield configuration}
For the initial validation, mock data are generated with \chem using the IllustrisTNG yield tables matching the training setup.
The ground-truth parameters are set to $\alpha_{\mathrm{IMF}} = -2.3$ and $\log_{10}N_{\mathrm{Ia}}=-2.89$.
Figure \ref{fig:sbi-chempy-TNG} compares the inference results obtained with \C against those from \cite{buck2025inferringgalacticparameterschemical} (hereafter "SBI") and a reimplementation of the hierarchical HMC model of \citet{Philcox_2019} applied to the same mock stars (hereafter "HMC"), when applied to this mock dataset.

Figure~\ref{fig:sbi-chempy-TNG} shows that both \C and SBI recover the ground-truth parameters, with posteriors that contract as $N_{\rm stars}$ increases. With $200$ stars, \C gives $\alpha_{\rm IMF} = -2.290 \pm 0.008$ and $\log_{10}N_{\rm Ia} = -2.876 \pm 0.011$ (relative deviations $\leq 0.5\%$). The truth lies $1.3\sigma$ from the mean in each parameter, with a Mahalanobis distance of $1.4$ (a multivariate measure of separation accounting for parameter uncertainties and correlations), comparable to SBI ($1.1\sigma$; $1.2$). The \C posterior is $1.4$--$1.6$ times wider than the SBI posterior. Given the calibration diagnostics in Appendix~\ref{sec:Results.Calibration.Posterior}, which show mildly under-confident \C posteriors for the global parameters, we interpret the larger width as conservative rather than as a more accurate uncertainty estimate.

The right-hand panel shows the combined estimates as a function of the number of stars. For matched yields, both SBI methods remain consistent with the truth while their uncertainties shrink, i.e. no significant bias accumulates up to $200$ stars (cf. Sec.~\ref{sec:COMPASS.Inference.MultiObs}).

As a sampling-based reference we use the hierarchical HMC model of \citet{Philcox_2019}, re-implemented in \texttt{NumPyro} \citep{Phan2019} with the No-U-Turn sampler \citep{Hoffman2014}. It samples the joint posterior of the global parameters, all per-star parameters and a per-element model-error term, and we apply it to exactly the same mock observations. With the original emulator of \citet{Philcox_2019}, which was trained on their TNG yield compilation including hypernova channels and is used with their priors (prior mean of $\log_{10}N_{\rm Ia}$ at $-2.75$), all chains converge ($\hat{R} \leq 1.01$). This is shown as the HMC results in Figs.~\ref{fig:sbi-chempy-TNG}, \ref{fig:sbi-chempy-alt} and \ref{fig:sbi-chempy-TNG-sim}. Because that yield compilation differs from the one used to generate our mocks, HMC with this emulator is itself misspecified on the matched-yield mock and gives $\alpha_{\rm IMF} = -2.340 \pm 0.006$. With the emulator of \citet{buck2025inferringgalacticparameterschemical}, which is trained on the exact same yields as SBI and \C\unskip, HMC runs recover the truth ($N = 10$: $\alpha_{\rm IMF} = -2.29 \pm 0.03$, $\log_{10}N_{\rm Ia} = -2.89 \pm 0.04$). However, we note that those chains, are metastable for $N \gtrsim 100$ on this mock, so we show only converged runs (open diamonds). In contrast, SBI uses the same simulator, yields, training proposal and mock data as \C\unskip.

Due to their amortized nature, \C and the previous SBI implementation can readily scale to larger numbers of observed stars with minimal additional computational overhead while HMC becomes computationally prohibitive for large datasets. 
This scalability is particularly crucial when analysing large datasets, where sample variances with small numbers of stars can otherwise dominate the uncertainty.
This confirms that \C is not only effective for model selection (as shown in the previous subsection \ref{sec:Results.Yields.Nissen}) but also an accurate tool for parameter inference when the model is well-specified.

\paragraph{Robustness to model misspecification}
Given that all tabulated yield sets likely deviate from astrophysical reality to some extent, and clarity on the optimal set is often lacking, it is crucial to test the inference method's performance under model misspecification. In particular, in Appendix~\ref{sec:model-misspecification} we investigate the sensitivity of \C to such misspecification by using mock data generated with yield sets different from those used for training (see Appendix~\ref{sec:model-missspecification-chempy}) and more severe misspecifications through using a completely different simulator model, i.e. a full hydrodynamical model from IllustrisTNG (see Appendix~\ref{sec:model-missspecification-ilustris}).  

\begin{table*}
    \centering
    \caption{Inferred global parameters from the \citet{Nissen_2020} data for the top six yield sets.}
\label{table:nissen-inference-results}
    \resizebox{\textwidth}{!}{%
    \begin{tabular}{lcccccc}
        \toprule
        \thead{AGB \\ CC-SN} & \thead{Nugrid {\cite{Ritter_2018}} \\ TNG {\cite{Pillepich2018}}} & \thead{Nugrid {\cite{Ritter_2018}} \\ Chieffi {\cite{Chieffi04}}} & \thead{Karakas {\cite{Karakas2016}} \\ Chieffi {\cite{Chieffi04}}} & \thead{Ventura {\cite{Ventura13}} \\ TNG {\cite{Pillepich2018}}} & \thead{Nugrid {\cite{Ritter_2018}} \\ Chieffi Net} & \thead{Nugrid {\cite{Ritter_2018}} \\ Nomoto Net {\cite{2013ARA&A..51..457N}}} \\
        \midrule
        $\alpha_{\rm IMF}$ & $-2.65 \pm 0.02$ & $-2.52 \pm 0.02$ & $-2.62 \pm 0.02$ & $-2.80 \pm 0.02$ & $-2.54 \pm 0.02$ & $-2.75 \pm 0.01$ \\
        $\log_{10}N_{\rm Ia}$ & $-2.81 \pm 0.02$ & $-2.68 \pm 0.02$ & $-2.88 \pm 0.03$ & $-3.10 \pm 0.02$ & $-2.70 \pm 0.02$ & $-2.88 \pm 0.02$ \\
        \bottomrule
    \end{tabular}}
    \tablefoot{Mean values and $\pm1\sigma$ uncertainties of $\alpha_{\rm IMF}$ and $\log_{10}(N_{\rm Ia})$ for the top six yield-set combinations identified in Sec.~\ref{sec:Results.Yields.Nissen}.}
\end{table*}

Under misspecification all methods are biased by several posterior standard deviations (Appendix~\ref{sec:model-misspecification}). \C is less biased than the NPE baseline in both tests, but its credible regions are not reliable in this regime. Since all tabulated yield sets are approximations, this motivates the explicit comparison of yield sets in Sec.~\ref{sec:Results.Yields} and caution in interpreting the absolute values of the inferred parameters.

\subsubsection{Inference from the \citet{Nissen_2020} data} \label{sec:Results.Params.Nissen}
\begin{figure}
    \centering
    \includegraphics[width=\linewidth]{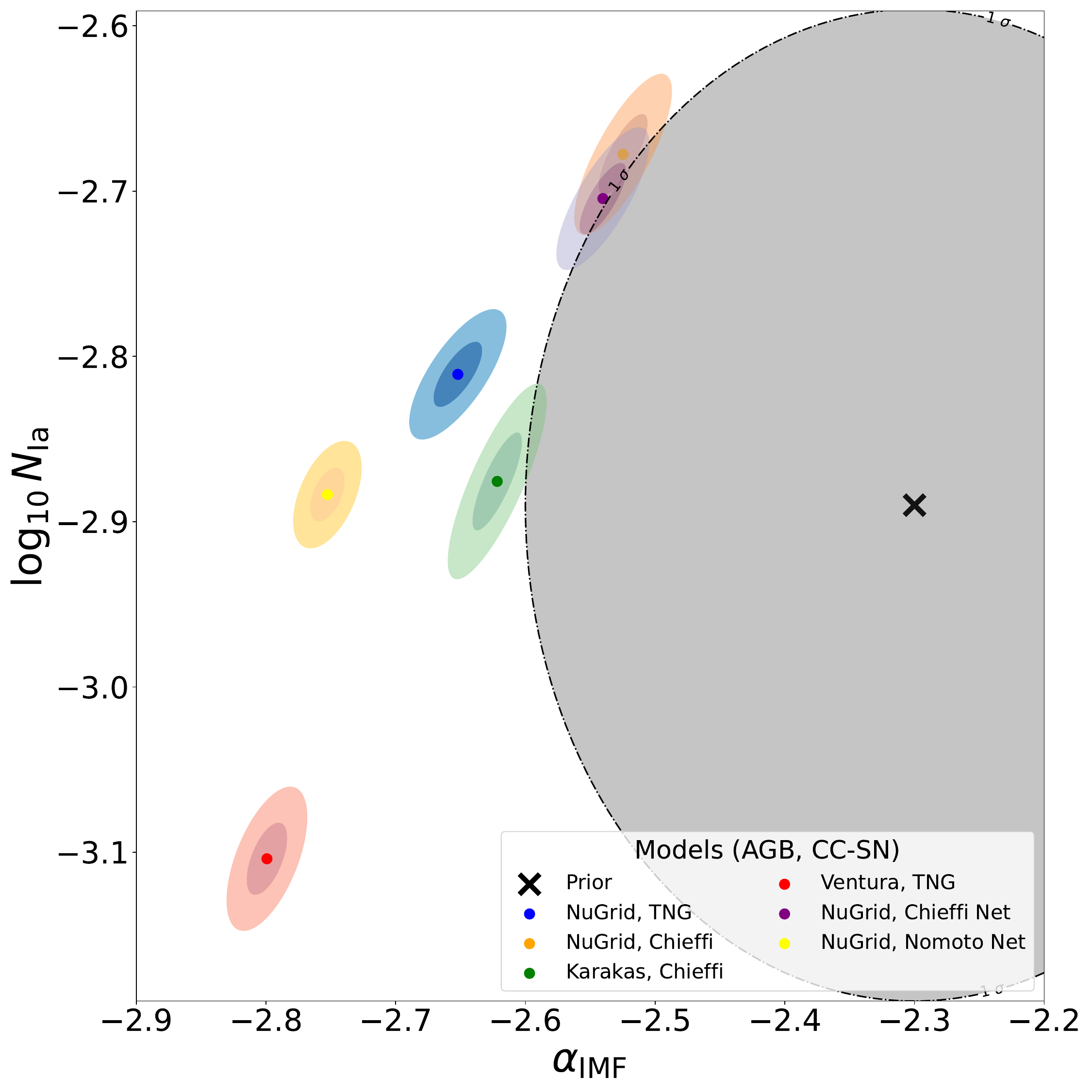}
    \vspace{-1em}
\caption{Inferred galactic parameters from observational data.  
Joint posterior distributions ($1\sigma \text{ and } 2\sigma$ contours) for the IMF high-mass slope $\alpha_{\mathrm{IMF}}$ and SN Ia normalization $\log_{10}N_{\mathrm{Ia}}$, inferred using \C from the \citet{Nissen_2020} data.
Each colored contour set corresponds to one of the top six yield set combinations identified in Section~\ref{sec:Results.Yields.Nissen} (see legend in figure and also Tab.~\ref{table:nissen-inference-results}).
The black 'x' marks the mean of the \chem prior distributions and the dashed line illustrates the $1\sigma$ standard deviation of this prior.}
\label{fig:Nissen-Inference}
\end{figure}

Having validated the accuracy of the \C framework in inferring galactic parameters in the previous subsection~\ref{sec:Results.Params.Chempy},  
we now apply it to the $69$ high-precision solar-type stars from \citet{Nissen_2020}, adopting the six highest-ranked yield combinations from Sec.~\ref{sec:Results.Yields.Nissen} (see Fig. \ref{fig:MTf-nissen}).
To maximize accuracy given the modest sample size of only $69$ stars, $1,000$ diffusion steps are used (acknowledging the findings from Figure \ref{fig:AccuracyVsTime} that inference accuracy still improves with more diffusion timesteps).

A crucial aspect of this analysis is \C's ability to handle observational data limitations. The \C framework naturally accommodates missing elements (He, N, Ne) via attention masking and stellar age information as an additional conditioning variable. 
\citet{Nissen_2020} only measured abundances for C, Fe, Mg, O and Si while our \chem runs additionally modeled He, N and Ne (upon which the SBI models were trained). 
\C's architecture (as described in Sec. \ref{sec:COMPASS.Inference}) allows it to process data with missing elements and still provide robust parameter inference for the observed set.
The observed stellar ages ($T_i$) are used as conditioning variables, fixing the birth time of each star.

Figure \ref{fig:Nissen-Inference} visualizes the joint posterior distributions for $\alpha_{\rm IMF}$ and $\log_{10}(N_{\rm Ia})$ for the top six yield set combinations. 
The corresponding mean parameter values and their uncertainties are numerically summarized in Table \ref{table:nissen-inference-results}.

Fig.~\ref{fig:Nissen-Inference} and Tab.~\ref{table:nissen-inference-results} show that the inferred global parameters depend strongly on the adopted yield set: the posteriors of the six highest-ranked combinations are mutually inconsistent at several $\sigma$, with $\alpha_{\rm IMF}$ ranging from $-2.52$ to $-2.80$ and $\log_{10}N_{\rm Ia}$ from $-2.68$ to $-3.10$, while the statistical uncertainties are $\approx 0.02$. The choice of yield set is thus the dominant source of uncertainty on these parameters.
The highest-ranked combination, featuring \citet{Ritter_2018} (NuGrid) AGB yields in combination with the CC-SN yields used in the IllustrisTNG simulation \cite{10.1093/mnras/stx2656} (\citet{Kobayashi_2006, portinari1997galacticchemicalenrichmentnew}; represented by blue contours in Fig.~\ref{fig:Nissen-Inference}), results in $\alpha_{\rm IMF} = -2.65 \pm 0.02$ and $\log_{10}N_{\rm Ia} = -2.81 \pm 0.02$.

All six yield sets favor an IMF slope steeper than the canonical Salpeter value ($-2.35$) and the Chabrier-like prior mean ($-2.3$), consistent with earlier studies of the solar neighborhood \citep[e.g.,][]{Rybizki2015}. The SN$_{\rm{Ia}}$ normalization, in contrast, scatters around the prior mean of $-2.89$ across yield sets.

This comprehensive analysis of the \citet{Nissen_2020} data, leveraging the optimal yield sets identified through Bayesian model comparison and the advanced capabilities of \C to incorporate stellar age constraints and handle incomplete elemental data, thus constrains the IMF slope and SN~Ia normalization conditional on the yield set, with the yield-set choice dominating the uncertainty.

\subsection{Model interpretability} \label{sec:Results.Interpretability}

Having demonstrated the accuracy and robustness of the \C framework for both model comparison and parameter inference, we now address a crucial question: does the model base its conclusions on physically meaningful relationships, or does it exploit spurious correlations in the simulator output?

The transformer architecture underlying \C provides a natural avenue for interpretability through its self-attention mechanism. Attention weights quantify how strongly each parameter query attends to individual observational features during inference. By analysing these attention patterns, we can trace the flow of information between elemental abundances and inferred parameters, thereby probing the model’s internal reasoning process to ensure that the results are not merely the product of a "black box" correlating simulator artifacts, but are grounded in physically meaningful relationships. Our analysis follows recent visualization approaches developed for diffusion transformers \citep[e.g.,][]{yeh2023attentionvizglobalviewtransformer}.

Since attention patterns evolve throughout the denoising trajectory, we inspect them at an intermediate diffusion time ($t=0.5$), where internal representations are known to be particularly structured and informative \citep{helbling2025conceptattentiondiffusiontransformerslearn}. Figures~\ref{fig:attention_avg} and~\ref{fig:attention_layer} in the Appendix present attention weights averaged over the full \citet{Nissen_2020} dataset for the preferred yield-set combination identified in Section~\ref{sec:Results.Yields.Nissen}.

\begin{figure}
    \centering
    \includegraphics[width=1\linewidth]{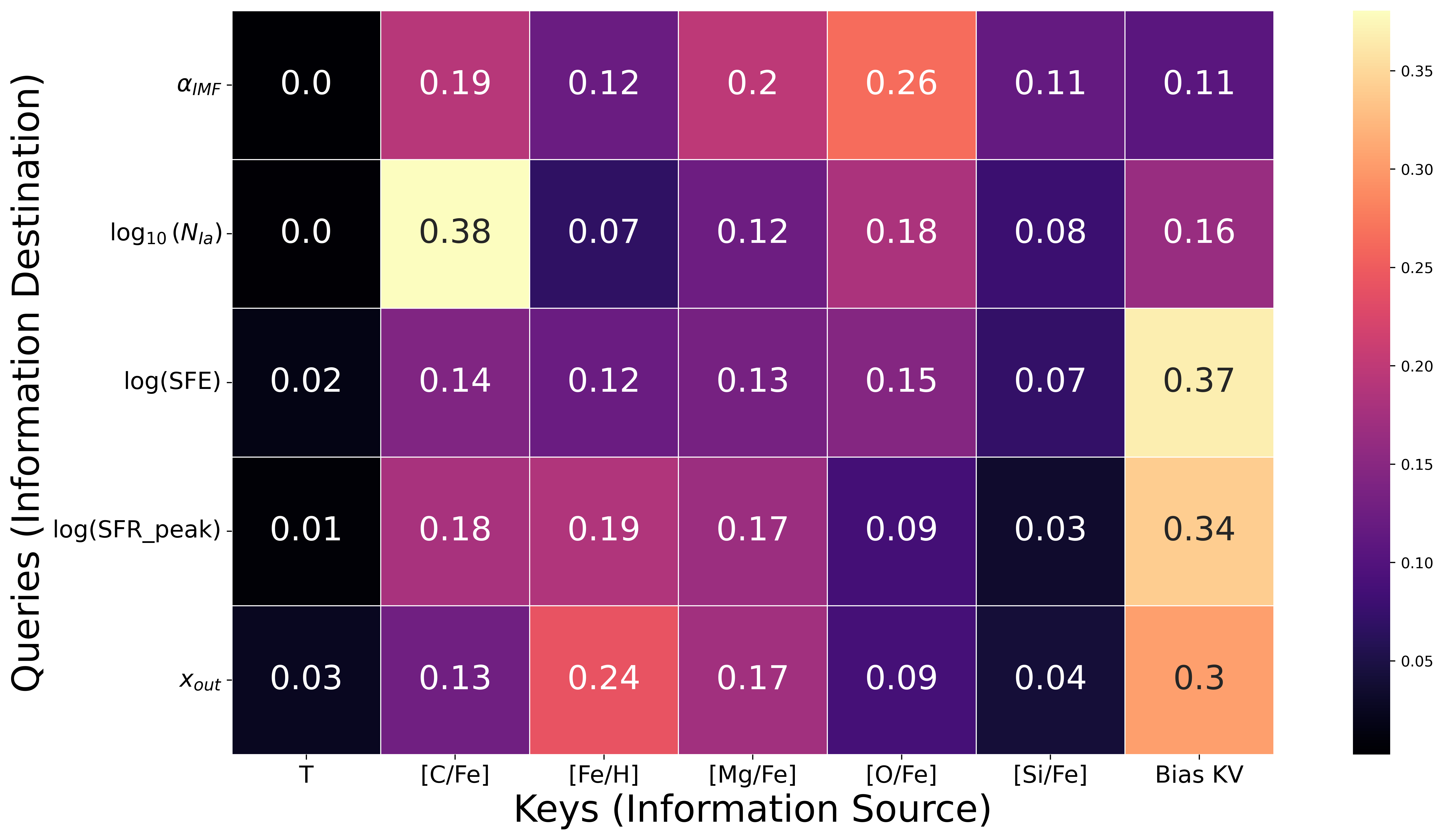}
    \vspace{-1em}
    \caption{Overall layer-averaged attention map. 
This figure summarizes the model's strategy by averaging the attention weights from Figure~\ref{fig:attention_layer} across all five layers. 
It highlights the most dominant information pathways integrated over the model's full depth.}
 \label{fig:attention_avg}
\end{figure}

\paragraph{Global information pathways.}
The layer-averaged attention map (Fig.~\ref{fig:attention_avg}) reveals clear and physically interpretable dependencies. 
The high-mass IMF slope $\alpha_{\rm IMF}$ predominantly attends to $\alpha$-elements such as Oxygen (O) and Magnesium (Mg), with additional contributions from Carbon (C) and Iron (Fe). This behavior is physically expected: massive stars, whose abundance is governed by $\alpha_{\rm IMF}$, are the primary production sites of $\alpha$-elements via core-collapse supernovae. A steeper IMF reduces the number of massive stars, directly lowering O and Mg yields. The model therefore correctly identifies $\alpha$-element abundances as the dominant constraints on $\alpha_{\rm IMF}$.

The normalization of Type Ia supernovae, $\log_{10}(N_{\rm Ia})$, shows its strongest average attention towards Carbon abundance ratios. This dependence is astrophysically meaningful: increasing the SN Ia rate enhances iron production, thereby altering relative abundance ratios such as [C/Fe]. Carbon thus acts as an indirect tracer of delayed iron enrichment, making it a sensitive diagnostic of SN Ia normalization.

Local ISM parameters exhibit a different pattern. Their attention is dominated by the learned bias term, with secondary contributions from Iron (Fe). This suggests that these parameters primarily affect the overall normalization and metal-retention efficiency of the system rather than imprinting sharp element-specific signatures. The sensitivity of $x_{\rm out}$ to Fe is physically plausible, as iron abundances respond strongly to the ISM’s ability to retain or lose enriched material.

Stellar age (T) exhibits only weak attention weights (see Fig.~\ref{fig:attention_layer}), primarily in early layers and mainly for local ISM parameters. This is consistent with the training setup: in \chem global parameters such as $\alpha_{\rm IMF}$ and $\log_{10}(N_{\rm Ia})$ are fixed in time and do not explicitly depend on cosmic epoch. The model therefore correctly refrains from attributing spurious predictive power to stellar age. This behavior provides additional evidence that \C captures genuine physical correlations rather than overfitting temporal structure absent from the simulator.

Overall, the attention analysis demonstrates that \C bases its inferences on physically interpretable relationships between nucleosynthetic channels and elemental abundance patterns. Rather than acting as a `black box', the model reproduces established chemical evolution reasoning within its learned internal representations. Hence, the alignment between learned attention pathways and established nucleosynthetic channels provides independent validation of the physical consistency of the inference framework.

\section{Discussion}
\label{sec:disucssion}

This work establishes conditional score-based diffusion models as a unified and scalable framework for simulation-based inference in GCE. By combining neural posterior estimation and neural likelihood estimation within a single amortized architecture, \C enables simultaneous parameter inference and Bayesian model comparison. Such capability has typically required separate and computationally intensive procedures. \C's ability to simultaneously perform Bayesian model comparison and parameter inference represents a significant methodological advance over previous approaches in GCE modeling.

Our results demonstrate that diffusion-based SBI can deliver conservative posteriors, competitive accuracy, and physically interpretable reasoning within a single coherent framework.

\subsection{Comparison to previous inference frameworks}
\label{sec:comp}

Traditional inference approaches in GCE, including Hamiltonian Monte Carlo \citep{Philcox_2018,Philcox_2019} and likelihood-free neural posterior estimation \citep{buck2025inferringgalacticparameterschemical}, have successfully constrained parameters such as the IMF slope and SN~Ia normalization. However, these techniques are often computationally demanding and limited in scalability when confronted with the large stellar samples of modern surveys.

In contrast, \C performs fully amortized inference: once trained, it evaluates posteriors and likelihoods at negligible additional computational cost. This makes it particularly well-suited for modern spectroscopic surveys containing large stellar samples and extended chemical abundance spaces.

Within the broader simulation-based inference literature \citep{Cranmer_2020,lemos2023samplingbasedaccuracytestingposterior}, \C extends recent diffusion-based approaches \citep{song2020,gloeckler2024allinonesimulationbasedinference} to astrophysical applications. Score-based generative models have been shown to provide improved stability and calibration relative to likelihood-ratio or density-ratio estimators. Importantly, the joint NPE/NLE functionality of \C allows direct model ranking without resorting to expensive nested sampling of evidences, which has limited earlier cosmological SBI studies \citep[e.g.][]{Alsing2018,Hahn2022}, at the price that the resulting weights are relative predictive weights rather than evidence ratios (Sec.~\ref{sec:COMPASS.Model-Comp}).

\subsection{Physical implications and consistency with literature}
Using high-precision elemental abundances from \cite{Nissen_2020}, \C favors a yield combination consisting of NuGrid AGB yields \citep{Ritter_2018} and the CC-SN yields adopted in IllustrisTNG \citep{Pillepich2018}. This preference aligns with previous studies indicating that hybrid prescriptions combining \cite{portinari1997galacticchemicalenrichmentnew} and \cite{Kobayashi_2006} best reproduce Galactic $\alpha$-element trends \citep[e.g.][]{Philcox_2019,buck2025inferringgalacticparameterschemical,Guenes2025}.

The inferred IMF slope, $\alpha_{\rm IMF} = -2.65 \pm 0.02$ for the preferred yield set ($-2.52$ to $-2.80$ across the best-ranked sets), is steeper than the canonical Salpeter value. Such a mildly bottom-heavy IMF is consistent with recent Galactic Archaeology analyses in the solar neighborhood \citep{Rybizki_2017,Li2023,buck2025inferringgalacticparameterschemical} and with evidence for IMF variations in external galaxies \citep{Martin2019,denBrok2024,Wainer2024}. 
Similarly, the inferred SN~Ia normalization ($\log N_{\rm Ia} \approx -2.81$ for the preferred yield set) is consistent with relatively efficient delayed iron enrichment, in agreement with the higher SN~Ia rates proposed by \cite{Rybizki2015}. Together, these findings reinforce a coherent picture in which efficient iron production and a moderately bottom-heavy IMF are required to reproduce the observed [$\alpha$/Fe]–[Fe/H] trends in the local Galactic disk.

\subsection{Current limitations and avenues for improvement}
Despite its strengths, several limitations highlight directions for future development.

First, the underlying \chem one-zone model neglects radial migration and time-dependent star-formation efficiencies. Such effects become increasingly important when analysing spatially resolved or cosmological datasets \citep[e.g.][]{Andrews2017}. Extending \C to multi-zone frameworks \citep[e.g. VICE;][]{Johnson2021} or hydrodynamical simulations \citep[e.g.][]{Buck2021,Buck2020c} will enable more realistic modeling of chemo-dynamical evolution.

Second, stellar yields are treated as discrete table choices. In reality, yields depend continuously on stellar physics such as rotation, mass loss, explosion energy, and metallicity \citep{Cote2017,Gronow2021}. Future extensions could interpolate across yield grids or introduce parametric yield uncertainties, potentially enabling direct inference of nucleosynthetic yields within the SBI framework.

Third, the IMF and SN~Ia delay-time distribution are fixed to specific parametric forms. Alternative functional shapes or break points \citep[e.g.][]{Dubay2024,Alexander2025} could introduce degeneracies with the inferred parameters. Joint marginalization over these functional degrees of freedom will be essential in follow-up studies.

Finally, the \cite{Nissen_2020} dataset contains only 69 solar-type stars with 13 elemental abundances measured (C, O, Na, Mg, Al, Si, Ca, Ti, Cr, Fe, Ni, Sr, and Y). However, for this study we focus on Fe and four alpha elements (C, O, Mg, and Si). Applying \C to large spectroscopic surveys such as GALAH \citep{Buder2024}, APOGEE \citep{apogee17}, LAMOST \citep{lamost2012}, and forthcoming SDSS-V \citep{Kollmeier2017}, WEAVE \citep{WEAVE2018} or 4MOST \citep{4MOST2019}  observations will allow spatially resolved and statistically more stringent constraints.

Beyond these physical limitations, our statistical treatment has limitations that we quantify in this work. (i) The model weights are Akaike weights at the MAP, not evidence ratios. However, the preferred yield set is confirmed by an independent classifier (Appendix~\ref{sec:classifier}), but Bayes factors, unlike these weights, depend on the parameter priors, notably those of the local ISM parameters, and a full evidence computation is left for future work. (ii) Single-star posteriors are combined under a Gaussian approximation, which lets shared systematic errors grow in significance with the number of stars. (iii) The \C posteriors are mildly under-confident (Appendix~\ref{sec:Results.Calibration.Posterior}). (iv) Under model misspecification all tested methods are biased by several $\sigma$ (Appendix~\ref{sec:model-misspecification}).

\subsection{Interpretability and physical transparency}

A key advantage of \C lies in its attention-based interpretability. The model associates $\alpha$-elements such as oxygen and magnesium with the inference of the IMF slope, while abundance ratios sensitive to iron enrichment dominate the determination of the SN~Ia normalization. This alignment between learned attention pathways and established nucleosynthetic channels provides internal validation that the model’s conclusions are physically grounded rather than driven by spurious correlations. 
In this sense, interpretability is not merely diagnostic but forms an integral part of the statistical validation of diffusion-based inference in astrophysics.

\section{Conclusions and outlook}
\label{sec:conclusions}

In this work we have introduced \C\unskip, a unified framework for joint parameter inference and Bayesian model comparison in Galactic Chemical Evolution based on conditional score-based diffusion models and transformer architectures. By combining neural posterior and likelihood estimation within a single amortized architecture, \C enables scalable and physically interpretable simulation-based inference in complex likelihood-free settings.

Our main findings can be summarized as follows:
\begin{itemize}
    \item In controlled mock experiments, \C recovers ground-truth parameters with accuracy comparable to existing SBI approaches, with conservative ($1.4$--$1.6$ times wider) posteriors, while reducing computational cost by orders of magnitude compared to traditional HMC-based techniques.
    \item \C ranks competing yield sets without requiring explicit evidence sampling, correctly identifying the true generative model in mock tests and favoring a hybrid NuGrid + TNG yield prescription when applied to Milky Way abundance data, a preference confirmed by an independent neural classifier.
    \item Applied to high-precision Milky Way abundance data, \C infers $\alpha_{\rm IMF} = -2.65 \pm 0.02$ and $\log(N_{\rm Ia}) = -2.81 \pm 0.02$ for the preferred yield set, with a larger systematic spread between the best-ranked yield sets, consistent with independent studies and supporting a moderately bottom-heavy IMF together with efficient delayed iron enrichment in the solar neighborhood.
    \item The transformer attention analysis demonstrates that the model’s inferences are anchored in physically meaningful abundance signatures, linking $\alpha$-elements to the IMF slope and iron-sensitive ratios to the SN~Ia normalization. This internal consistency mitigates concerns about `black-box' behavior and establishes interpretability as a core strength of the framework.
\end{itemize}

Beyond these specific results, \C illustrates the broader potential of diffusion-based simulation-based inference for data-driven model testing in astrophysics. Its amortized nature makes it particularly well suited for the data volume and dimensionality of current and upcoming spectroscopic surveys (e.g., SDSS-V, WEAVE, 4MOST), while its modular design enables integration with multi-zone GCE models, cosmological hydrodynamical simulations, and hierarchical population-level inference.

Future work will extend the framework to more flexible IMF and SN~Ia delay-time parameterizations, continuous yield modeling, and spatially resolved Galactic datasets. By unifying model comparison, parameter inference, and interpretability within a single probabilistic architecture, \C provides a scalable pathway towards statistically rigorous and physically transparent constraints on galactic enrichment.

In this sense, diffusion-based SBI does not merely offer a computational improvement, but represents a conceptual step towards closing the loop between stellar nucleosynthesis theory, galaxy formation modeling, and precision chemical-abundance observations.

\section*{Data availability}
\label{sec:appendix_code}

We publicly release our code via Github: \url{https://github.com/bGuenes/COMPASS}. Our trained model weights can be found on zenodo: \url{https://zenodo.org/records/18718146}

\begin{acknowledgements}
      This project was made possible by funding from the Carl-Zeiss-Stiftung. This work was supported by the Deutsche Forschungsgemeinschaft (DFG, German Research Foundation)--573580088.
      We acknowledge the usage of the AI clusters Tom and Jerry, funded by Field of Focus~2 of Heidelberg University. 
      This work benefited from discussions at the Lorentz Center workshop “Machine Learning in Astrochemistry”.
      
\end{acknowledgements}

\section*{AI Disclosure}
\label{sec:AI}
We used Anthropic's Claude, Google's Gemini and OpenAI's ChatGPT to aid in code development and to proof-read text and formulae. The authors reviewed and edited all generated text and take full responsibility for the final content. All AI-generated code was independently tested, verified, and supervised by the authors.

\bibliographystyle{aa}
\bibliography{bibtex/references}

\begin{thebibliography}{140}
\expandafter\ifx\csname natexlab\endcsname\relax\def\natexlab#1{#1}\fi

\bibitem[{{Abdurro'uf} {et~al.}(2022){Abdurro'uf}, {Accetta}, {Aerts}, {Silva Aguirre}, {Ahumada}, {Ajgaonkar}, {Filiz Ak}, {Alam}, {Allende Prieto}, {Almeida}, {Anders}, {Anderson}, {Andrews}, {Anguiano}, {Aquino-Ort{\'\i}z}, {Arag{\'o}n-Salamanca}, {Argudo-Fern{\'a}ndez}, {Ata}, {Aubert}, {Avila-Reese}, {Badenes}, {Barb{\'a}}, {Barger}, {Barrera-Ballesteros}, {Beaton}, {Beers}, {Belfiore}, {Bender}, {Bernardi}, {Bershady}, {Beutler}, {Bidin}, {Bird}, {Bizyaev}, {Blanc}, {Blanton}, {Boardman}, {Bolton}, {Boquien}, {Borissova}, {Bovy}, {Brandt}, {Brown}, {Brownstein}, {Brusa}, {Buchner}, {Bundy}, {Burchett}, {Bureau}, {Burgasser}, {Cabang}, {Campbell}, {Cappellari}, {Carlberg}, {Wanderley}, {Carrera}, {Cash}, {Chen}, {Chen}, {Cherinka}, {Chiappini}, {Choi}, {Chojnowski}, {Chung}, {Clerc}, {Cohen}, {Comerford}, {Comparat}, {da Costa}, {Covey}, {Crane}, {Cruz-Gonzalez}, {Culhane}, {Cunha}, {Dai}, {Damke}, {Darling}, {Davidson}, {Davies}, {Dawson}, {De Lee}, {Diamond-Stanic}, {Cano-D{\'\i}az}, {S{\'a}nchez},
  {Donor}, {Duckworth}, {Dwelly}, {Eisenstein}, {Elsworth}, {Emsellem}, {Eracleous}, {Escoffier}, {Fan}, {Farr}, {Feng}, {Fern{\'a}ndez-Trincado}, {Feuillet}, {Filipp}, {Fillingham}, {Frinchaboy}, {Fromenteau}, {Galbany}, {Garc{\'\i}a}, {Garc{\'\i}a-Hern{\'a}ndez}, {Ge}, {Geisler}, {Gelfand}, {G{\'e}ron}, {Gibson}, {Goddy}, {Godoy-Rivera}, {Grabowski}, {Green}, {Greener}, {Grier}, {Griffith}, {Guo}, {Guy}, {Hadjara}, {Harding}, {Hasselquist}, {Hayes}, {Hearty}, {Hern{\'a}ndez}, {Hill}, {Hogg}, {Holtzman}, {Horta}, {Hsieh}, {Hsu}, {Hsu}, {Huber}, {Huertas-Company}, {Hutchinson}, {Hwang}, {Ibarra-Medel}, {Chitham}, {Ilha}, {Imig}, {Jaekle}, {Jayasinghe}, {Ji}, {Johnson}, {Jones}, {J{\"o}nsson}, {Katkov}, {Khalatyan}, {Kinemuchi}, {Kisku}, {Knapen}, {Kneib}, {Kollmeier}, {Kong}, {Kounkel}, {Kreckel}, {Krishnarao}, {Lacerna}, {Lane}, {Langgin}, {Lavender}, {Law}, {Lazarz}, {Leung}, {Leung}, {Lewis}, {Li}, {Li}, {Lian}, {Liang}, {Lin}, {Lin}, {Lin}, {Lintott}, {Long}, {Longa-Pe{\~n}a}, {L{\'o}pez-Cob{\'a}}, {Lu},
  {Lundgren}, {Luo}, {Mackereth}, {de la Macorra}, {Mahadevan}, {Majewski}, {Manchado}, {Mandeville}, {Maraston}, {Margalef-Bentabol}, {Masseron}, {Masters}, {Mathur}, {McDermid}, {Mckay}, {Merloni}, {Merrifield}, {Meszaros}, {Miglio}, {Di Mille}, {Minniti}, {Minsley}, \& {Monachesi}}]{apogee17}
{Abdurro'uf}, {Accetta}, K., {Aerts}, C., {et~al.} 2022, \apjs, 259, 35

\bibitem[{{Agertz} {et~al.}(2013){Agertz}, {Kravtsov}, {Leitner}, \& {Gnedin}}]{Agertz2013}
{Agertz}, O., {Kravtsov}, A.~V., {Leitner}, S.~N., \& {Gnedin}, N.~Y. 2013, \apj, 770, 25

\bibitem[{{Agertz} {et~al.}(2021){Agertz}, {Renaud}, {Feltzing}, {Read}, {Ryde}, {Andersson}, {Rey}, {Bensby}, \& {Feuillet}}]{Agertz2021}
{Agertz}, O., {Renaud}, F., {Feltzing}, S., {et~al.} 2021, \mnras, 503, 5826

\bibitem[{Akaike(1978)}]{AKAIKE19781877}
Akaike, H. 1978, IFAC Proceedings Volumes, 11, 1877, 7th Triennial World Congress of the IFAC on A Link Between Science and Applications of Automatic Control, Helsinki, Finland, 12-16 June

\bibitem[{AKAIKE(1979)}]{AKAIKE1979}
AKAIKE, H. 1979, Biometrika, 66, 237

\bibitem[{Akiba {et~al.}(2019)Akiba, Sano, Yanase, Ohta, \& Koyama}]{akiba2019optunanextgenerationhyperparameteroptimization}
Akiba, T., Sano, S., Yanase, T., Ohta, T., \& Koyama, M. 2019, Optuna: A Next-generation Hyperparameter Optimization Framework

\bibitem[{{Alexander} \& {Vincenzo}(2025)}]{Alexander2025}
{Alexander}, R.~K. \& {Vincenzo}, F. 2025, \mnras, 538, 1127

\bibitem[{{Alsing} {et~al.}(2018){Alsing}, {Wandelt}, \& {Feeney}}]{Alsing2018}
{Alsing}, J., {Wandelt}, B., \& {Feeney}, S. 2018, \mnras, 477, 2874

\bibitem[{Anderson(1982)}]{ANDERSON1982313}
Anderson, B.~D. 1982, Stochastic Processes and their Applications, 12, 313

\bibitem[{{Andrews} {et~al.}(2017{\natexlab{a}}){Andrews}, {Weinberg}, {Sch{\"o}nrich}, \& {Johnson}}]{Andrews2017}
{Andrews}, B.~H., {Weinberg}, D.~H., {Sch{\"o}nrich}, R., \& {Johnson}, J.~A. 2017{\natexlab{a}}, \apj, 835, 224

\bibitem[{{Andrews} {et~al.}(2017{\natexlab{b}}){Andrews}, {Weinberg}, {Sch{\"o}nrich}, \& {Johnson}}]{2017ApJ...835..224A}
{Andrews}, B.~H., {Weinberg}, D.~H., {Sch{\"o}nrich}, R., \& {Johnson}, J.~A. 2017{\natexlab{b}}, \apj, 835, 224

\bibitem[{{Asplund} {et~al.}(2009){Asplund}, {Grevesse}, {Sauval}, \& {Scott}}]{2009ARA&A..47..481A}
{Asplund}, M., {Grevesse}, N., {Sauval}, A.~J., \& {Scott}, P. 2009, \araa, 47, 481

\bibitem[{{Astropy Collaboration} {et~al.}(2013){Astropy Collaboration}, {Robitaille}, {Tollerud}, {Greenfield}, {Droettboom}, {Bray}, {Aldcroft}, {Davis}, {Ginsburg}, {Price-Whelan}, {Kerzendorf}, {Conley}, {Crighton}, {Barbary}, {Muna}, {Ferguson}, {Grollier}, {Parikh}, {Nair}, {Unther}, {Deil}, {Woillez}, {Conseil}, {Kramer}, {Turner}, {Singer}, {Fox}, {Weaver}, {Zabalza}, {Edwards}, {Azalee Bostroem}, {Burke}, {Casey}, {Crawford}, {Dencheva}, {Ely}, {Jenness}, {Labrie}, {Lim}, {Pierfederici}, {Pontzen}, {Ptak}, {Refsdal}, {Servillat}, \& {Streicher}}]{astropy:2013}
{Astropy Collaboration}, {Robitaille}, T.~P., {Tollerud}, E.~J., {et~al.} 2013, \aap, 558, A33

\bibitem[{Austin {et~al.}(2017)Austin, West, \& Heger}]{West17}
Austin, S.~M., West, C., \& Heger, A. 2017, \apjl, 839, L9

\bibitem[{{Bastian} {et~al.}(2010){Bastian}, {Covey}, \& {Meyer}}]{2010ARA&A..48..339B}
{Bastian}, N., {Covey}, K.~R., \& {Meyer}, M.~R. 2010, \araa, 48, 339

\bibitem[{{Bigiel} {et~al.}(2008){Bigiel}, {Leroy}, {Walter}, {Brinks}, {de Blok}, {Madore}, \& {Thornley}}]{2008AJ....136.2846B}
{Bigiel}, F., {Leroy}, A., {Walter}, F., {et~al.} 2008, \aj, 136, 2846

\bibitem[{Bozdogan(1987)}]{Bozdogan_1987}
Bozdogan, H. 1987, Psychometrika, 52, 353–356

\bibitem[{{Buck}(2020)}]{Buck2020}
{Buck}, T. 2020, \mnras, 491, 5435

\bibitem[{{Buck} {et~al.}(2025){Buck}, {G{\"u}nes}, {Viterbo}, {Oliver}, \& {Buder}}]{buck2025inferringgalacticparameterschemical}
{Buck}, T., {G{\"u}nes}, B., {Viterbo}, G., {Oliver}, W.~H., \& {Buder}, S. 2025, \aap, 702, A184

\bibitem[{{Buck} {et~al.}(2020){Buck}, {Obreja}, {Macci{\`o}}, {Minchev}, {Dutton}, \& {Ostriker}}]{Buck2020c}
{Buck}, T., {Obreja}, A., {Macci{\`o}}, A.~V., {et~al.} 2020, \mnras, 491, 3461

\bibitem[{{Buck} {et~al.}(2023){Buck}, {Obreja}, {Ratcliffe}, {Lu}, {Minchev}, \& {Macci{\`o}}}]{Buck2023}
{Buck}, T., {Obreja}, A., {Ratcliffe}, B., {et~al.} 2023, \mnras, 523, 1565

\bibitem[{{Buck} {et~al.}(2021){Buck}, {Rybizki}, {Buder}, {Obreja}, {Macci{\`o}}, {Pfrommer}, {Steinmetz}, \& {Ness}}]{Buck2021}
{Buck}, T., {Rybizki}, J., {Buder}, S., {et~al.} 2021, \mnras, 508, 3365

\bibitem[{{Buder} {et~al.}(2025{\natexlab{a}}){Buder}, {Buck}, {Sk{\'u}lad{\'o}ttir}, {Ness}, {McKenzie}, \& {Monty}}]{Buder2025b}
{Buder}, S., {Buck}, T., {Sk{\'u}lad{\'o}ttir}, {\'A}., {et~al.} 2025{\natexlab{a}}, \mnras, in press [\eprint[arXiv]{2510.20233}]

\bibitem[{{Buder} {et~al.}(2026){Buder}, {Buck}, {Sk{\'u}lad{\'o}ttir}, {Ness}, {McKenzie}, \& {Monty}}]{Buder2025}
{Buder}, S., {Buck}, T., {Sk{\'u}lad{\'o}ttir}, {\'A}., {et~al.} 2026, \mnras, 549, stag1090

\bibitem[{{Buder} {et~al.}(2025{\natexlab{b}}){Buder}, {Kos}, {Wang}, {McKenzie}, {Howell}, {Martell}, {Hayden}, {Zucker}, {Nordlander}, {Montet}, {Traven}, {Bland-Hawthorn}, {De Silva}, {Freeman}, {Lewis}, {Lind}, {Sharma}, {Simpson}, {Stello}, {Zwitter}, {Amarsi}, {Armstrong}, {Banks}, {Beavis}, {Beeson}, {Chen}, {Ciuc{\u{a}}}, {Da Costa}, {de Grijs}, {Martin}, {Nataf}, {Ness}, {Rains}, {Scarr}, {Vogrin{\v{c}}i{\v{c}}}, {Wang}, {Wittenmyer}, {Xie}, \& {The GALAH Collaboration}}]{Buder2024}
{Buder}, S., {Kos}, J., {Wang}, E.~X., {et~al.} 2025{\natexlab{b}}, \pasa, 42, e051

\bibitem[{{Buder} {et~al.}(2024){Buder}, {Mijnarends}, \& {Buck}}]{Buder2024a}
{Buder}, S., {Mijnarends}, L., \& {Buck}, T. 2024, \mnras, 532, 1010

\bibitem[{{Buder} {et~al.}(2021){Buder}, {Sharma}, {Kos}, {Amarsi}, {Nordlander}, {Lind}, {Martell}, {Asplund}, {Bland-Hawthorn}, {Casey}, {de Silva}, {D'Orazi}, {Freeman}, {Hayden}, {Lewis}, {Lin}, {Schlesinger}, {Simpson}, {Stello}, {Zucker}, {Zwitter}, {Beeson}, {Buck}, {Casagrande}, {Clark}, {{\v{C}}otar}, {da Costa}, {de Grijs}, {Feuillet}, {Horner}, {Kafle}, {Khanna}, {Kobayashi}, {Liu}, {Montet}, {Nandakumar}, {Nataf}, {Ness}, {Spina}, {Tepper-Garc{\'\i}a}, {Ting}, {Traven}, {Vogrin{\v{c}}i{\v{c}}}, {Wittenmyer}, {Wyse}, {{\v{Z}}erjal}, {{\v{Z}}erjal}, \& {Galah Collaboration}}]{Buder2021}
{Buder}, S., {Sharma}, S., {Kos}, J., {et~al.} 2021, \mnras, 506, 150

\bibitem[{Burnham \& Anderson(2002)}]{burnham2002model}
Burnham, K. \& Anderson, D. 2002, Model selection and multimodel inference: a practical information-theoretic approach (Springer Verlag)

\bibitem[{{Chabrier}(2003)}]{2003PASP..115..763C}
{Chabrier}, G. 2003, \pasp, 115, 763

\bibitem[{{Chabrier} {et~al.}(2014){Chabrier}, {Hennebelle}, \& {Charlot}}]{Chabrier2014}
{Chabrier}, G., {Hennebelle}, P., \& {Charlot}, S. 2014, \apj, 796, 75

\bibitem[{{Chiappini} {et~al.}(1997){Chiappini}, {Matteucci}, \& {Gratton}}]{Chiappini1997}
{Chiappini}, C., {Matteucci}, F., \& {Gratton}, R. 1997, \apj, 477, 765

\bibitem[{{Chieffi} \& {Limongi}(2004)}]{Chieffi04}
{Chieffi}, A. \& {Limongi}, M. 2004, \apj, 608, 405

\bibitem[{{Clarke} {et~al.}(2019){Clarke}, {Debattista}, {Nidever}, {Loebman}, {Simons}, {Kassin}, {Du}, {Ness}, {Fisher}, {Quinn}, {Wadsley}, {Freeman}, \& {Popescu}}]{2019MNRAS.484.3476C}
{Clarke}, A.~J., {Debattista}, V.~P., {Nidever}, D.~L., {et~al.} 2019, \mnras, 484, 3476

\bibitem[{{C{\^o}t{\'e}} {et~al.}(2017){C{\^o}t{\'e}}, {O'Shea}, {Ritter}, {Herwig}, \& {Venn}}]{Cote2017}
{C{\^o}t{\'e}}, B., {O'Shea}, B.~W., {Ritter}, C., {Herwig}, F., \& {Venn}, K.~A. 2017, \apj, 835, 128

\bibitem[{Cranmer {et~al.}(2020)Cranmer, Brehmer, \& Louppe}]{Cranmer_2020}
Cranmer, K., Brehmer, J., \& Louppe, G. 2020, Proceedings of the National Academy of Sciences, 117, 30055–30062

\bibitem[{{Cristallo} {et~al.}(2015){Cristallo}, {Straniero}, {Piersanti}, \& {Gobrecht}}]{Cristallo2015}
{Cristallo}, S., {Straniero}, O., {Piersanti}, L., \& {Gobrecht}, D. 2015, \apjs, 219, 40

\bibitem[{{Dalton} {et~al.}(2018){Dalton}, {Trager}, {Abrams}, {Bonifacio}, {Aguerri}, {Vallenari}, {Middleton}, {Benn}, {Dee}, {Say{\`e}de}, {Lewis}, {Pragt}, {Pic{\'o}}, {Walton}, {Rey}, {Allende}, {Lhom{\'e}}, {Terrett}, {Brock}, {Gilbert}, {Ridings}, {Verheijen}, {Tosh}, {Steele}, {Stuik}, {Kroes}, {Tromp}, {Kragt}, {Lesman}, {Mottram}, {Bates}, {Gribbin}, {Burgal}, {Herreros}, {Delgado}, {Martin}, {Cano}, {Navarro}, {Irwin}, {Lewis}, {Gonzales Solares}, {O'Mahony}, {Bianco}, {Zurita}, {ter Horst}, {Molinari}, {Lodi}, {Guerra}, {Baruffolo}, {Carrasco}, {Farkas}, {Schallig}, {Hill}, {Smith}, {Drew}, {Poggianti}, {Pieri}, {Jin}, {Dominquez Palmero}, {Fari{\~n}a}, {Martin}, {Worley}, {Murphy}, {Hidalgo}, {Mignot}, {Bishop}, {Guest}, {Elswijk}, {de Haan}, {Hanenburg}, {Salasnich}, {Mayya}, {Izazaga-P{\'e}rez}, \& {Peralta de Arriba}}]{WEAVE2018}
{Dalton}, G., {Trager}, S., {Abrams}, D.~C., {et~al.} 2018, SPIE, 10702, 1B

\bibitem[{{Dax} {et~al.}(2025){Dax}, {Green}, {Gair}, {Gupte}, {P{\"u}rrer}, {Raymond}, {Wildberger}, {Macke}, {Buonanno}, \& {Sch{\"o}lkopf}}]{Dax_2025}
{Dax}, M., {Green}, S.~R., {Gair}, J., {et~al.} 2025, \nat, 639, 49

\bibitem[{{de Jong} {et~al.}(2019){de Jong}, {Agertz}, {Berbel}, {Aird}, {Alexander}, {Amarsi}, {Anders}, {Andrae}, {Ansarinejad}, {Ansorge}, {Antilogus}, {Anwand -Heerwart}, {Arentsen}, {Arnadottir}, {Asplund}, {Auger}, {Azais}, {Baade}, {Baker}, {Baker}, {Balbinot}, {Baldry}, {Banerji}, {Barden}, {Barklem}, {Barth{\'e}l{\'e}my-Mazot}, {Battistini}, {Bauer}, {Bell}, {Bellido-Tirado}, {Bellstedt}, {Belokurov}, {Bensby}, {Bergemann}, {Bestenlehner}, {Bielby}, {Bilicki}, {Blake}, {Bland-Hawthorn}, {Boeche}, {Boland}, {Boller}, {Bongard}, {Bongiorno}, {Bonifacio}, {Boudon}, {Brooks}, {Brown}, {Brown}, {Br{\"u}ggen}, {Brynnel}, {Brzeski}, {Buchert}, {Buschkamp}, {Caffau}, {Caillier}, {Carrick}, {Casagrande}, {Case}, {Casey}, {Cesarini}, {Cescutti}, {Chapuis}, {Chiappini}, {Childress}, {Christlieb}, {Church}, {Cioni}, {Cluver}, {Colless}, {Collett}, {Comparat}, {Cooper}, {Couch}, {Courbin}, {Croom}, {Croton}, {Daguis{\'e}}, {Dalton}, {Davies}, {Davis}, {de Laverny}, {Deason}, {Dionies}, {Disseau}, {Doel},
  {D{\"o}scher}, {Driver}, {Dwelly}, {Eckert}, {Edge}, {Edvardsson}, {Youssoufi}, {Elhaddad}, {Enke}, {Erfanianfar}, {Farrell}, {Fechner}, {Feiz}, {Feltzing}, {Ferreras}, {Feuerstein}, {Feuillet}, {Finoguenov}, {Ford}, {Fotopoulou}, {Fouesneau}, {Frenk}, {Frey}, {Gaessler}, {Geier}, {Fusillo}, {Gerhard}, {Giannantonio}, {Giannone}, {Gibson}, {Gillingham}, {Gonz{\'a}lez-Fern{\'a}ndez}, {Gonzalez-Solares}, {Gottloeber}, {Gould}, {Grebel}, {Gueguen}, {Guiglion}, {Haehnelt}, {Hahn}, {Hansen}, {Hartman}, {Hauptner}, {Hawkins}, {Haynes}, {Haynes}, {Heiter}, {Helmi}, {Aguayo}, {Hewett}, {Hinton}, {Hobbs}, {Hoenig}, {Hofman}, {Hook}, {Hopgood}, {Hopkins}, {Hourihane}, {Howes}, {Howlett}, {Huet}, {Irwin}, {Iwert}, {Jablonka}, {Jahn}, {Jahnke}, {Jarno}, {Jin}, {Jofre}, {Johl}, {Jones}, {J{\"o}nsson}, {Jordan}, {Karovicova}, {Khalatyan}, {Kelz}, {Kennicutt}, {King}, {Kitaura}, {Klar}, {Klauser}, {Kneib}, {Koch}, {Koposov}, {Kordopatis}, {Korn}, {Kosmalski}, {Kotak}, {Kovalev}, {Kreckel}, {Kripak}, {Krumpe}, {Kuijken},
  {Kunder}, {Kushniruk}, {Lam}, {Lamer}, {Laurent}, {Lawrence}, {Lehmitz}, {Lemasle}, {Lewis}, {Li}, {Lidman}, {Lind}, {Liske}, {Lizon}, {Loveday}, {Ludwig}, {McDermid}, {Maguire}, {Mainieri}, {Mali}, {Mandel}, {Mandel}, {Mannering}, {Martell}, {Martinez Delgado}, {Matijevic}, {McGregor}, {McMahon}, {McMillan}, {Mena}, {Merloni}, {Meyer}, {Michel}, {Micheva}, {Migniau}, {Minchev}, {Monari}, {Muller}, {Murphy}, {Muthukrishna}, {Nandra}, {Navarro}, {Ness}, {Nichani}, {Nichol}, {Nicklas}, {Niederhofer}, {Norberg}, {Obreschkow}, {Oliver}, {Owers}, {Pai}, {Pankratow}, {Parkinson}, {Paschke}, {Paterson}, {Pecontal}, {Parry}, {Phillips}, {Pillepich}, {Pinard}, {Pirard}, {Piskunov}, {Plank}, {Pl{\"u}schke}, {Pons}, {Popesso}, {Power}, {Pragt}, {Pramskiy}, {Pryer}, {Quattri}, {Queiroz}, {Quirrenbach}, {Rahurkar}, {Raichoor}, {Ramstedt}, {Rau}, {Recio-Blanco}, {Reiss}, {Renaud}, {Revaz}, {Rhode}, {Richard}, {Richter}, {Rix}, {Robotham}, {Roelfsema}, {Romaniello}, {Rosario}, {Rothmaier}, {Roukema}, {Ruchti},
  {Rupprecht}, {Rybizki}, {Ryde}, {Saar}, {Sadler}, {Sahl{\'e}n}, {Salvato}, {Sassolas}, {Saunders}, {Saviauk}, {Sbordone}, {Schmidt}, {Schnurr}, {Scholz}, {Schwope}, {Seifert}, {Shanks}, {Sheinis}, {Sivov}, {Sk{\'u}lad{\'o}ttir}, {Smartt}, {Smedley}, {Smith}, {Smith}, {Sorce}, {Spitler}, {Starkenburg}, {Steinmetz}, {Stilz}, {Storm}, {Sullivan}, {Sutherland}, {Swann}, {Tamone}, {Taylor}, {Teillon}, {Tempel}, {ter Horst}, {Thi}, {Tolstoy}, {Trager}, {Traven}, {Tremblay}, {Tresse}, {Valentini}, {van de Weygaert}, {van den Ancker}, {Veljanoski}, {Venkatesan}, {Wagner}, {Wagner}, {Walcher}, {Waller}, {Walton}, {Wang}, {Winkler}, {Wisotzki}, {Worley}, {Worseck}, {Xiang}, {Xu}, {Yong}, {Zhao}, {Zheng}, {Zscheyge}, \& {Zucker}}]{4MOST2019}
{de Jong}, R.~S., {Agertz}, O., {Berbel}, A.~A., {et~al.} 2019, The Messenger, 175, 3

\bibitem[{Defazio {et~al.}(2024)Defazio, Yang, Mehta, Mishchenko, Khaled, \& Cutkosky}]{defazio2024roadscheduled}
Defazio, A., Yang, X.~A., Mehta, H., {et~al.} 2024, in Advances in Neural Information Processing Systems (NeurIPS), Vol.~37

\bibitem[{{den Brok} {et~al.}(2024){den Brok}, {Krajnovi{\'c}}, {Emsellem}, {Mercier}, {Steinmetz}, \& {Weilbacher}}]{denBrok2024}
{den Brok}, M., {Krajnovi{\'c}}, D., {Emsellem}, E., {et~al.} 2024, \mnras, 530, 3278

\bibitem[{{Doherty} {et~al.}(2014){Doherty}, {Gil-Pons}, {Lau}, {Lattanzio}, \& {Siess}}]{2014MNRAS.437..195D}
{Doherty}, C.~L., {Gil-Pons}, P., {Lau}, H. H.~B., {Lattanzio}, J.~C., \& {Siess}, L. 2014, \mnras, 437, 195

\bibitem[{{Dubay} {et~al.}(2024){Dubay}, {Johnson}, \& {Johnson}}]{Dubay2024}
{Dubay}, L.~O., {Johnson}, J.~A., \& {Johnson}, J.~W. 2024, \apj, 973, 55

\bibitem[{{Fishlock} {et~al.}(2014){Fishlock}, {Karakas}, {Lugaro}, \& {Yong}}]{2014ApJ...797...44F}
{Fishlock}, C.~K., {Karakas}, A.~I., {Lugaro}, M., \& {Yong}, D. 2014, \apj, 797, 44

\bibitem[{{Font} {et~al.}(2020){Font}, {McCarthy}, {Poole-Mckenzie}, {Stafford}, {Brown}, {Schaye}, {Crain}, {Theuns}, \& {Schaller}}]{Font2020}
{Font}, A.~S., {McCarthy}, I.~G., {Poole-Mckenzie}, R., {et~al.} 2020, \mnras, 498, 1765

\bibitem[{{Freeman} \& {Bland-Hawthorn}(2002)}]{Freeman2002}
{Freeman}, K. \& {Bland-Hawthorn}, J. 2002, \araa, 40, 487

\bibitem[{Frischknecht {et~al.}(2015)Frischknecht, Hirschi, Pignatari, Maeder, Meynet, Chiappini, Thielemann, Rauscher, Georgy, \& Ekström}]{Frischknecht_2015}
Frischknecht, U., Hirschi, R., Pignatari, M., {et~al.} 2015, \mnras, 456, 1803–1825

\bibitem[{{Geffner} {et~al.}(2022){Geffner}, {Papamakarios}, \& {Mnih}}]{geffner2023compositionalscoremodelingsimulationbased}
{Geffner}, T., {Papamakarios}, G., \& {Mnih}, A. 2022, arXiv e-prints, arXiv:2209.14249

\bibitem[{Gloeckler {et~al.}(2024)Gloeckler, Deistler, Weilbach, Wood, \& Macke}]{gloeckler2024allinonesimulationbasedinference}
Gloeckler, M., Deistler, M., Weilbach, C., Wood, F., \& Macke, J.~H. 2024, in PMLR, Vol. 235, Proceedings of the 41st International Conference on Machine Learning (ICML)

\bibitem[{{Grand} {et~al.}(2018){Grand}, {Bustamante}, {G{\'o}mez}, {Kawata}, {Marinacci}, {Pakmor}, {Rix}, {Simpson}, {Sparre}, \& {Springel}}]{2018MNRAS.474.3629G}
{Grand}, R.~J.~J., {Bustamante}, S., {G{\'o}mez}, F.~A., {et~al.} 2018, \mnras, 474, 3629

\bibitem[{{Griffith} {et~al.}(2025){Griffith}, {Hogg}, {Hasselquist}, {Johnson}, {Price-Whelan}, {Sit}, {Stone-Martinez}, \& {Weinberg}}]{Griffith2025}
{Griffith}, E.~J., {Hogg}, D.~W., {Hasselquist}, S., {et~al.} 2025, \aj, 169, 280

\bibitem[{{Griffith} {et~al.}(2022){Griffith}, {Weinberg}, {Buder}, {Johnson}, {Johnson}, \& {Vincenzo}}]{Griffith2022}
{Griffith}, E.~J., {Weinberg}, D.~H., {Buder}, S., {et~al.} 2022, \apj, 931, 23

\bibitem[{{Gronow} {et~al.}(2021){Gronow}, {C{\^o}t{\'e}}, {Lach}, {Seitenzahl}, {Collins}, {Sim}, \& {R{\"o}pke}}]{Gronow2021}
{Gronow}, S., {C{\^o}t{\'e}}, B., {Lach}, F., {et~al.} 2021, \aap, 656, A94

\bibitem[{{Gutcke} \& {Springel}(2019)}]{2019MNRAS.482..118G}
{Gutcke}, T.~A. \& {Springel}, V. 2019, \mnras, 482, 118

\bibitem[{{Günes} {et~al.}(2025){Günes}, {Buder}, \& {Buck}}]{Guenes2025}
{Günes}, B., {Buder}, S., \& {Buck}, T. 2025, Machine Learning for Astrophysics Workshop (ML4Astro 2025), Vancouver [\eprint[arXiv]{2507.05060}]

\bibitem[{{Hahn} \& {Melchior}(2022)}]{Hahn2022}
{Hahn}, C. \& {Melchior}, P. 2022, \apj, 938, 11

\bibitem[{{Hayden} {et~al.}(2015){Hayden}, {Bovy}, {Holtzman}, {Nidever}, {Bird}, {Weinberg}, {Andrews}, {Majewski}, {Allende Prieto}, {Anders}, {Beers}, {Bizyaev}, {Chiappini}, {Cunha}, {Frinchaboy}, {Garc{\'\i}a-Her{\'n}andez}, {Garc{\'\i}a P{\'e}rez}, {Girardi}, {Harding}, {Hearty}, {Johnson}, {M{\'e}sz{\'a}ros}, {Minchev}, {O'Connell}, {Pan}, {Robin}, {Schiavon}, {Schneider}, {Schultheis}, {Shetrone}, {Skrutskie}, {Steinmetz}, {Smith}, {Wilson}, {Zamora}, \& {Zasowski}}]{Hayden2015}
{Hayden}, M.~R., {Bovy}, J., {Holtzman}, J.~A., {et~al.} 2015, \apj, 808, 132

\bibitem[{Helbling {et~al.}(2025)Helbling, Meral, Hoover, Yanardag, \& Chau}]{helbling2025conceptattentiondiffusiontransformerslearn}
Helbling, A., Meral, T. H.~S., Hoover, B., Yanardag, P., \& Chau, D.~H. 2025, in PMLR, Vol. 267, Proceedings of the 42nd International Conference on Machine Learning (ICML)

\bibitem[{{Hoffman} \& {Gelman}(2014)}]{Hoffman2014}
{Hoffman}, M.~D. \& {Gelman}, A. 2014, Journal of Machine Learning Research, 15, 1593

\bibitem[{{Hopkins} {et~al.}(2018){Hopkins}, {Wetzel}, {Kere{\v{s}}}, {Faucher-Gigu{\`e}re}, {Quataert}, {Boylan-Kolchin}, {Murray}, {Hayward}, {Garrison-Kimmel}, {Hummels}, {Feldmann}, {Torrey}, {Ma}, {Angl{\'e}s-Alc{\'a}zar}, {Su}, {Orr}, {Schmitz}, {Escala}, {Sanderson}, {Grudi{\'c}}, {Hafen}, {Kim}, {Fitts}, {Bullock}, {Wheeler}, {Chan}, {Elbert}, \& {Narayanan}}]{Hopkins2018}
{Hopkins}, P.~F., {Wetzel}, A., {Kere{\v{s}}}, D., {et~al.} 2018, \mnras, 480, 800

\bibitem[{{Jeffrey} \& {Wandelt}(2024)}]{Jeffrey2024}
{Jeffrey}, N. \& {Wandelt}, B.~D. 2024, Machine Learning: Science and Technology, 5, 015008

\bibitem[{{Jim{\'e}nez} {et~al.}(2015){Jim{\'e}nez}, {Tissera}, \& {Matteucci}}]{2015ApJ...810..137J}
{Jim{\'e}nez}, N., {Tissera}, P.~B., \& {Matteucci}, F. 2015, \apj, 810, 137

\bibitem[{{Johnson} {et~al.}(2021){Johnson}, {Weinberg}, {Vincenzo}, {Bird}, {Loebman}, {Brooks}, {Quinn}, {Christensen}, \& {Griffith}}]{Johnson2021}
{Johnson}, J.~W., {Weinberg}, D.~H., {Vincenzo}, F., {et~al.} 2021, \mnras, 508, 4484

\bibitem[{{Karakas}(2010)}]{2010MNRAS.403.1413K}
{Karakas}, A.~I. 2010, \mnras, 403, 1413

\bibitem[{{Karakas} \& {Lugaro}(2016)}]{Karakas2016}
{Karakas}, A.~I. \& {Lugaro}, M. 2016, \apj, 825, 26

\bibitem[{{Karakas} {et~al.}(2014){Karakas}, {Marino}, \& {Nataf}}]{Karakas2014}
{Karakas}, A.~I., {Marino}, A.~F., \& {Nataf}, D.~M. 2014, \apj, 784, 32

\bibitem[{{Kn{\"o}ll} {et~al.}(2026){Kn{\"o}ll}, {Buck}, {Branca}, \& {Viterbo}}]{Knoell2026}
{Kn{\"o}ll}, N., {Buck}, T., {Branca}, L., \& {Viterbo}, G. 2026, ICML 2026 Workshop AI4Physics, accepted [\eprint[arXiv]{2606.10762}]

\bibitem[{{Kobayashi} {et~al.}(2020){Kobayashi}, {Karakas}, \& {Lugaro}}]{Kobayashi2020}
{Kobayashi}, C., {Karakas}, A.~I., \& {Lugaro}, M. 2020, \apj, 900, 179

\bibitem[{{Kobayashi} {et~al.}(2006){Kobayashi}, {Umeda}, {Nomoto}, {Tominaga}, \& {Ohkubo}}]{2006ApJ...653.1145K}
{Kobayashi}, C., {Umeda}, H., {Nomoto}, K., {Tominaga}, N., \& {Ohkubo}, T. 2006, \apj, 653, 1145

\bibitem[{Kobayashi {et~al.}(2006)Kobayashi, Umeda, Nomoto, Tominaga, \& Ohkubo}]{Kobayashi_2006}
Kobayashi, C., Umeda, H., Nomoto, K., Tominaga, N., \& Ohkubo, T. 2006, \apj, 653, 1145–1171

\bibitem[{{Kollmeier} {et~al.}(2017){Kollmeier}, {Zasowski}, {Rix}, {Johns}, {Anderson}, {Drory}, {Johnson}, {Pogge}, {Bird}, {Blanc}, {Brownstein}, {Crane}, {De Lee}, {Klaene}, {Kreckel}, {MacDonald}, {Merloni}, {Ness}, {O'Brien}, {Sanchez-Gallego}, {Sayres}, {Shen}, {Thakar}, {Tkachenko}, {Aerts}, {Blanton}, {Eisenstein}, {Holtzman}, {Maoz}, {Nandra}, {Rockosi}, {Weinberg}, {Bovy}, {Casey}, {Chaname}, {Clerc}, {Conroy}, {Eracleous}, {G{\"a}nsicke}, {Hekker}, {Horne}, {Kauffmann}, {McQuinn}, {Pellegrini}, {Schinnerer}, {Schlafly}, {Schwope}, {Seibert}, {Teske}, \& {van Saders}}]{Kollmeier2017}
{Kollmeier}, J.~A., {Zasowski}, G., {Rix}, H.-W., {et~al.} 2017, arXiv e-prints, arXiv:1711.03234

\bibitem[{Lemos {et~al.}(2023)Lemos, Coogan, Hezaveh, \& Perreault-Levasseur}]{lemos2023samplingbasedaccuracytestingposterior}
Lemos, P., Coogan, A., Hezaveh, Y., \& Perreault-Levasseur, L. 2023, Sampling-Based Accuracy Testing of Posterior Estimators for General Inference

\bibitem[{{Li} {et~al.}(2023){Li}, {Liu}, {Zhang}, {Tian}, {Fu}, {Li}, \& {Yan}}]{Li2023}
{Li}, J., {Liu}, C., {Zhang}, Z.-Y., {et~al.} 2023, \nat, 613, 460

\bibitem[{{Liang} {et~al.}(2023){Liang}, {Gjergo}, \& {Fan}}]{Liang2023}
{Liang}, J., {Gjergo}, E., \& {Fan}, X. 2023, \mnras, 522, 863

\bibitem[{Limongi \& Chieffi(2018)}]{Limongi_2018}
Limongi, M. \& Chieffi, A. 2018, \apjs, 237, 13

\bibitem[{{Lovell} {et~al.}(2025){Lovell}, {Starkenburg}, {Ho}, {Angl{\'e}s-Alc{\'a}zar}, {Dav{\'e}}, {Gabrielpillai}, {Iyer}, {Matthews}, {Roper}, {Somerville}, {Sommovigo}, \& {Villaescusa-Navarro}}]{Lovell2025}
{Lovell}, C.~C., {Starkenburg}, T., {Ho}, M., {et~al.} 2025, \mnras, 544, 3949

\bibitem[{Lu {et~al.}(2022)Lu, Zhou, Bao, Chen, Li, \& Zhu}]{lu2022dpmsolverfastodesolver}
Lu, C., Zhou, Y., Bao, F., {et~al.} 2022, in Advances in Neural Information Processing Systems (NeurIPS), Vol.~35

\bibitem[{{Lu} {et~al.}(2022){Lu}, {Buck}, {Minchev}, \& {Ness}}]{Lu2022}
{Lu}, Y., {Buck}, T., {Minchev}, I., \& {Ness}, M.~K. 2022, \mnras, 515, L34

\bibitem[{{Lu} {et~al.}(2024){Lu}, {Minchev}, {Buck}, {Khoperskov}, {Steinmetz}, {Libeskind}, {Cescutti}, {Freeman}, \& {Ratcliffe}}]{Lu2024}
{Lu}, Y.~L., {Minchev}, I., {Buck}, T., {et~al.} 2024, \mnras, 535, 392

\bibitem[{{Mackereth} {et~al.}(2018){Mackereth}, {Crain}, {Schiavon}, {Schaye}, {Theuns}, \& {Schaller}}]{2018MNRAS.477.5072M}
{Mackereth}, J.~T., {Crain}, R.~A., {Schiavon}, R.~P., {et~al.} 2018, \mnras, 477, 5072

\bibitem[{{Maoz} \& {Mannucci}(2012)}]{2012PASA...29..447M}
{Maoz}, D. \& {Mannucci}, F. 2012, \pasa, 29, 447

\bibitem[{{Maoz} {et~al.}(2012){Maoz}, {Mannucci}, \& {Brandt}}]{2012MNRAS.426.3282M}
{Maoz}, D., {Mannucci}, F., \& {Brandt}, T.~D. 2012, \mnras, 426, 3282

\bibitem[{{Maoz} {et~al.}(2014){Maoz}, {Mannucci}, \& {Nelemans}}]{Maoz2014}
{Maoz}, D., {Mannucci}, F., \& {Nelemans}, G. 2014, \araa, 52, 107

\bibitem[{{Maoz} {et~al.}(2010){Maoz}, {Sharon}, \& {Gal-Yam}}]{2010ApJ...722.1879M}
{Maoz}, D., {Sharon}, K., \& {Gal-Yam}, A. 2010, \apj, 722, 1879

\bibitem[{{Mart{\'\i}n-Navarro} {et~al.}(2019){Mart{\'\i}n-Navarro}, {Lyubenova}, {van de Ven}, {Falc{\'o}n-Barroso}, {Coccato}, {Corsini}, {Gadotti}, {Iodice}, {La Barbera}, {McDermid}, {Pinna}, {Sarzi}, {Viaene}, {de Zeeuw}, \& {Zhu}}]{Martin2019}
{Mart{\'\i}n-Navarro}, I., {Lyubenova}, M., {van de Ven}, G., {et~al.} 2019, \aap, 626, A124

\bibitem[{Matteucci(2016)}]{Matteucci_2016}
Matteucci, F. 2016, Journal of Physics: Conference Series, 703, 012004

\bibitem[{{McWilliam}(1997)}]{1997ARA&A..35..503M}
{McWilliam}, A. 1997, \araa, 35, 503

\bibitem[{{Minchev} {et~al.}(2013){Minchev}, {Chiappini}, \& {Martig}}]{2013A&A...558A...9M}
{Minchev}, I., {Chiappini}, C., \& {Martig}, M. 2013, \aap, 558, A9

\bibitem[{{Moll{\'a}} {et~al.}(2015){Moll{\'a}}, {Cavichia}, {Gavil{\'a}n}, \& {Gibson}}]{2015MNRAS.451.3693M}
{Moll{\'a}}, M., {Cavichia}, O., {Gavil{\'a}n}, M., \& {Gibson}, B.~K. 2015, \mnras, 451, 3693

\bibitem[{{Nelson} {et~al.}(2019){Nelson}, {Springel}, {Pillepich}, {Rodriguez-Gomez}, {Torrey}, {Genel}, {Vogelsberger}, {Pakmor}, {Marinacci}, {Weinberger}, {Kelley}, {Lovell}, {Diemer}, \& {Hernquist}}]{2019ComAC...6....2N}
{Nelson}, D., {Springel}, V., {Pillepich}, A., {et~al.} 2019, Computational Astrophysics and Cosmology, 6, 2

\bibitem[{{Ness} {et~al.}(2018){Ness}, {Rix}, {Hogg}, {Casey}, {Holtzman}, {Fouesneau}, {Zasowski}, {Geisler}, {Shetrone}, {Minniti}, {Frinchaboy}, \& {Roman-Lopes}}]{Ness2018}
{Ness}, M., {Rix}, H.~W., {Hogg}, D.~W., {et~al.} 2018, \apj, 853, 198

\bibitem[{Nissen {et~al.}(2020)Nissen, Christensen-Dalsgaard, Mosumgaard, Silva~Aguirre, Spitoni, \& Verma}]{Nissen_2020}
Nissen, P.~E., Christensen-Dalsgaard, J., Mosumgaard, J.~R., {et~al.} 2020, \aap, 640, A81

\bibitem[{{Nomoto} {et~al.}(2013){Nomoto}, {Kobayashi}, \& {Tominaga}}]{2013ARA&A..51..457N}
{Nomoto}, K., {Kobayashi}, C., \& {Tominaga}, N. 2013, \araa, 51, 457

\bibitem[{Papamakarios \& Murray(2016)}]{Papamakarios2016}
Papamakarios, G. \& Murray, I. 2016, Fast $\epsilon$-free Inference of Simulation Models with Bayesian Conditional Density Estimation

\bibitem[{Papamakarios {et~al.}(2019)Papamakarios, Sterratt, \& Murray}]{99aad490fd134b849b5336473b4342c3}
Papamakarios, G., Sterratt, D., \& Murray, I. 2019, in Proceedings of the 22nd International Conference on Artificial Intelligence and Statistics (AISTATS) 2019, Proceedings of Machine Learning Research (PMLR), 837--848, 22nd International Conference on Artificial Intelligence and Statistics, AISTATS 2019 ; Conference date: 16-04-2019 Through 18-04-2019

\bibitem[{Peebles \& Xie(2023)}]{peebles2023scalablediffusionmodelstransformers}
Peebles, W. \& Xie, S. 2023, in Proceedings of the IEEE/CVF International Conference on Computer Vision (ICCV), 4195--4205

\bibitem[{{Phan} {et~al.}(2019){Phan}, {Pradhan}, \& {Jankowiak}}]{Phan2019}
{Phan}, D., {Pradhan}, N., \& {Jankowiak}, M. 2019, NeurIPS 2019 Workshop on Program Transformations for Machine Learning [\eprint{1912.11554}]

\bibitem[{Philcox {et~al.}(2018)Philcox, Rybizki, \& Gutcke}]{Philcox_2018}
Philcox, O., Rybizki, J., \& Gutcke, T.~A. 2018, \apj, 861, 40

\bibitem[{Philcox \& Rybizki(2019)}]{Philcox_2019}
Philcox, O. H.~E. \& Rybizki, J. 2019, \apj, 887, 9

\bibitem[{{Pillepich} {et~al.}(2018){Pillepich}, {Nelson}, {Hernquist}, {Springel}, {Pakmor}, {Torrey}, {Weinberger}, {Genel}, {Naiman}, {Marinacci}, \& {Vogelsberger}}]{2018MNRAS.475..648P}
{Pillepich}, A., {Nelson}, D., {Hernquist}, L., {et~al.} 2018, \mnras, 475, 648

\bibitem[{Pillepich {et~al.}(2017{\natexlab{a}})Pillepich, Springel, Nelson, Genel, Naiman, Pakmor, Hernquist, Torrey, Vogelsberger, Weinberger, \& Marinacci}]{Pillepich2018}
Pillepich, A., Springel, V., Nelson, D., {et~al.} 2017{\natexlab{a}}, \mnras, 473, 4077

\bibitem[{Pillepich {et~al.}(2017{\natexlab{b}})Pillepich, Springel, Nelson, Genel, Naiman, Pakmor, Hernquist, Torrey, Vogelsberger, Weinberger, \& Marinacci}]{10.1093/mnras/stx2656}
Pillepich, A., Springel, V., Nelson, D., {et~al.} 2017{\natexlab{b}}, \mnras, 473, 4077

\bibitem[{{Planck Collaboration} {et~al.}(2016){Planck Collaboration}, {Ade}, {Aghanim}, {Arnaud}, {Ashdown}, {Aumont}, {Baccigalupi}, {Banday}, {Barreiro}, {Bartlett}, {Bartolo}, {Battaner}, {Battye}, {Benabed}, {Beno{\^\i}t}, {Benoit-L{\'e}vy}, {Bernard}, {Bersanelli}, {Bielewicz}, {Bock}, {Bonaldi}, {Bonavera}, {Bond}, {Borrill}, {Bouchet}, {Boulanger}, {Bucher}, {Burigana}, {Butler}, {Calabrese}, {Cardoso}, {Catalano}, {Challinor}, {Chamballu}, {Chary}, {Chiang}, {Chluba}, {Christensen}, {Church}, {Clements}, {Colombi}, {Colombo}, {Combet}, {Coulais}, {Crill}, {Curto}, {Cuttaia}, {Danese}, {Davies}, {Davis}, {de Bernardis}, {de Rosa}, {de Zotti}, {Delabrouille}, {D{\'e}sert}, {Di Valentino}, {Dickinson}, {Diego}, {Dolag}, {Dole}, {Donzelli}, {Dor{\'e}}, {Douspis}, {Ducout}, {Dunkley}, {Dupac}, {Efstathiou}, {Elsner}, {En{\ss}lin}, {Eriksen}, {Farhang}, {Fergusson}, {Finelli}, {Forni}, {Frailis}, {Fraisse}, {Franceschi}, {Frejsel}, {Galeotta}, {Galli}, {Ganga}, {Gauthier}, {Gerbino}, {Ghosh}, {Giard},
  {Giraud-H{\'e}raud}, {Giusarma}, {Gjerl{\o}w}, {Gonz{\'a}lez-Nuevo}, {G{\'o}rski}, {Gratton}, {Gregorio}, {Gruppuso}, {Gudmundsson}, {Hamann}, {Hansen}, {Hanson}, {Harrison}, {Helou}, {Henrot- Versill{\'e}}, {Hern{\'a}ndez-Monteagudo}, {Herranz}, {Hildebrandt}, {Hivon}, {Hobson}, {Holmes}, {Hornstrup}, {Hovest}, {Huang}, {Huffenberger}, {Hurier}, {Jaffe}, {Jaffe}, {Jones}, {Juvela}, {Keih{\"a}nen}, {Keskitalo}, {Kisner}, {Kneissl}, {Knoche}, {Knox}, {Kunz}, {Kurki-Suonio}, {Lagache}, {L{\"a}hteenm{\"a}ki}, {Lamarre}, {Lasenby}, {Lattanzi}, {Lawrence}, {Leahy}, {Leonardi}, {Lesgourgues}, {Levrier}, {Lewis}, {Liguori}, {Lilje}, {Linden-V{\o}rnle}, {L{\'o}pez-Caniego}, {Lubin}, {Mac{\'\i}as-P{\'e}rez}, {Maggio}, {Maino}, {Mandolesi}, {Mangilli}, {Marchini}, {Maris}, {Martin}, {Martinelli}, {Mart{\'\i}nez-Gonz{\'a}lez}, {Masi}, {Matarrese}, {McGehee}, {Meinhold}, {Melchiorri}, {Melin}, {Mendes}, {Mennella}, {Migliaccio}, {Millea}, {Mitra}, {Miville-Desch{\^e}nes}, {Moneti}, {Montier}, {Morgante}, {Mortlock},
  {Moss}, {Munshi}, {Murphy}, {Naselsky}, {Nati}, {Natoli}, {Netterfield}, {N{\o}rgaard-Nielsen}, {Noviello}, {Novikov}, {Novikov}, {Oxborrow}, {Paci}, {Pagano}, {Pajot}, {Paladini}, {Paoletti}, {Partridge}, {Pasian}, {Patanchon}, {Pearson}, {Perdereau}, {Perotto}, {Perrotta}, {Pettorino}, {Piacentini}, {Piat}, {Pierpaoli}, {Pietrobon}, {Plaszczynski}, {Pointecouteau}, {Polenta}, {Popa}, {Pratt}, {Pr{\'e}zeau}, {Prunet}, {Puget}, {Rachen}, {Reach}, {Rebolo}, {Reinecke}, {Remazeilles}, {Renault}, {Renzi}, {Ristorcelli}, {Rocha}, {Rosset}, {Rossetti}, {Roudier}, {Rouill{\'e} d'Orfeuil}, {Rowan-Robinson}, {Rubi{\~n}o-Mart{\'\i}n}, {Rusholme}, {Said}, {Salvatelli}, {Salvati}, {Sandri}, {Santos}, {Savelainen}, {Savini}, {Scott}, {Seiffert}, {Serra}, {Shellard}, {Spencer}, {Spinelli}, {Stolyarov}, {Stompor}, {Sudiwala}, {Sunyaev}, {Sutton}, {Suur-Uski}, {Sygnet}, {Tauber}, {Terenzi}, {Toffolatti}, {Tomasi}, {Tristram}, {Trombetti}, {Tucci}, {Tuovinen}, {T{\"u}rler}, {Umana}, {Valenziano}, {Valiviita}, {Van Tent},
  {Vielva}, {Villa}, {Wade}, {Wandelt}, {Wehus}, {White}, {White}, {Wilkinson}, {Yvon}, {Zacchei}, \& {Zonca}}]{planck2015}
{Planck Collaboration}, {Ade}, P.~A.~R., {Aghanim}, N., {et~al.} 2016, \aap, 594, A13

\bibitem[{Portinari {et~al.}(1997)Portinari, Chiosi, \& Bressan}]{portinari1997galacticchemicalenrichmentnew}
Portinari, L., Chiosi, C., \& Bressan, A. 1997, Galactic chemical enrichment with new metallicity dependent yields

\bibitem[{{Portinari} {et~al.}(1998){Portinari}, {Chiosi}, \& {Bressan}}]{portinari}
{Portinari}, L., {Chiosi}, C., \& {Bressan}, A. 1998, \aap, 334, 505

\bibitem[{{Price-Whelan} {et~al.}(2018){Price-Whelan}, {Sip{'{o}}cz}, {G{"u}nther}, {Lim}, {Crawford}, {Conseil}, {Shupe}, {Craig}, {Dencheva}, {Ginsburg}, {VanderPlas}, {Bradley}, {P{'e}rez-Su{'a}rez}, {de Val-Borro}, {Paper Contributors}, {Aldcroft}, {Cruz}, {Robitaille}, {Tollerud}, {Coordination Committee}, {Ardelean}, {Babej}, {Bach}, {Bachetti}, {Bakanov}, {Bamford}, {Barentsen}, {Barmby}, {Baumbach}, {Berry}, {Biscani}, {Boquien}, {Bostroem}, {Bouma}, {Brammer}, {Bray}, {Breytenbach}, {Buddelmeijer}, {Burke}, {Calderone}, {Cano Rodr{'i}guez}, {Cara}, {Cardoso}, {Cheedella}, {Copin}, {Corrales}, {Crichton}, {D{ extquoteright}Avella}, {Deil}, {Depagne}, {Dietrich}, {Donath}, {Droettboom}, {Earl}, {Erben}, {Fabbro}, {Ferreira}, {Finethy}, {Fox}, {Garrison}, {Gibbons}, {Goldstein}, {Gommers}, {Greco}, {Greenfield}, {Groener}, {Grollier}, {Hagen}, {Hirst}, {Homeier}, {Horton}, {Hosseinzadeh}, {Hu}, {Hunkeler}, {Ivezi{'c}}, {Jain}, {Jenness}, {Kanarek}, {Kendrew}, {Kern}, {Kerzendorf}, {Khvalko}, {King},
  {Kirkby}, {Kulkarni}, {Kumar}, {Lee}, {Lenz}, {Littlefair}, {Ma}, {Macleod}, {Mastropietro}, {McCully}, {Montagnac}, {Morris}, {Mueller}, {Mumford}, {Muna}, {Murphy}, {Nelson}, {Nguyen}, {Ninan}, {N{"o}the}, {Ogaz}, {Oh}, {Parejko}, {Parley}, {Pascual}, {Patil}, {Patil}, {Plunkett}, {Prochaska}, {Rastogi}, {Reddy Janga}, {Sabater}, {Sakurikar}, {Seifert}, {Sherbert}, {Sherwood-Taylor}, {Shih}, {Sick}, {Silbiger}, {Singanamalla}, {Singer}, {Sladen}, {Sooley}, {Sornarajah}, {Streicher}, {Teuben}, {Thomas}, {Tremblay}, {Turner}, {Terr{'o}n}, {van Kerkwijk}, {de la Vega}, {Watkins}, {Weaver}, {Whitmore}, {Woillez}, {Zabalza}, \& {Contributors}}]{astropy:2018}
{Price-Whelan}, A.~M., {Sip{'{o}}cz}, B.~M., {G{"u}nther}, H.~M., {et~al.} 2018, \aj, 156, 123

\bibitem[{{Ratcliffe} {et~al.}(2025){Ratcliffe}, {Khoperskov}, {Minchev}, {Lee}, {Buck}, {Marques}, {Bernaldez}, {Lu}, \& {Steinmetz}}]{Ratcliffe2025}
{Ratcliffe}, B., {Khoperskov}, S., {Minchev}, I., {et~al.} 2025, \aap, 698, A267

\bibitem[{{Ratcliffe} {et~al.}(2023){Ratcliffe}, {Minchev}, {Anders}, {Khoperskov}, {Guiglion}, {Buck}, {Cunha}, {Queiroz}, {Nitschelm}, {Meszaros}, {Steinmetz}, {de Jong}, {Nepal}, {Lane}, \& {Sobeck}}]{Ratcliffe2023}
{Ratcliffe}, B., {Minchev}, I., {Anders}, F., {et~al.} 2023, \mnras, 525, 2208

\bibitem[{Ritter {et~al.}(2018)Ritter, Herwig, Jones, Pignatari, Fryer, \& Hirschi}]{Ritter_2018}
Ritter, C., Herwig, F., Jones, S., {et~al.} 2018, \mnras, 480, 538–571

\bibitem[{{Romano} {et~al.}(2005){Romano}, {Chiappini}, {Matteucci}, \& {Tosi}}]{2005A&A...430..491R}
{Romano}, D., {Chiappini}, C., {Matteucci}, F., \& {Tosi}, M. 2005, \aap, 430, 491

\bibitem[{{Romano} {et~al.}(2010){Romano}, {Karakas}, {Tosi}, \& {Matteucci}}]{2010A&A...522A..32R}
{Romano}, D., {Karakas}, A.~I., {Tosi}, M., \& {Matteucci}, F. 2010, \aap, 522, A32

\bibitem[{{Rybizki} \& {Just}(2015)}]{Rybizki2015}
{Rybizki}, J. \& {Just}, A. 2015, \mnras, 447, 3880

\bibitem[{Rybizki {et~al.}(2017)Rybizki, Just, \& Rix}]{Rybizki_2017}
Rybizki, J., Just, A., \& Rix, H.-W. 2017, \aap, 605, A59

\bibitem[{{Sante} {et~al.}(2026){Sante}, {Font}, {Kawata}, {Makinen}, \& {Grand}}]{Sante2026}
{Sante}, A., {Font}, A.~S., {Kawata}, D., {Makinen}, T.~L., \& {Grand}, R. J.~J. 2026, \mnras

\bibitem[{{Sawala} {et~al.}(2016){Sawala}, {Frenk}, {Fattahi}, {Navarro}, {Bower}, {Crain}, {Dalla Vecchia}, {Furlong}, {Helly}, {Jenkins}, {Oman}, {Schaller}, {Schaye}, {Theuns}, {Trayford}, \& {White}}]{Sawala2016}
{Sawala}, T., {Frenk}, C.~S., {Fattahi}, A., {et~al.} 2016, \mnras, 457, 1931

\bibitem[{{Sch{\"o}nrich} \& {Binney}(2009)}]{2009MNRAS.396..203S}
{Sch{\"o}nrich}, R. \& {Binney}, J. 2009, \mnras, 396, 203

\bibitem[{Seitenzahl {et~al.}(2012)Seitenzahl, Ciaraldi-Schoolmann, Röpke, Fink, Hillebrandt, Kromer, Pakmor, Ruiter, Sim, \& Taubenberger}]{Seitenzahl2013}
Seitenzahl, I.~R., Ciaraldi-Schoolmann, F., Röpke, F.~K., {et~al.} 2012, \mnras, 429, 1156

\bibitem[{Song \& Ermon(2020)}]{song2020improvedtechniquestrainingscorebased}
Song, Y. \& Ermon, S. 2020, in Advances in Neural Information Processing Systems (NeurIPS), Vol.~33

\bibitem[{{Song} {et~al.}(2020){Song}, {Sohl-Dickstein}, {Kingma}, {Kumar}, {Ermon}, \& {Poole}}]{song2020}
{Song}, Y., {Sohl-Dickstein}, J., {Kingma}, D.~P., {et~al.} 2020, arXiv e-prints, arXiv:2011.13456

\bibitem[{{Spitoni} {et~al.}(2022){Spitoni}, {Aguirre B{\o}rsen-Koch}, {Verma}, \& {Stokholm}}]{Spitoni2022}
{Spitoni}, E., {Aguirre B{\o}rsen-Koch}, V., {Verma}, K., \& {Stokholm}, A. 2022, \aap, 663, A174

\bibitem[{Talts {et~al.}(2020)Talts, Betancourt, Simpson, Vehtari, \& Gelman}]{talts2020validatingbayesianinferencealgorithms}
Talts, S., Betancourt, M., Simpson, D., Vehtari, A., \& Gelman, A. 2020, Validating Bayesian Inference Algorithms with Simulation-Based Calibration

\bibitem[{{Thielemann} {et~al.}(2003){Thielemann}, {Argast}, {Brachwitz}, {Hix}, {H{\"o}flich}, {Liebend{\"o}rfer}, {Martinez-Pinedo}, {Mezzacappa}, {Panov}, \& {Rauscher}}]{2003NuPhA.718..139T}
{Thielemann}, F.~K., {Argast}, D., {Brachwitz}, F., {et~al.} 2003, \nphysa, 718, 139

\bibitem[{{Ting} \& {Weinberg}(2022)}]{Ting2022}
{Ting}, Y.-S. \& {Weinberg}, D.~H. 2022, \apj, 927, 209

\bibitem[{{van Dokkum} {et~al.}(2013){van Dokkum}, {Leja}, {Nelson}, {Patel}, {Skelton}, {Momcheva}, {Brammer}, {Whitaker}, {Lundgren}, {Fumagalli}, {Conroy}, {F{\"o}rster Schreiber}, {Franx}, {Kriek}, {Labb{\'e}}, {Marchesini}, {Rix}, {van der Wel}, \& {Wuyts}}]{2013ApJ...771L..35V}
{van Dokkum}, P.~G., {Leja}, J., {Nelson}, E.~J., {et~al.} 2013, \apjl, 771, L35

\bibitem[{Ventura {et~al.}(2013)Ventura, Di~Criscienzo, Carini, \& D’Antona}]{Ventura13}
Ventura, P., Di~Criscienzo, M., Carini, R., \& D’Antona, F. 2013, \mnras, 431, 3642

\bibitem[{{Vincenzo} {et~al.}(2015){Vincenzo}, {Matteucci}, {Recchi}, {Calura}, {McWilliam}, \& {Lanfranchi}}]{2015MNRAS.449.1327V}
{Vincenzo}, F., {Matteucci}, F., {Recchi}, S., {et~al.} 2015, \mnras, 449, 1327

\bibitem[{{Vincenzo} {et~al.}(2019){Vincenzo}, {Spitoni}, {Calura}, {Matteucci}, {Silva Aguirre}, {Miglio}, \& {Cescutti}}]{Vincenzo2019}
{Vincenzo}, F., {Spitoni}, E., {Calura}, F., {et~al.} 2019, \mnras, 487, L47

\bibitem[{Virtanen {et~al.}(2020)Virtanen, Gommers, Oliphant, Haberland, Reddy, Cournapeau, Burovski, Peterson, Weckesser, Bright, {van der Walt}, Brett, Wilson, Millman, Mayorov, Nelson, Jones, Kern, Larson, Carey, Polat, Feng, Moore, {VanderPlas}, Laxalde, Perktold, Cimrman, Henriksen, Quintero, Harris, Archibald, Ribeiro, Pedregosa, {van Mulbregt}, \& {SciPy 1.0 Contributors}}]{2020SciPy-NMeth}
Virtanen, P., Gommers, R., Oliphant, T.~E., {et~al.} 2020, Nature Methods, 17, 261

\bibitem[{{Viterbo} {et~al.}(2026){Viterbo}, {Arruda}, \& {Buck}}]{Viterbo2026b}
{Viterbo}, G., {Arruda}, J., \& {Buck}, T. 2026, arXiv e-prints, arXiv:2607.20725

\bibitem[{{Viterbo} \& {Buck}(2024)}]{Viterbo2024}
{Viterbo}, G. \& {Buck}, T. 2024, arXiv e-prints, arXiv:2411.17269

\bibitem[{{Viterbo} \& {Buck}(2026)}]{Viterbo2026}
{Viterbo}, G. \& {Buck}, T. 2026, \aap, 707, A363

\bibitem[{Wagenmakers \& Farrell(2004)}]{Wagenmakers2004}
Wagenmakers, E.-J. \& Farrell, S. 2004, Psychonomic Bulletin \& Review, 11, 192

\bibitem[{{Wainer} {et~al.}(2024){Wainer}, {Williams}, {Johnson}, {Weisz}, {Dalcanton}, {Seth}, {Dolphin}, {Durbin}, {Bell}, {Chen}, {Guhathakurta}, {Koch}, {Lindberg}, {Rosolowsky}, {Sandstrom}, {Skillman}, {Smercina}, \& {TorresVillanueva}}]{Wainer2024}
{Wainer}, T.~M., {Williams}, B.~F., {Johnson}, L.~C., {et~al.} 2024, \aj, 168, 86

\bibitem[{{Wang} {et~al.}(2024){Wang}, {Carrillo}, {Ness}, \& {Buck}}]{Wang2024}
{Wang}, K., {Carrillo}, A., {Ness}, M.~K., \& {Buck}, T. 2024, \mnras, 527, 321

\bibitem[{{Weinberg} {et~al.}(2024){Weinberg}, {Griffith}, {Johnson}, \& {Thompson}}]{Weinberg2024}
{Weinberg}, D.~H., {Griffith}, E.~J., {Johnson}, J.~W., \& {Thompson}, T.~A. 2024, \apj, 973, 122

\bibitem[{{Weinberg} {et~al.}(2019){Weinberg}, {Holtzman}, {Hasselquist}, {et~al.}}]{2019ApJ...874..102W}
{Weinberg}, D.~H., {Holtzman}, J.~A., {Hasselquist}, S., {et~al.} 2019, \apj, 874, 102

\bibitem[{{Weisz} {et~al.}(2015){Weisz}, {Johnson}, {Foreman-Mackey}, {Dolphin}, {Beerman}, {Williams}, {Dalcanton}, {Rix}, {Hogg}, {Fouesneau}, {Johnson}, {Bell}, {Boyer}, {Gouliermis}, {Guhathakurta}, {Kalirai}, {Lewis}, {Seth}, \& {Skillman}}]{2015ApJ...806..198W}
{Weisz}, D.~R., {Johnson}, L.~C., {Foreman-Mackey}, D., {et~al.} 2015, \apj, 806, 198

\bibitem[{Yeh {et~al.}(2024)Yeh, Chen, Wu, Chen, Vi{\'e}gas, \& Wattenberg}]{yeh2023attentionvizglobalviewtransformer}
Yeh, C., Chen, Y., Wu, A., {et~al.} 2024, IEEE Transactions on Visualization and Computer Graphics, 30, 262

\bibitem[{{Zhao} {et~al.}(2012){Zhao}, {Zhao}, {Chu}, {Jing}, \& {Deng}}]{lamost2012}
{Zhao}, G., {Zhao}, Y.-H., {Chu}, Y.-Q., {Jing}, Y.-P., \& {Deng}, L.-C. 2012, Research in Astronomy and Astrophysics, 12, 723

\bibitem[{{Zhou} {et~al.}(2025){Zhou}, {Radev}, {Oliver}, {Obreja}, {Jin}, \& {Buck}}]{Zhou2025}
{Zhou}, L., {Radev}, S.~T., {Oliver}, W.~H., {et~al.} 2025, \aap, 701, A44

\end{thebibliography}

\begin{appendix}
\nolinenumbers

\section{The \C engine} \label{sec:COMPASS.Core}
At the heart of \C lies the \texttt{ScoreBasedInferenceModel} (\texttt{SBIm}), an implementation built upon the principles of stochastic differential equations (SDEs) for generative modeling.
This section details the forward and reverse diffusion process, the architecture of the score-predicting \texttt{ConditionTransformer}, and its training and sampling procedures.

\subsection{The diffusion process} \label{sec:COMPASS.Core.Diffusion}
The foundation of the generative model is a forward diffusion process that gradually perturbs input data $\mathbf{z}_0$ (representing the joint vector $(\theta,x)$) into a simple noise distribution over a continuous time variable $t \in [0, 1]$. 
This process is defined by the stochastic differential equation (SDE):
\begin{align} 
d\mathbf{z} = \mathbf{f}(\mathbf{z},t) dt + g(t) d\mathbf{w} 
\end{align}
where $\mathbf{f}(\cdot, t): \mathbb{R}^D \to \mathbb{R}^D$ is the drift coefficient, $g(t) \in \mathbb{R}$ is the diffusion coefficient, and $\mathbf{w}$ represents standard Brownian motion. 
\C employs a Variance Exploding SDE (VESDE) \cite{song2020improvedtechniquestrainingscorebased}, where the drift and diffusion coefficients are specifically:
\begin{align} 
\mathbf{f}(\mathbf{z}, t) = \mathbf{0} \quad \text{and} \quad g(t) = \sigma^t
\end{align}
Here, $\sigma > 1$ is a hyperparameter controlling the diffusion scale.
The perturbed data distribution $p_{0t}(\mathbf{z}_t|\mathbf{z}_0)$ at time $t$, given an initial state $\mathbf{z}_0$, is a Gaussian:
\begin{align} 
p_{0t}(\mathbf{z}_t|\mathbf{z}_0) = \mathcal{N} \left( \mathbf{z}_t | \mathbf{z}_0, \frac{\sigma^{2t}-1}{2\ln \sigma}\mathbf{I} \right)
\end{align}
Thus, the variance of the added noise at time $t$ is $\sigma_t^2 = \frac{\sigma^{2t}-1}{2\ln \sigma}$, which grows exponentially in the diffusion process.

The generative process involves reversing this diffusion guided by the corresponding reverse-time SDE \cite{ANDERSON1982313}:
\begin{align} 
d\mathbf{z} = \left[\mathbf{f}(\mathbf{z}, t) - g^2(t)\nabla_{\mathbf{z}}\log p_t(\mathbf{z}) \right] dt + g(t) d\bar{\mathbf{w}} 
\end{align}
where $\bar{\mathbf{w}}$ is Brownian motion in reverse time, and $dt$ is an infinitesimal negative time step. 
For the chosen VESDE ($ \mathbf{f}(\mathbf{z}, t) = \mathbf{0}, g(t) = \sigma^t $), this simplifies to:
\begin{align} 
d\mathbf{z} = \left[ -\sigma^{2t}\nabla_{\mathbf{z}_t}\log p_t(\mathbf{z}_t) \right] dt + \sigma^t d\bar{\mathbf{w}} \label{eq:ReverseSDE}
\end{align}
Generation of samples from the target data distribution $p_0(\mathbf{z})$ starting from a simple Gaussian prior $p_1(\mathbf{z})$ (conventionally $T=1$) requires estimating the score function, $\nabla_{\mathbf{z}_t}\log p_t(\mathbf{z}_t)$, at each time $t$.

\subsection{Conditional transformer architecture} \label{sec:COMPASS.Core.Transformer}

\begin{figure}
    \centering
    \includegraphics[width=.8\linewidth]{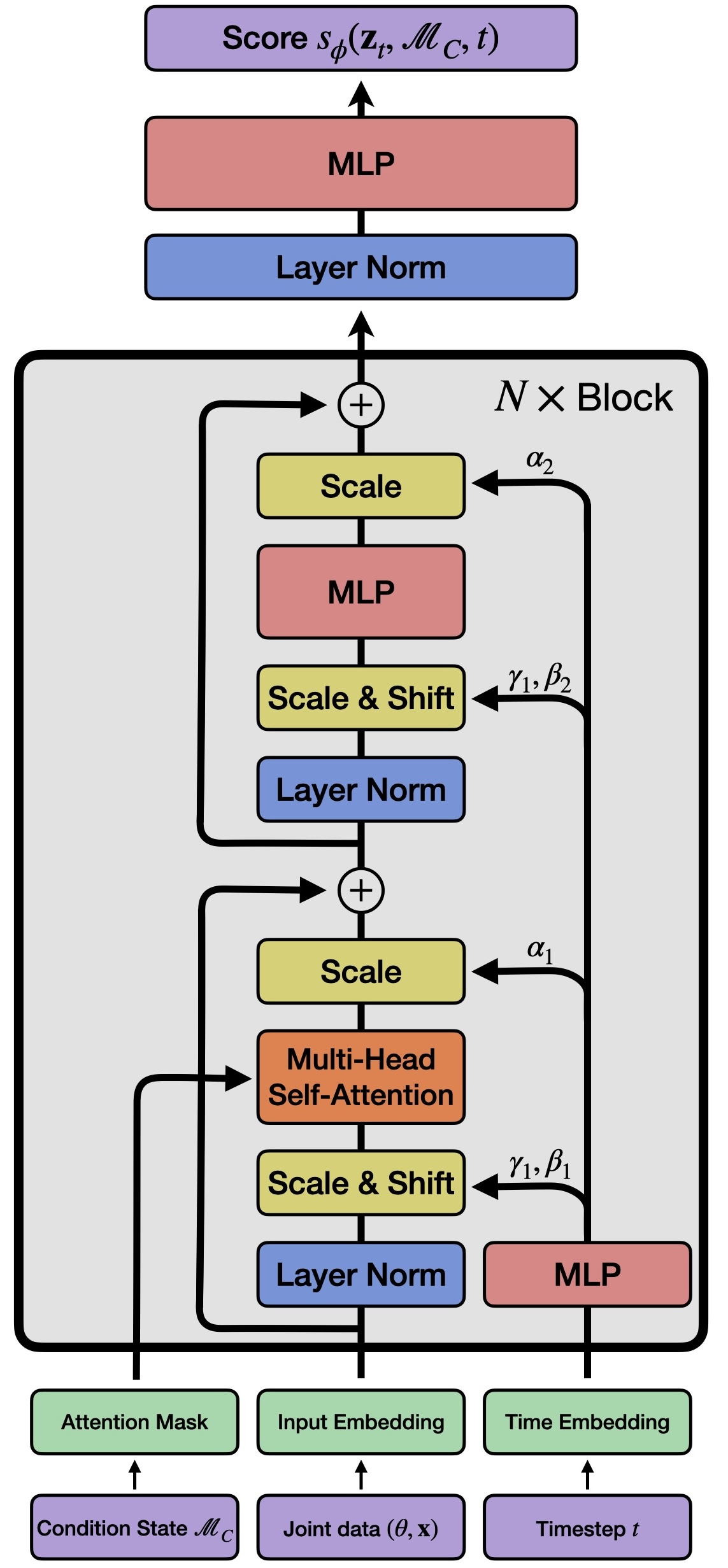}
    \caption{Condition transformer architecture. 
The network, built upon a transformer model, is designed to predict the conditional score. The joint data $(\theta,x)$ are provided as input tokens.
Conditioning is achieved in two ways: the diffusion timestep $t$ is incorporated via adaptive layer normalization (Scale \& Shift), and the condition mask $\mathcal{M}_C$ generates a self-attention mask to govern the information flow between tokens.}
\label{fig:ConditionTransformer}
\end{figure}

The \texttt{ConditionTransformer} in \C approximates the time-dependent conditional score function in Equation \ref{eq:ReverseSDE} using a neural network with parameters $\phi$. 
\begin{align}
 \nabla_{\mathbf{z}_t}\log p_t(\mathbf{z}_t) \approx s_{\phi}(\mathbf{z}_t, \mathcal{M}_C, t)
\end{align}

The architecture is inspired by recent advancements in generative modeling \cite{peebles2023scalablediffusionmodelstransformers, gloeckler2024allinonesimulationbasedinference}, introducing key modifications for flexible and accurate conditional density estimation.
\C employs a transformer architecture based on the Diffusion Transformer (DiT) \cite{peebles2023scalablediffusionmodelstransformers}, specifically utilizing adaptive layer normalization initialized at zero (adaLN-Zero) and incorporating the timestep $t$ as additional condition context. \\

The input to the \C transformer is a joint vector $\mathbf{z} = (\theta, x)$, where $\theta \in \mathbb{R}^{D_\theta}$ are the model parameters and $x \in \mathbb{R}^{D_x}$ are the corresponding observations. 
The total dimensionality of the input is $D = D_\theta + D_x$.
Crucially, a binary condition mask $\mathcal{M}_C \in \{0, 1\}^D$ is also provided as input. This mask dictates the role of each element in $\mathbf{z}$:
\begin{itemize}
    \item $\mathcal{M}_C^{(j)} = 1$: The $j$-th element $z^{(j)}$ is an observed or conditional value.
    \item $\mathcal{M}_C^{(j)} = 0$: The $j$-th element $z^{(j)}$ is a latent value to be inferred or generated.
\end{itemize}

The input data $\mathbf{z}$ are first projected into a high-dimensional embedding space and the diffusion timestep $t$ is embedded using a sinusoidal timestep embedding.
These embeddings, along with the condition mask $\mathcal{M}_C$, are processed by a series of transformer blocks, as illustrated in Figure \ref{fig:ConditionTransformer}.
The Multi-Head Self-Attention layers within these blocks are modified with a custom attention mask derived from $\mathcal{M}_C$.
This attention mask ensures that latent tokens (where $\mathcal{M}_C^{(j)} = 0$) can only attend to observed/conditional tokens (where $\mathcal{M}_C^{(k)} = 1$), but not to other latent tokens.
This focused attention mechanism allows the model to effectively learn the conditional score function.

This conditional architecture allows \C to serve dual roles:
\begin{itemize}
    \item Neural Posterior Estimator (NPE): To estimate $P(\theta|\mathbf{x}_{\text{obs}})$, $\mathbf{x}_{\text{obs}}$ is set as the condition ($\mathcal{M}_C$ elements for $\mathbf{x}$ are 1, for $\theta$ are 0). The model then learns to denoise and generate samples of $\theta$.
    \item Neural Likelihood Estimator (NLE): To estimate $P(\mathbf{x}|\theta_{\text{cond}})$, $\theta_{\text{cond}}$ is set as the condition ($\mathcal{M}_C$ elements for $\theta$ are 1, for $\mathbf{x}$ are 0). The model learns to denoise and generate samples of $\mathbf{x}$.
\end{itemize}

\subsection{Training \C} \label{sec:COMPASS.Core.Training}
\begin{minipage}[t]{\linewidth}
\centering
\begin{algorithm}[H]
    \algnewcommand\algorithmicsample{\textbf{Sample:}}
    \algnewcommand\Sample{\item[\algorithmicsample]}
    \algnewcommand\algorithmicupdate{\textbf{Update:}}
    \algnewcommand\Update{\item[\algorithmicupdate]}
    
    \caption{Training \C} \label{alg:Training}
    \begin{algorithmic}
    \Require $z_0 = (\theta,x)$
    \State \vphantom{$\mathcal{M}_C \gets (\{1\}^{D_{\theta}},\{0\}^{D_{x}})$}
    \Sample $\epsilon \sim \mathcal{N}(0,\mathbf{I})$
    \Sample $t \sim \text{Uniform}(0,1)$
    \Sample $\mathcal{M}_C \sim \text{Bernoulli}(1/3)$
    \State \vphantom{$\mathcal{M}_C \gets (\{1\}^{D_{\theta}},\{0\}^{D_{x}})$}
    \State $\sigma_t \gets  \sqrt{\frac{\sigma^{2t}-1}{2\ln \sigma}}$%
    \State $z_t \gets \mathcal{M}_C \cdot z_0 + (1-\mathcal{M}_C) \cdot \sigma_t \epsilon$
    \State $\hat s \gets s_{\phi}(\mathbf{z}_t, \mathcal{M}_C, t)$
    \State $\mathcal{L}(\phi) \gets \sigma_ t^2 \Vert (1-\mathcal{M}_C) \cdot (\epsilon - \sigma_ t \hat s) \Vert ^2$
    \State \vphantom{$\mathcal{N}(0,\mathbf{I})$}
    \Update $\phi \gets \phi - \eta \nabla_{\phi} \mathcal{L}_{avg}$ \vphantom{$\sqrt{2\sigma^2}$}
    \end{algorithmic}
\end{algorithm}
\end{minipage}

The \texttt{ScoreBasedInferenceModel} $s_{\phi}(\mathbf{z}_t, \mathcal{M}_C, t)$ is trained by minimizing a weighted sum of denoising score matching objectives.
Specifically, the objective is to learn the score of the data distribution conditioned on the observed values specified by $\mathcal{M}_C$.
During the training process, the condition mask $\mathcal{M}_C$ is sampled randomly at every training batch from a Bernoulli distribution, as proposed by \citet{gloeckler2024allinonesimulationbasedinference}, to learn the correlation between all tokens.

\C is implemented in PyTorch and designed for efficient training on multiple GPUs.
The \texttt{AdamWScheduleFree} optimizer from the \texttt{schedulefree} package \citep{defazio2024roadscheduled} is employed, which replaces the momentum of an underlying optimizer (AdamW) with a combination of interpolation and averaging.
This approach eliminates the need for manual learning rate schedule tuning while maintaining state-of-the-art performance. 
The training routine of a batch is detailed in Algorithm \ref{alg:Training}.

\subsection{Sampling with \C} \label{sec:COMPASS.Core.Sampling}
\begin{minipage}[t]{\linewidth}
\centering
\begin{algorithm}[H]
    \algnewcommand\algorithmicsample{\textbf{Sample:}}
    \algnewcommand\Sample{\item[\algorithmicsample]}
    \algnewcommand\algorithmicforb{\textbf{for}}
    \algnewcommand\forb{\item[\algorithmicforb]}
    
    \caption{Sampling \C} \label{alg:Sampling}
    \begin{algorithmic}
    \Require $z_{0;\theta} = \theta$ \textbf{or} $z_{0;x} = x$ 
    \State \textbf{if} $z_{0;\theta}$ \textbf{then} $\mathcal{M}_C \gets (\{1\}^{D_{\theta}},\{0\}^{D_{x}})$
    \State \textbf{if} $z_{0;x}$ \textbf{then} $\mathcal{M}_C \gets (\{0\}^{D_{\theta}},\{1\}^{D_{x}})$
    \State
    \Sample $\epsilon \sim \mathcal{N}(0,\mathbf{I})$
    \State $z_1 \gets (1-\mathcal{M}_C) \cdot \epsilon + \mathcal{M}_C \cdot z_{0;\{\theta \text{ or } x\}}$ 
    \State \vphantom{$\sigma_t \gets  \sqrt{\frac{\sigma^{2t}-1}{2\ln \sigma}}$}
    \forb $t=1$ \textbf{to} $0$
    \State $z_{t'} \gets z_t + (1-\mathcal{M}_C) \sigma_t s_{\phi}(\mathbf{z}_t, \mathcal{M}_C, t) dt $
    \State \textbf{if} Corrector Step  \textbf{then}
        \State $\quad n \sim \mathcal{N}(0,\mathbf{I})$
        \State $\quad  z_{t'} \gets z_{t'} + (1-\mathcal{M}_C) \cdot ( \delta_L \sigma^2_{t'} s_{\phi} + \sqrt{2 \delta_L \sigma^2_{t'}} \cdot n) $
    \end{algorithmic}
\end{algorithm}
\end{minipage}

Once the score model \texttt{SBIm} $s_{\phi}$ is trained, samples can be generated from the target conditional distribution $p_0(\mathbf{z}_{\text{latent}} | \mathbf{z}_{\text{observed}})$ by numerically solving the reverse SDE (see Sec. \ref{sec:COMPASS.Core.Diffusion}).

\C primarily utilizes the DPM-Solver proposed in \citet{lu2022dpmsolverfastodesolver}, a family of high-order solvers that generally offers better accuracy and efficiency compared to the simpler Euler-Maruyama method for a given number of function evaluations.
The DPM-Solver leverages the semi-linearity of diffusion ODEs and it directly approximates a simplified formulation of the exact solution, which consists of an exponentially weighted integral of the noise prediction model \cite{lu2022dpmsolverfastodesolver}.
The deterministic part of the reverse SDE for the employed VESDE is $-\sigma^{2t} s_{\phi}(\mathbf{z}, \mathcal{M}_C, t)$.
The DPM-Solver integrates this term, often by leveraging the corresponding probability flow ODE \cite{song2020}.
For a time step form $t$ to $t'$ (where $dt = t-t' > 0$) with the condition mask $\mathcal{M}_C$, the first order DPM-Solver update in the \texttt{SBIm} is:
\begin{align}
    \mathbf{z}_{t'} = \mathbf{z}_t - (1-\mathcal{M}_C) \sigma_t s_{\phi}(\mathbf{z}_t, \mathcal{M}_C, t) dt 
\end{align}
Higher-order versions (2nd and 3rd), also implemented in \C, utilize intermediate evaluations of the score function and weighted averages to achieve greater accuracy per step.

To further enhance sample quality, \texttt{COMPASS}'s implementation incorporates optional Langevin corrector steps. 
Following a predictor step a few corrector steps can be applied:
\begin{align}
\mathbf{z}_{t'} = \mathbf{z}_{t'} + (1-\mathcal{M}_C) \cdot \Big( \delta_L \sigma^2_{t'} s_{\phi}(\mathbf{z}_{t'}, \mathcal{M}_C, t') + \sqrt{2 \delta_L \sigma^2_{t'}} \cdot \mathbf{n} \Big)
\end{align}
where $\mathbf{n} \sim \mathcal{N}(0, \mathbf{I})$, $t'$ is the next timestep after the last predictor step, and $\delta_L$ is the signal-to-noise ratio.
These corrector steps inject appropriately scaled noise and re-apply a score-based update, helping to refine samples,  improve adherence to the data manifold, and prevent potential mode collapse. 
In \C, these Langevin corrector steps are applied periodically (e.g., every $5$ main diffusion timesteps by default) and for a fixed number of iterations (e.g., $5$ corrector steps) each time they are triggered.

The combination of the DPM-Solver (predictor) with periodic Langevin correction allows for a more accurate and robust evolution of the diffusion process.
The ODE solver efficiently traverses the probability space, while the stochastic Langevin refinement helps to explore it more thoroughly.
Algorithm \ref{alg:Sampling} details the sampling procedure for generating conditional samples using a first order DPM-Solver coupled with Langevin corrector steps.

\section{Calibration} \label{sec:Results.Calibration}

The performance of score-based inference models is highly sensitive to their hyperparameter configurations. 
This part presents a systematic evaluation of two key components that govern the accuracy and computational efficiency of the model:
the architecture of the neural network used for score estimation and the temporal discretization of the reverse diffusion process. 
Both aspects were optimized independently to balance inference quality against resource demands.

\subsection{Network architecture \& hyperparameter selection} \label{sec:optuna}
The network architecture defines the inductive bias and capacity of the model used to approximate the score function in the reverse diffusion process. 
The performance of the score-based inference model is highly sensitive to the architectural choices that define its inductive bias and capacity.
To identify a robust and high-performing configuration, we conducted a comprehensive hyperparameter optimization study using the \texttt{optuna} framework \citep{akiba2019optunanextgenerationhyperparameteroptimization}.
The optimization trials were targeting the predictive accuracy and the posterior calibration.
The accuracy is measured by the negative log-posterior ($-\log P(\theta|x)$), evaluated at the ground-truth parameters for a set of mock observations.
Lower values indicate that the true parameters fall in regions of higher posterior probability, signifying a more accurate model.
The calibration of the posterior is measured by the maximum deviation of the TARP diagnostic ($\Delta_{\rm max} \rm{TARP}$) from the ideal diagonal line \citep{lemos2023samplingbasedaccuracytestingposterior}.
Lower values indicate better statistical coverage and more reliable uncertainty estimates.
The key architectural hyperparameters explored in this study were the number of transformer blocks (\texttt{depth}), the dimensionality of the token embeddings (\texttt{hidden size}), the expansion factor for the hidden layers in the MLP blocks (\texttt{mlp ratio}), the number of attention heads in the Multi-Head Self-Attention mechanism (\texttt{num heads}), the batch size (\texttt{batch size}) and the maximum noise level in the VESDE schedule (\texttt{sigma}).

Table \ref{table:optuna-results} summarizes the top-performing configurations for each of the two objectives on the GCE example from Section \ref{sec:Results.Yields}.
Trial $275$ emerged as a uniquely strong candidate, as it was the only hyperparameter combination to appear in the top-$10$ list for both predictive accuracy and posterior calibration.
This configuration strikes an excellent balance between generating sharp, accurate posteriors and achieving good calibration on the optimization set (see Appendix~\ref{sec:Results.Calibration.Posterior} for the per-parameter assessment).
We therefore selected the architecture from Trial $275$ as the default for the GCE experiments.

\begin{table}[!htb]
\caption{Top-performing hyperparameter configurations of the \texttt{optuna} study.} \label{table:optuna-results}
\resizebox{\columnwidth}{!}{
\begin{tabular}{c r r | r r r r r r}
    \toprule
    Trial & $-\log P(\theta|x)$ & $\Delta_{\rm max}$TARP & Batch Size & Sigma & Depth & Heads & Hidden Size & MLP Ratio \\
    \midrule
    113 & -1.76 & 0.08 & 862 &  2.5 &   5 &   1 &  36 &  3 \\
    174 & -1.60 & 0.09 & 862 &  2.5 &   5 &   1 &  36 &  3 \\
    215 & -1.90 & 0.10 & 862 &  2.5 &  11 &   1 &  65 &  3 \\
    223 & -1.97 & 0.10 & 375 &  2.5 &   5 &   1 &  36 &  1 \\
    260 & -1.94 & 0.10 & 125 &  2.5 &   5 &   1 &  34 &  1 \\
    242 & -1.76 & 0.10 & 842 &  2.5 &   5 &   1 &  28 &  3 \\
    249 & -1.87 & 0.10 & 862 &  2.5 &   5 &   1 &  34 &  3 \\
    275$^\dagger$ & -2.20 & 0.10 & 125 &  2.5 &   5 &   1 &  65 &  3 \\
    287 & -1.92 & 0.10 & 608 &  2.5 &   5 &   1 &  65 &  3 \\
    277 & -1.74 & 0.11 & 842 &  2.5 &   6 &   1 &  28 &  3 \\
    \midrule
    193 & -2.26 & 0.15 & 125 & 22.2 &  11 &   1 &  65 &  3 \\
    233 & -2.22 & 0.16 & 125 & 23.0 &   3 &   1 &  65 & 10 \\
    213 & -2.22 & 0.12 & 485 &  2.5 &   5 &   1 & 167 &  1 \\
    275$^\dagger$ & -2.20 & 0.10 & 125 &  2.5 &   5 &   1 &  65 &  3 \\
    272 & -2.17 & 0.12 & 485 &  2.5 &   3 &   1 & 167 &  2 \\
    263 & -2.16 & 0.16 & 203 & 27.0 &  11 &   1 &  34 &  3 \\
    256 & -2.15 & 0.11 & 125 &  2.5 &   7 &   1 &  65 &  1 \\
    290 & -2.15 & 0.16 & 161 & 23.0 &   4 &   1 &  65 &  9 \\
    284 & -2.14 & 0.13 & 125 &  3.8 &   3 &   1 & 106 & 10 \\
    195 & -2.10 & 0.16 & 203 & 20.1 &  11 &   1 &  27 &  8 \\
    \bottomrule
\end{tabular}}
\tablefoot{Results of a $1000$-trial \texttt{optuna} study on the GCE task, minimizing both the negative predictive accuracy ($-\log P(\theta|x)$) and the maximal deviation of the TARP diagnostic ($\Delta_{\rm max}$TARP). The upper section lists the top $10$ trials ranked by the $\Delta_{\rm max}$TARP objective, the lower section the top $10$ trials ranked by the $-\log P(\theta|x)$ objective. Trial $275$ (marked with $^\dagger$) is the only configuration to appear in both lists and was selected as the default architecture.}
\end{table}

A crucial insight from the \texttt{optuna} study was the overwhelming importance of the diffusion noise schedule parameter, \texttt{sigma} ($\sigma$). 
The choice of $\sigma$ defines the variance of the terminal noise distribution, $p_T(z) = \mathcal N(0, \sigma^2 )$, into which all data points are perturbed during the forward process. 
For the reverse generative process to be effective, this terminal distribution must fully contain the support of the initial data distribution $p_0(z)$. 
If the initial data contains values that fall outside the typical range of the noise distribution (i.e., if the data variance is larger than the noise variance), the model will fail to generate them accurately.
\texttt{optuna}'s internal feature importance analysis confirmed this theoretical sensitivity, identifying \texttt{sigma} as the most influential hyperparameter.
This highlights that ensuring the validity of the diffusion process's core assumption is more critical to performance than the specific architectural details of the denoiser network.

\subsection{Diffusion time} \label{sec:Results.Calibration.Time}
The accuracy and efficiency of sampling in score-based diffusion models critically depend on the choice of time discretization.
Selecting appropriate time steps is essential to faithfully approximate the reverse SDE and avoid numerical artifacts.
This part aims to find the best trade-off between accuracy and inference speed, as well as compare the Euler-Maruyama scheme to the DPM-Solver (see \ref{sec:COMPASS.Core.Sampling}).

\begin{figure}
    \centering
    \includegraphics[width=1\linewidth]{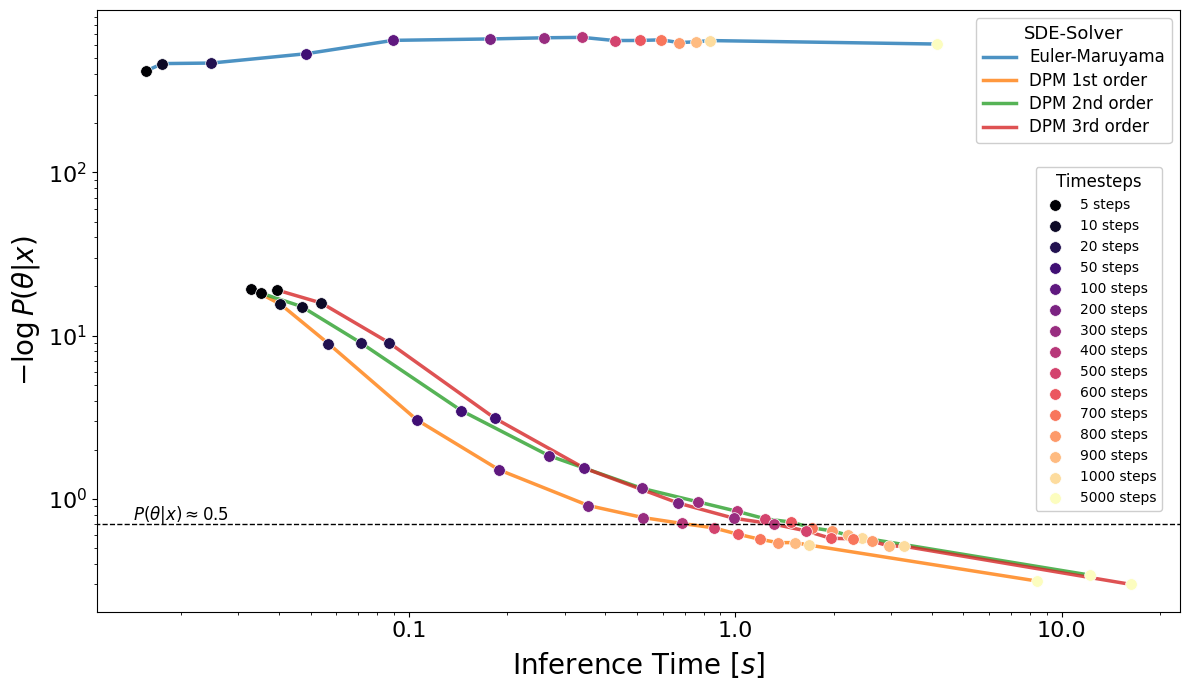}
    \vspace{-1em}
\caption{Accuracy of SDE solvers. 
Comparison of SDE solver performance in terms of predictive accuracy ($-\log P(\theta|x)$) versus inference time per sample. 
Euler-Maruyama and DPM-Solvers (1st, 2nd, and 3rd order) were tested with varying numbers of diffusion steps (indicated by colored points, ranging from 5 to 5000 steps).
Accuracy is averaged over $1000$ mock observations. The dashed horizontal line at $-\log P(\theta|x) \approx 0.693$ represents a posterior probability of $P(\theta|x) = 0.5$ for the true parameter. 
Lower $-\log P(\theta|x)$ values indicate higher accuracy. 
The DPM-Solver (1st order) with $500$ steps offers a good balance of accuracy and computational efficiency.}
\label{fig:AccuracyVsTime}
\end{figure}

Figure \ref{fig:AccuracyVsTime} shows the accuracy of the solver over the inference time per sample. 
The predictive accuracy of the network is measured by inferring $1000$ samples from a test set observation $x$, over which a Gaussian kernel density estimator fits the posterior function, which is then evaluated at the parameters $\theta$ of the test set.
For this test a total of $1000$ mock observations have been evaluated and the figure shows the average inference time and accuracy per sample.
Each stochastic differential equation (SDE) solver is tested on $15$ different diffusion timesteps and the inference time is averaged over the samples.
The evaluation has been conducted on $8$ Nvidia RTX 2080 Ti GPUs in parallel.

The dashed horizontal line in Figure \ref{fig:AccuracyVsTime} is set at $-\log P(\theta|x) \approx 0.693$, corresponding to a posterior probability density value of $P(\theta|x) = 0.5$ evaluated at the true parameter $\theta$. 
This threshold serves as a practical benchmark: values below this line indicate that the true parameter falls within a region of relatively high posterior density, suggesting a reasonably well-concentrated and informative posterior.
Based on the results in figure \ref{fig:AccuracyVsTime}, it can be deduced that the DPM-Solver, across all tested orders (1st, 2nd and 3rd), consistently achieves significantly higher accuracy (lower $-\log P(\theta|x)$ values) compared to the Euler-Maruyama scheme for any given inference time or number of steps.
This makes DPM-Solvers the unequivocally preferred choice for this application. 
While higher-order DPM-Solvers can offer marginal accuracy improvements, these often come with increased computational demands. 
The first order DPM-Solver demonstrates a robust balance, achieving accuracy comparable to its higher-order counterparts but with greater computational efficiency within the targeted accuracy regime.
With $500$ diffusion timesteps, the first order DPM-Solver yields an average $-\log P(\theta|x)$ value comfortably below the $P(\theta|x) = 0.5$ (i.e., $-\log P(\theta|x) \approx 0.693$) acceptability threshold, indicating that the resulting posteriors are well-concentrated around the true parameter values.
The trajectory of the DPM-Solver curves in Figure \ref{fig:AccuracyVsTime} show that increasing the number of timesteps beyond approximately $500$ results in progressively smaller gains in accuracy. 
The curve begins to flatten, signifying that the additional computational expense of more steps yields diminishing returns in terms of reducing $-\log P(\theta|x)$.
An average inference time of approximately $1$ second per sample, as achieved with the first order DPM-Solver and $500$ steps, is deemed a practical and efficient duration for the scope of this work.
This allows for extensive model testing and application without prohibitive computational overhead.
Therefore, the first order DPM-Solver configured with $500$ diffusion timesteps is selected as the default for subsequent inference experiments. This configuration offers a compelling balance, delivering high inference accuracy that meets practical acceptability criteria, while maintaining reasonable computational efficiency and avoiding the diminishing returns associated with a significantly larger number of timesteps.

However, for the Bayesian model comparison process we relax this timestep choice a little to reduce computational cost. 
Bayesian model comparison requires estimating the model posterior $P(\mathcal{M}|x)$. 
This typically involves inferring the parameters $\theta$ for each model $\mathcal{M}$ and then evaluating the likelihood $P(x|\theta, \mathcal{M})$, which can be computationally intensive if full parameter inference is performed for every comparison.
For model comparison we therefore reduced the number of diffusion steps to $50$, which lowers the inference time roughly tenfold, to $\approx 0.1$\,s per sample (Fig.~\ref{fig:AccuracyVsTime}). In the controlled comparison (Sec.~\ref{sec:Results.Yields}) this setting identifies the generating model in all three cases, and on the \citet{Nissen_2020} data the resulting preferred yield set is confirmed independently (Appendix~\ref{sec:classifier}).

\subsection{Posterior calibration} \label{sec:Results.Calibration.Posterior}
\begin{figure}[!htb]
    \centering
    \includegraphics[width=\linewidth]{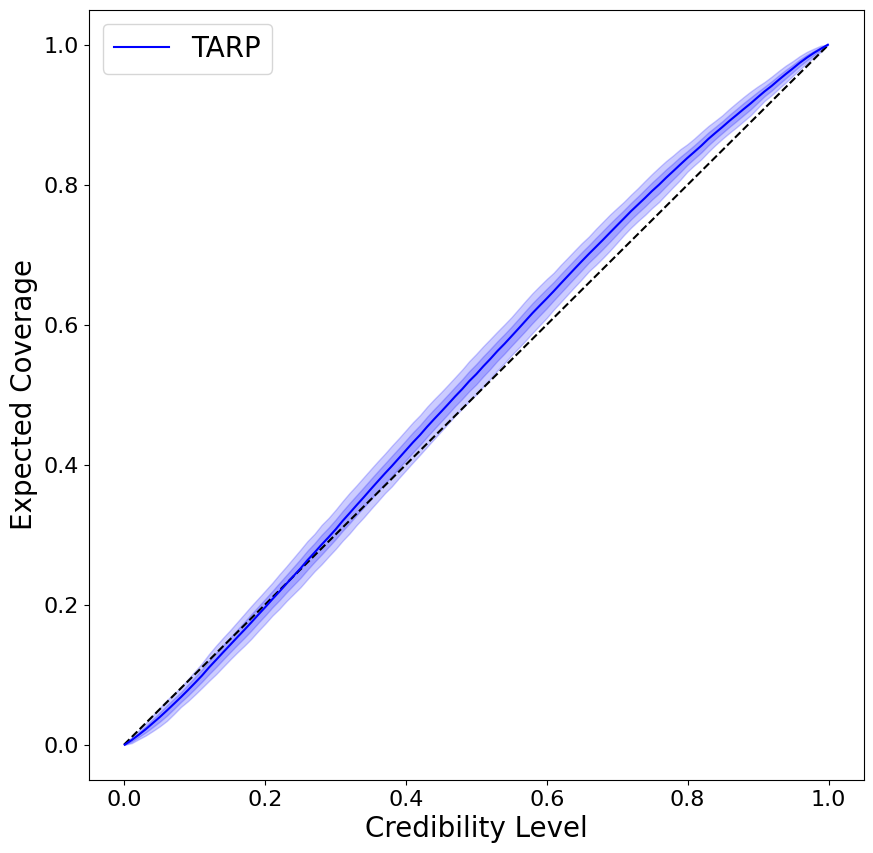}
    \vspace{-1em}
\caption{TARP plot.
The TARP plot shows the expected coverage probability vs. the ground truth parameters under the inferred posterior samples credibility level $\alpha$.}
\label{fig:TARP}
\end{figure}
Following the optimization of the network architecture and the selection of an efficient SDE-solver and diffusion timestep configuration, it is crucial to verify that the resulting inference pipeline produces well-calibrated posterior distributions. 
A calibrated posterior accurately reflects the true uncertainty: an X\% credibility interval should contain the true parameter value X\% of the time over many different observations. 
This part evaluates the calibration using diagnostic plots, primarily TARP \citep{lemos2023samplingbasedaccuracytestingposterior} and true-versus-predicted parameter plots. 

Figure \ref{fig:TARP} presents an aggregate TARP plot, which assesses the overall calibration of the inference procedure across all parameters and test samples.
The expected coverage is plotted against the credibility level.
The aggregate curve lies close to the diagonal, i.e. on average the coverage of the credible regions is approximately correct.

\begin{figure*}
\centering
    \includegraphics[width=1\linewidth]{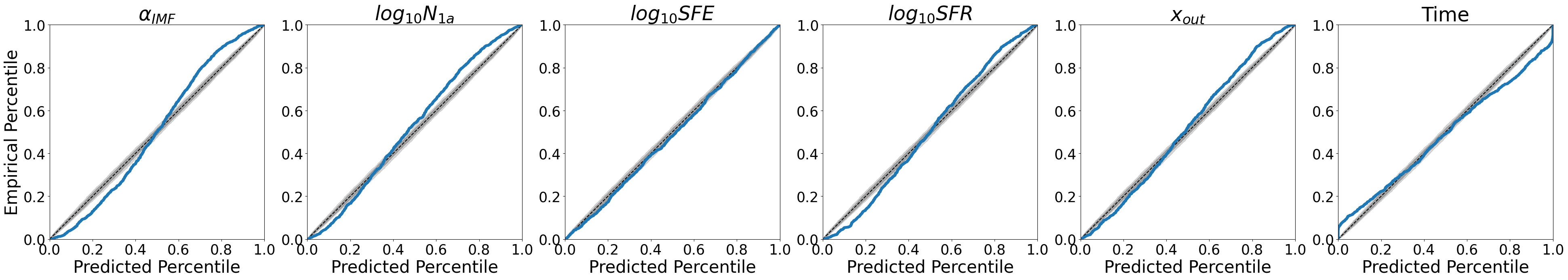}
    \includegraphics[width=1\linewidth]{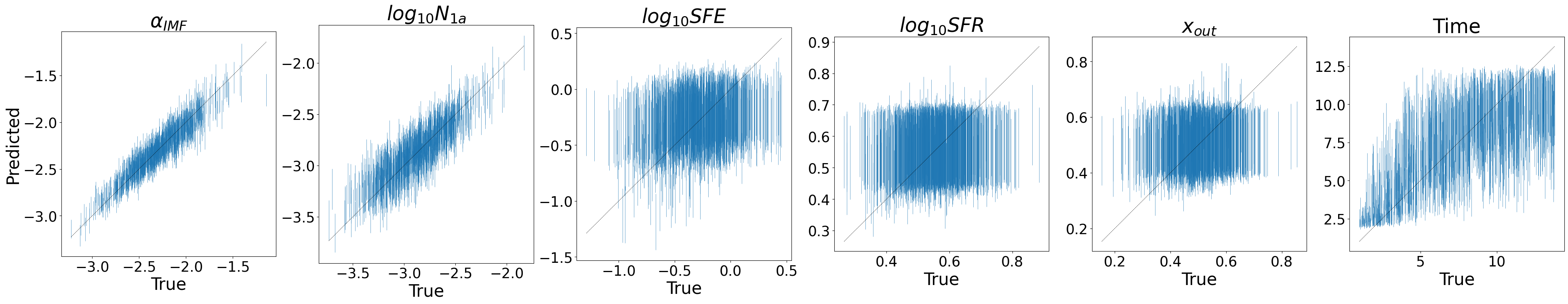}
\caption{Posterior calibration. 
Posterior calibration diagnostics for $1000$ inferred parameters from a set of test observations.
(Top) The individual TARP plots for each parameter. 
(Bottom) True parameter values against the mean of their inferred posterior distributions.
}
\label{fig:Calibration}
\end{figure*}

Figure \ref{fig:Calibration} provides more detailed information on the posterior calibration for each of the six \chem parameters:
the high-mass slope of the IMF $\alpha_{\rm IMF}$, the normalization of SN Type Ia $\log_{10}(N_{\rm Ia})$, the star formation efficiency $\log_{10}(\rm{SFE})$, the peak of the star formation rate $\log_{10}(\rm{SFR}_{\rm peak})$, the stellar feedback fraction $x_{\rm out}$, and the stellar birth time $T$.
For a perfectly calibrated model, these plots should align closely with the $y=x$ diagonal. 

The per-parameter TARP curves (Fig.~\ref{fig:Calibration}, top) reveal deviations that the aggregate curve averages out. For $\alpha_{\rm IMF}$, $\log_{10}(N_{\rm Ia})$ and $\log_{10}(\rm{SFR}_{\rm peak})$ the curves deviate from the diagonal in the direction of under-confidence: the credible regions contain the true value more often than their nominal level, i.e. the posteriors are over-dispersed. We traced part of this over-dispersion to how our first-order solver implementation parameterizes its steps (uniformly in diffusion time rather than in noise scale), which inflates posterior widths. The posterior means of the global parameters, which are of primary interest for this work, correlate tightly with the true values (Fig.~\ref{fig:Calibration}, bottom).
In contrast, the local ISM parameters ($\log_{10}(\rm{SFE})$, $\log_{10}(\rm{SFR}_{\rm peak})$, $x_{\rm out}$) and the stellar birth time ($T$) show considerably more scatter and wider posterior distributions relative to the true values.
While the posteriors for these parameters still generally encompass the true values, the precision is lower. 
This difference in calibration quality is anticipated, as the observational data used are expected to provide stronger constraints on the global parameters that dictate the overall chemical evolution trends, rather than on the specifics of individual star-forming regions or individual stellar ages which have more localized effects or degeneracies.

In summary, the inference is approximately calibrated in aggregate but mildly under-confident for the global parameters. The resulting credible regions are therefore conservative, which should be kept in mind when comparing posterior widths with other methods (Sec.~\ref{sec:Results.Params.Chempy}).
The lower precision for local parameters is acknowledged as a characteristic of the inference problem given the available data constraints.

\begin{figure*}
    \centering
    \includegraphics[width=\linewidth]{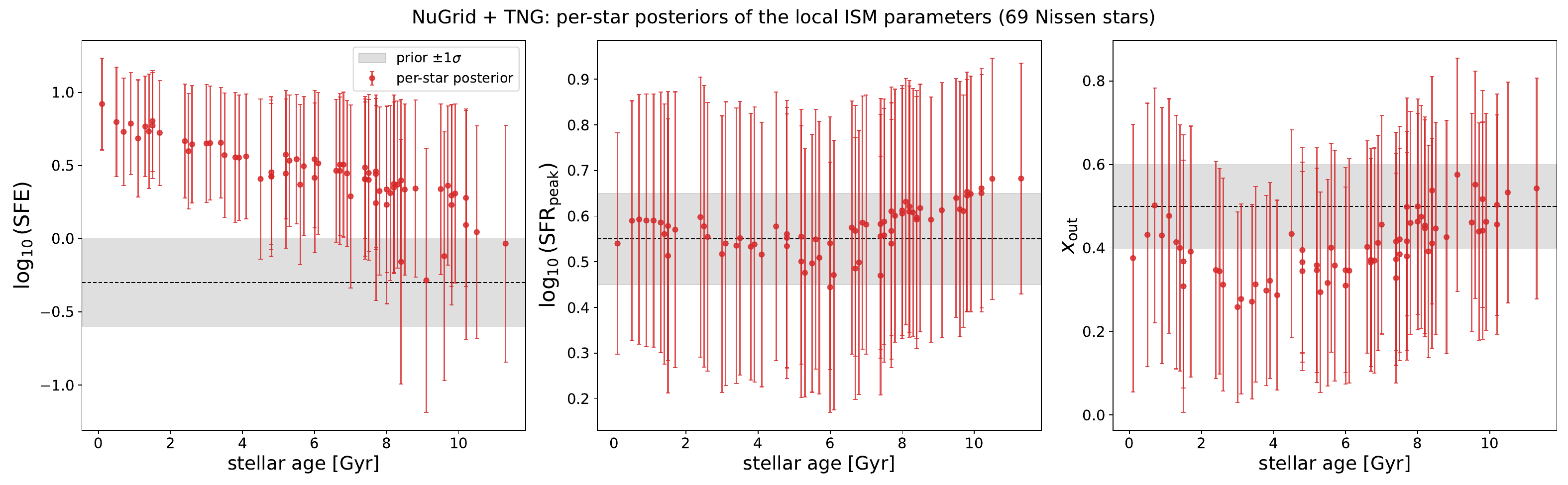}
    \vspace{-1em}
\caption{Local ISM parameters. Single-star posteriors (mean $\pm 1\sigma$) of $\log_{10}({\rm SFE})$, $\log_{10}({\rm SFR}_{\rm peak})$ and $x_{\rm out}$ against stellar age for the $69$ \citet{Nissen_2020} stars, for the preferred NuGrid + TNG yield set. The gray band shows the prior mean $\pm 1\sigma$ (Table~\ref{tab:GCEpriors}).}
\label{fig:local-params}
\end{figure*}

Figure~\ref{fig:local-params} shows the single-star posteriors of the local ISM parameters for the preferred yield set and the \citet{Nissen_2020} stars. Their median widths are $1.7$ ($\log_{10}({\rm SFE})$), $2.8$ ($\log_{10}({\rm SFR}_{\rm peak})$) and $2.6$ ($x_{\rm out}$) times the corresponding prior widths, so individual stars constrain them only weakly. This is why we focus on the global parameters, which are shared by all stars and are constrained by combining them.

\subsection{Layer-wise attention masks} \label{sec:layer-wise-attention}

\begin{figure}
    \centering
    \includegraphics[width=1\linewidth]{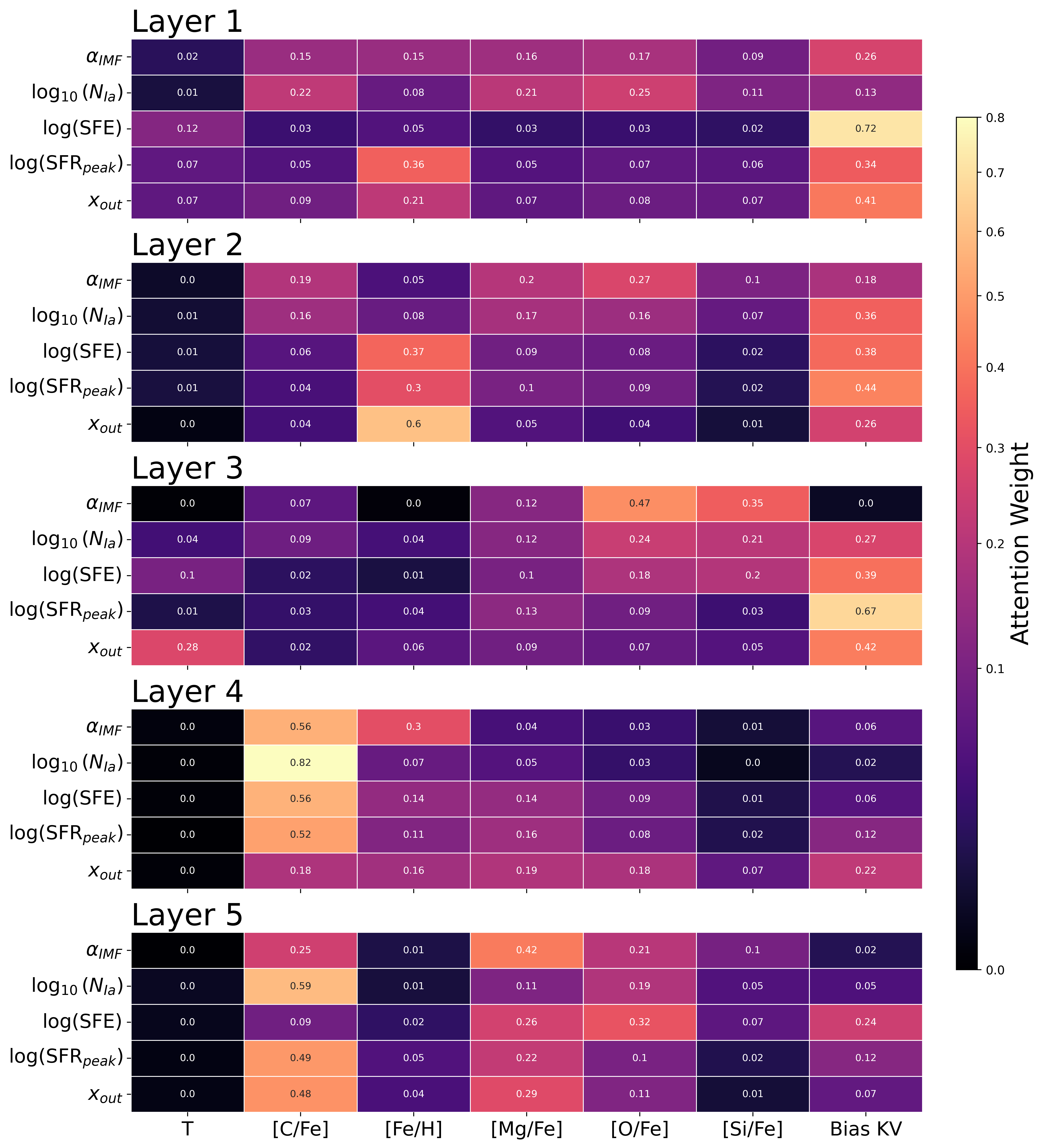}
    \vspace{-1em}
    \caption{Evolution of mean attention across transformer layers. 
Mean attention weights of the \C transformer, visualized for each of its five layers. The weights are averaged over all stars in the observational sample at a fixed diffusion timestep of $t=0.5$. 
The x-axis represents the information sources (keys), including stellar age ($T$), elemental abundances, and a learned bias term.
The y-axis represents the parameters being inferred (queries) which contain the global parameters $\alpha_{\rm IMF}$ and $\log_{10}(N_{\rm Ia})$ as well as local ISM parameters ($\log(\rm{SFE}), \log({\rm{SFR_{peak)}}}$, and $x_{\rm out}$).}
\label{fig:attention_layer}
\end{figure}

Figure \ref{fig:attention_layer} reveals a clear and dynamic evolution of the model's reasoning through its five layers.
In the early layers (1-2), attention from all parameter queries is broadly distributed across multiple observational features and the learned bias term (bias kv).
A key initial finding is the heavy reliance of local ISM parameters, particularly $\log(\rm{SFE})$, on the learned bias term with slight reliance on stellar age.
A significant shift occurs in layer 3, where attention for $\alpha_{\rm IMF}$ sharpens to a value of $0.56$ on Oxygen ([O/Fe]), and attention for local ISM parameters further concentrate on the bias.
In the final layers (4 \& 5), the model builds upon this foundation to form a more sophisticated, multi-variate understanding.
For $\alpha_{\rm IMF}$, the attention remains focused on key alpha-elements like Magnesium ([Mg/Fe]) and Oxygen ([O/Fe]), but also incorporates Carbon ([C/Fe]) and Iron ([Fe/H]).
The $\log_{10}(N_{\rm Ia})$ query develops an interestingly dominant and persistent focus on [C/Fe], reaching a peak attention weight of $0.66$ in Layer 4.

This progression from broad feature aggregation to targeted, physically meaningful associations suggests that the model constructs hierarchical representations rather than relying on single-element shortcuts.
\section{Derivation of model posterior}
\label{sec:AIC}
Model comparison typically relies on model evidence. 
However, in simulator-based settings where the likelihood function is often intractable, calculating model evidence directly is not feasible.
An alternative approach for model comparison, particularly suitable for non-nested models, is to use the Akaike Information Criterion (AIC) \cite{AKAIKE19781877}. 
AIC estimates the prediction error and thereby the relative quality of statistical models for a given set of data.

Given the maximized log-likelihood for model $\mathcal{M}_i$, denoted as $\ln\mathcal{L}(\mathbf{x}|\hat{\theta}_i, \mathcal{M}_i)$ (where $\hat{\theta}_i$ are the parameter values that maximize the likelihood for model $\mathcal{M}_i$), the AIC is defined as:
\begin{align}
    \text{AIC}_i = -2\ln\mathcal{L}(\mathbf{x}|\hat \theta_i, \mathcal{M}_i) + 2k_i \label{eq:aic_def}
\end{align}
where $k_i$ is the number of estimable parameters in model $\mathcal{M}_i$. 
The model with the lowest AIC is generally preferred.

To compare a set of $N$ models, we first calculate the AIC difference for each model $\mathcal{M}_i$ relative to the model with the minimum AIC in the set ($\text{AIC}_{\min}$):
\cite{Wagenmakers2004, AKAIKE19781877, AKAIKE1979, Bozdogan_1987}:
\begin{align}
    \Delta_i(\text{AIC}) &= \text{AIC}_i - \text{AIC}_{\min} \label{eq:delta_aic}
\end{align}
The likelihood of model $\mathcal{M}_i$ being the best model (in the Kullback-Leibler information sense), given the data $\mathbf{x}$, can be estimated relative to the other models using these AIC differences. 
The relative likelihood of model $\mathcal{M}_i$, sometimes called an "Akaike weight", is given by \cite{Wagenmakers2004, burnham2002model}:
\begin{align}
    \mathcal{L}_{\rm rel}(\mathcal{M}_i|\mathbf{x}) &\propto \exp\left(-\frac{1}{2} \Delta_i(\text{AIC}) \right)
\end{align}
To obtain normalized model weights (Akaike weights) for each model, $\mathcal{P}(\mathcal{M}_i|\mathbf{x})$, these relative likelihoods are normalized by summing over all models in the candidate set:
\begin{align}
    \mathcal{P}(\mathcal{M}_i|\mathbf{x}) = \frac{\exp\left(-\frac{1}{2} \Delta_i(\text{AIC})\right)}{\sum_{j=1}^{N} \exp\left(-\frac{1}{2} \Delta_j(\text{AIC})\right)} \label{eq:posterior_prob_delta_aic}
\end{align}
Substituting Eq. \eqref{eq:delta_aic} into Eq. \eqref{eq:posterior_prob_delta_aic}:
\begin{align}
\mathcal{P}(\mathcal{M}_i|\mathbf{x}) &= \frac{\exp\left(-\frac{1}{2} (\text{AIC}_i - \text{AIC}_{\min})\right)}{\sum_{j=1}^{N} \exp\left(-\frac{1}{2} (\text{AIC}_j - \text{AIC}_{\min})\right)} \\
&= \frac{\exp\left(-\frac{1}{2} \text{AIC}_i\right)}{\sum_{j=1}^{N} \exp\left(-\frac{1}{2} \text{AIC}_j\right)} \label{eq:posterior_prob_aic_i}
\end{align}
Now, substituting the definition of $\text{AIC}_i$ from Eq. \eqref{eq:aic_def} and under consideration that all models in \texttt{COMPASS} under comparison have the same number of parameters, i.e., $k_i = k_j = k$ for all $i,j$, then the additional parameter term is common to the numerator and all terms in the sum in the denominator, and thus cancels out. 
In this specific scenario, the posterior model probability simplifies to a direct ratio of the maximized likelihoods of the data given each model:
\begin{align}
    \mathcal{P}(\mathcal{M}_i|\mathbf{x}) &= \frac{\mathcal{L}(\mathbf{x}|\hat \theta_i, \mathcal{M}_i)}
    {\sum_{j=1}^{N} \mathcal{L}(\mathbf{x}|\hat \theta_j, \mathcal{M}_j )} \quad (\text{if } k_i = k_j \text{ for all } i,j)
\end{align}
The model with the highest weight $\mathcal{P}(\mathcal{M}_i|\mathbf{x})$ is then considered the best-supported model by the data, according to this criterion.
For equal $k_i$, the same holds for the BIC (Sec.~\ref{sec:COMPASS.Model-Comp}).

\subsection{Independent check with a neural classifier} \label{sec:classifier}
\begin{figure}
    \centering
    \includegraphics[width=\linewidth]{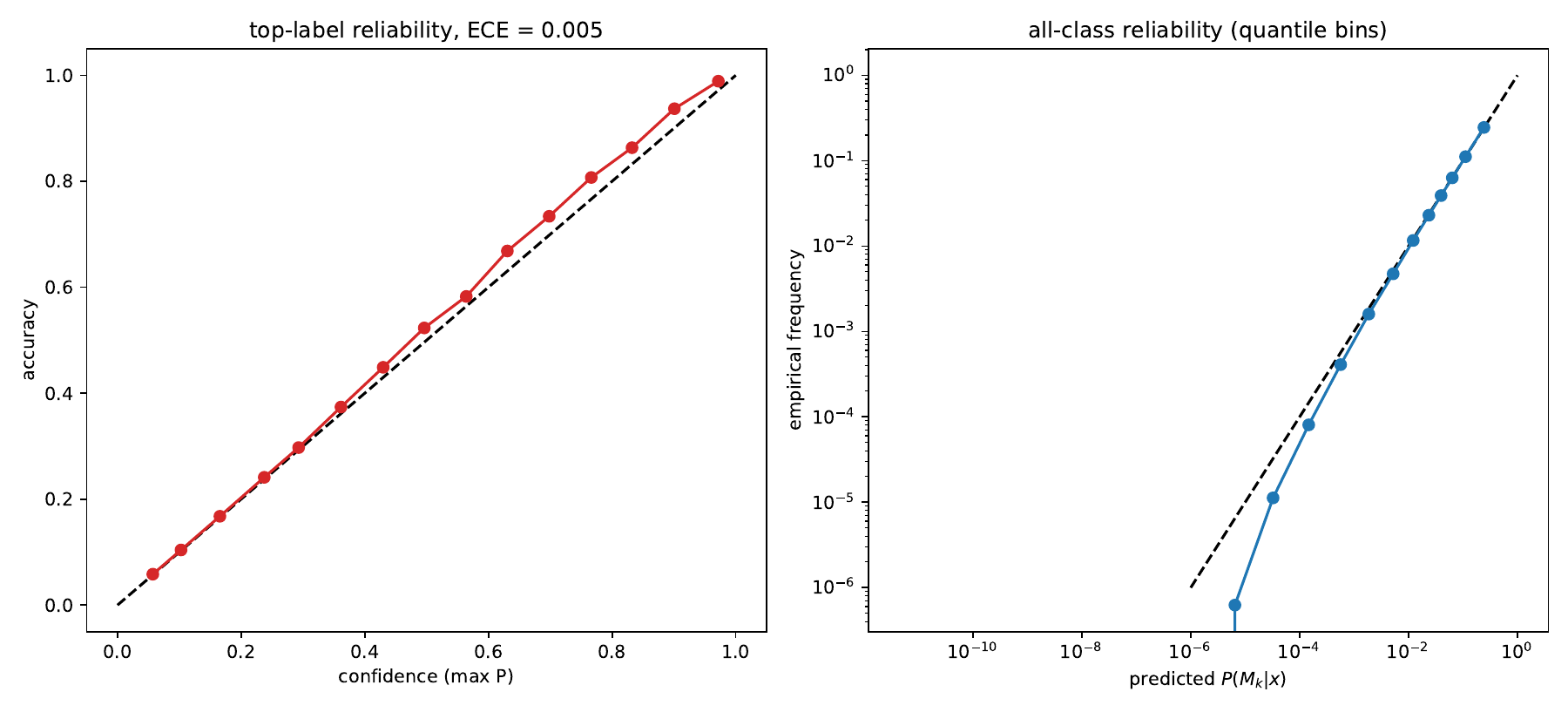}
    \vspace{-1em}
\caption{Calibration of the yield-set classifier on $8\times10^5$ validation simulations. (Left) Accuracy against confidence of the top-ranked class. (Right) Predicted class probabilities $P(\mathcal{M}_k|x)$ against the empirical frequency of the true class, in quantile bins over all classes. The dashed lines show perfect calibration.}
\label{fig:classifier-calibration}
\end{figure}

To test the yield-set ranking independently of the \C likelihood estimate, the MAP and the AIC, we trained a neural classifier to identify which of the $40$ yield sets generated a star \citep[cf.][]{Jeffrey2024}. The classifier is a four-layer multilayer perceptron ($256$ units per layer, GELU activations, dropout $0.1$) trained on $2\times10^5$ \chem simulations per yield set. Its inputs are the observables available for the \citet{Nissen_2020} stars (age and the C, Fe, Mg, O and Si abundances), with the same $0.05$\,dex Gaussian noise as used for \C\unskip. For balanced classes, its softmax output estimates $P(\mathcal{M}_i \mid x)$ under equal model priors and the simulation prior. On $8\times10^5$ simulations from the validation split, which were not used in training, the classifier is well calibrated (Fig.~\ref{fig:classifier-calibration}), i.e. the expected calibration error of the top-ranked class is $0.005$, and predicted probabilities match empirical frequencies across the range $10^{-4}$--$0.25$. Its accuracy is $23.7\%$, against a chance level of $2.5\%$, reflecting that many yield sets predict similar abundance patterns for individual stars.

For the $69$ \citet{Nissen_2020} stars we marginalize the observational errors of each star with $300$ Monte Carlo realizations and combine the stars as in Eq.~\ref{eq:combined-likelihood}. We find that the classifier prefers the NuGrid + TNG combination with the next best model (TNG + TNG) being lower by $\Delta \ln P = -109$. Its ranking correlates with Table~\ref{table:yield-comparison-results} (Spearman $\rho = 0.89$; 7 of the top 10 models in common).

\section{Model misspecification tests}
\label{sec:model-misspecification}

Here we investigate the sensitivity of \C to misspecification by using mock data generated with yield sets different from those used for training. 

\subsection{Mock \chem data}
\label{sec:model-missspecification-chempy}
The mock data generation follows the same procedure as previously described, using \chem with an alternative yield set combination consisting of SN Ia yields taken from \citet{2003NuPhA.718..139T}, CC-SN yields from \citet{2013ARA&A..51..457N} and AGB yields from \citet{Karakas2016}. 
The global parameters are fixed again at $\alpha_{\mathrm{IMF}} = -2.3$ and $\log_{10}N_{\mathrm{Ia}}=-2.89$ while local ISM parameters were drawn from their priors (see Tab.~1 in \citealt{buck2025inferringgalacticparameterschemical}) effectively leading to a multi-zone model.
The chosen alternative yield sets ensure that the contributions from SN Ia, CC-SN and AGB nucleosynthesis differ by approximately $\mathcal{O}(10\%)$ from the training yields, providing a significant test of robustness.

\begin{figure}
    \centering
    \includegraphics[width=0.49\linewidth]{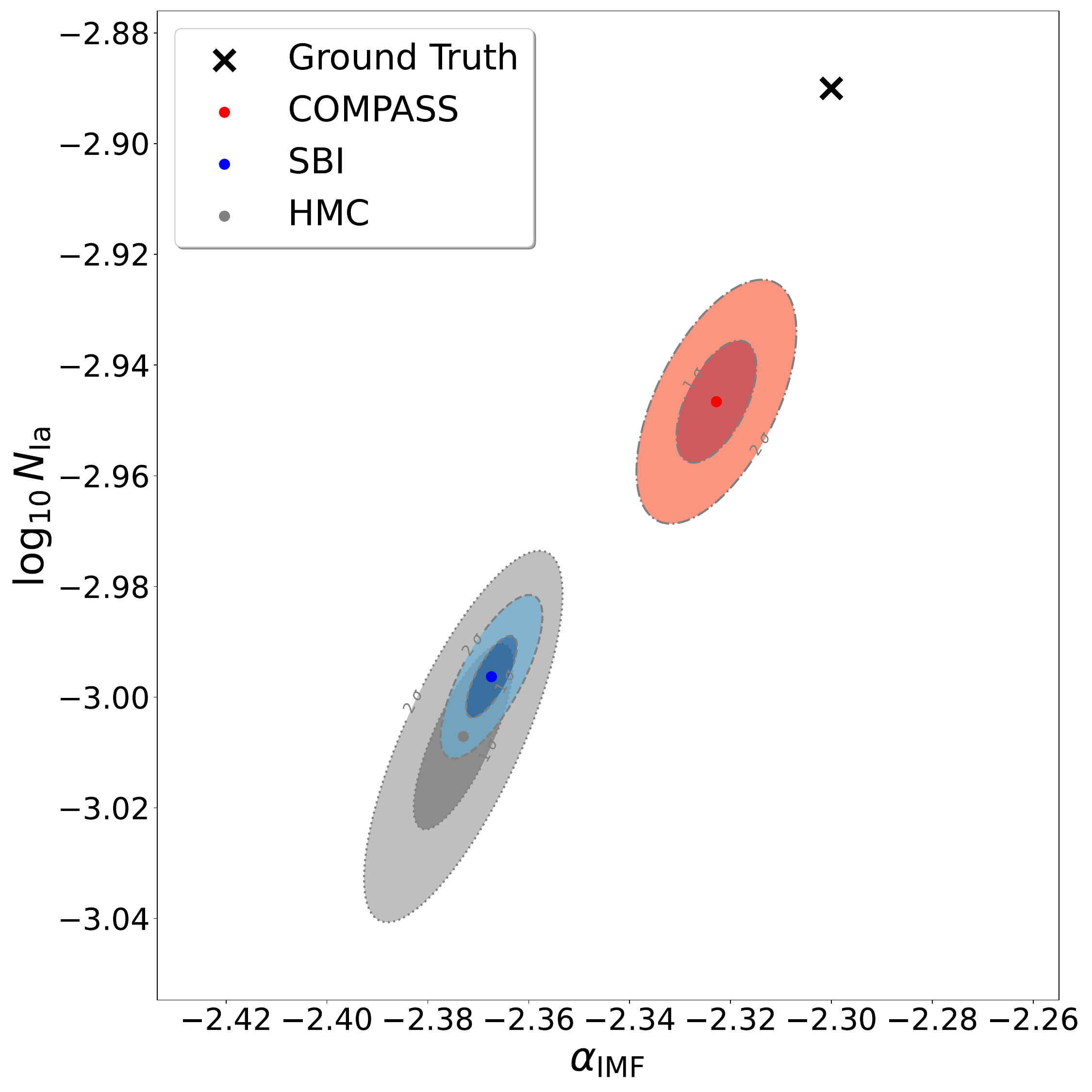}
    \includegraphics[width=0.49\linewidth]{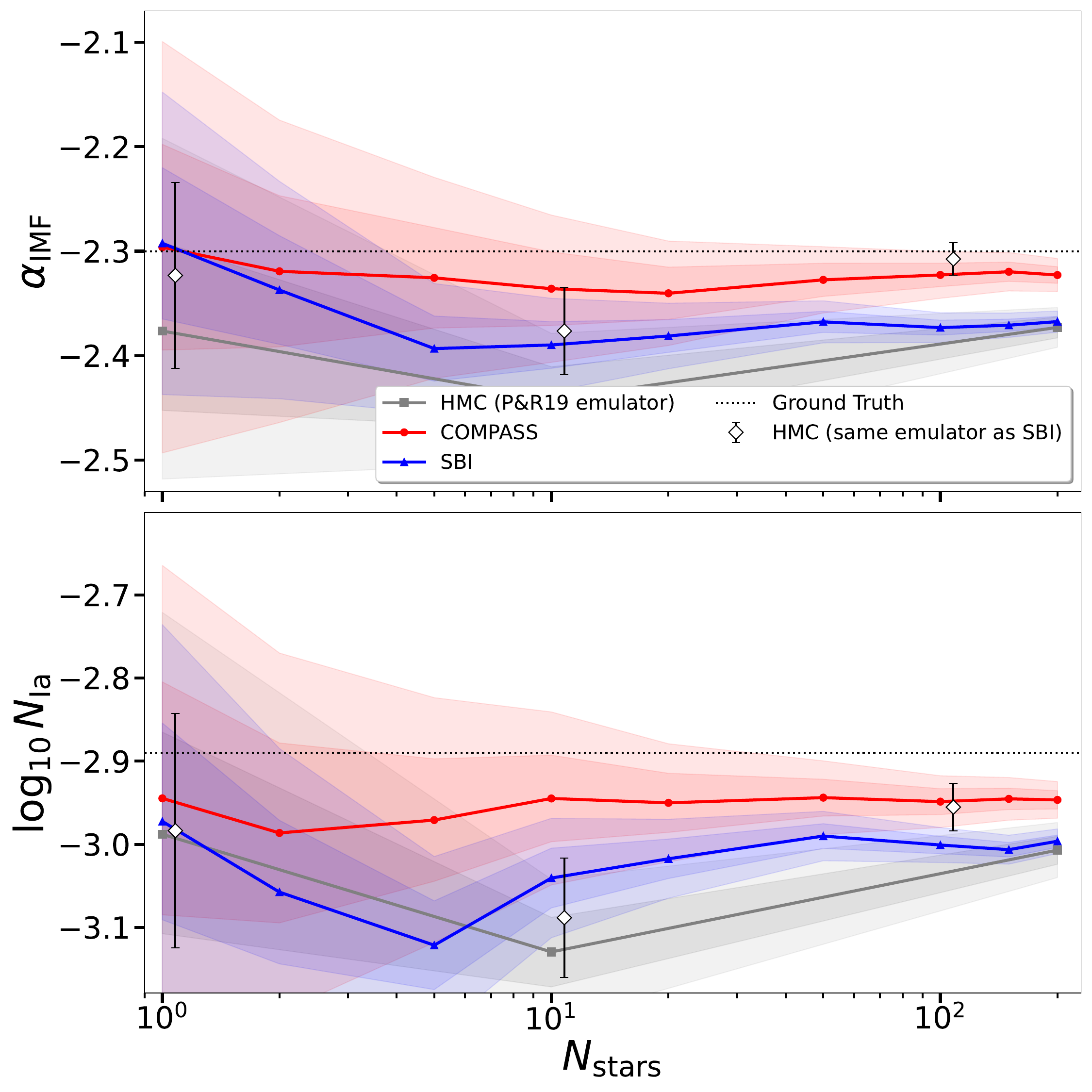}
\caption{Parameter inference for alternative yields.
Comparison of inference results for the global galactic parameters using \C (red), SBI (blue) and HMC (gray) on mock data generated with alternative yield sets that do not match the yields used for training.
(Left) Joint posterior distribution of $\alpha_{\mathrm{IMF}}$ and $\log_{10}N_{\mathrm{Ia}}$ inferred from $200$ stars; methods and contours as in Fig.~\ref{fig:sbi-chempy-TNG}.
(Right) Convergence of the inferred parameters as a function of $N_{\rm stars}$; open diamonds as in Fig.~\ref{fig:sbi-chempy-TNG}. }
\label{fig:sbi-chempy-alt}
\end{figure}

Figure \ref{fig:sbi-chempy-alt} presents the parameter inference results under these misspecified yield sets conditions.
Model misspecification biases all methods. With $200$ stars, \C gives $\alpha_{\rm IMF} = -2.323 \pm 0.008$ and $\log_{10}N_{\rm Ia} = -2.947 \pm 0.011$, i.e. biases of $-2.9\sigma$ and $-5.1\sigma$ (Mahalanobis distance $5.1$; $1.0\%$ and $2.0\%$). SBI is biased by $-13\sigma$ and $-14\sigma$ ($2.9\%$ and $3.7\%$), and HMC with the \citet{Philcox_2019} emulator by $-7.4\sigma$ and $-7.0\sigma$. HMC with the same emulator as SBI is less biased ($N = 100$: $-0.5\sigma$ and $-2.3\sigma$, against $-2.0\sigma$ and $-3.8\sigma$ for \C at $N = 100$), because its per-element model-error term widens its posterior. \C is thus considerably less biased than the NPE baseline, but its posterior does not cover the truth. This means inferences under misspecified yields must not be interpreted at face value, which underscores the importance of selecting the yield set first. The right-hand panel shows that the bias in units of $\sigma$ grows as the posterior contracts with $N_{\rm stars}$.

\subsection{IllustrisTNG hydrodynamical simulation}
\label{sec:model-missspecification-ilustris}
The one-zone GCE model \chem inherently simplifies complex ISM physics, such as star formation, feedback and metal mixing or even radial migration of stars between different zones. 
While \chem allows for star-specific ISM environments effectively creating multi-zone models, it does not model gas or star exchange between these environments or episodic events like sudden star formation bursts. 
Hence this section investigates the potential biases introduced by these simplifications when inferring SSP parameters from more realistic hydrodynamical galaxy simulations. 
In order to explore these effects, we analyze a more complex model of the formation and chemical enrichment of a Milky Way type galaxy taken from the IllustrisTNG simulations. Unlike \chem, IllustrisTNG includes spatially resolved star formation, feedback and metal mixing and mergers and radial migration.

In detail, we use a galaxy with subhalo ID $523071$ and mass close to $10^{12}\rm{M}_{\odot}$ from the $z=0$ snapshot of highest-resolution TNG100-1 simulation\footnote{This particular TNG galaxy was chosen by \citet{Philcox_2019} for its distinct high-$\alpha$ and low-$\alpha$ sequences mimicking the chemical abundance patterns of the Milky Way \citep[e.g.]{Hayden2015}, providing a challenging test case.
The origin of such bimodalities is actively debated, potentially arising from gas-rich mergers, varied star formation modes \citep[e.g.][]{Chiappini1997,2018MNRAS.474.3629G,2018MNRAS.477.5072M,2019MNRAS.484.3476C,Buck2020,Buck2023}  or chemo-dynamical processes like radial migration \citep[e.g.,][]{2009MNRAS.396..203S,2013A&A...558A...9M,2017ApJ...835..224A}.}. 
From approximately $40,000$ stellar particles in this subhalo, $1,000$ are randomly selected. 
Each particle, with a mass of $\sim 1.4\times 10^6 \mathrm{M}_\odot$ \citep{2019ComAC...6....2N}, serves as a proxy for a distinct stellar environment, providing elemental mass fractions $\{d_i^j\}$ and a cosmological scale factor $a_i$ at its formation time. 
These are converted to [X/Fe] abundance ratios and stellar birth times $T_i$ using \citet{2009ARA&A..47..481A} solar abundances and \texttt{astropy} \citep{astropy:2013,astropy:2018} with a $\Lambda$CDM cosmology (\citet{planck2015} parameters), consistent with TNG \citep{2018MNRAS.475..648P}. 
Observational errors ($5\%$ Gaussian) are added to mimic the \chem mock data. 
As in \citet{Philcox_2019}, particles with $T_i\notin[2,12.8]\,\mathrm{Gyr}$ (approx. $5\%$ of stars) are excluded to avoid neural network extrapolation errors near training age boundaries. 

\begin{figure}
    \centering
    \includegraphics[width=0.49\linewidth]{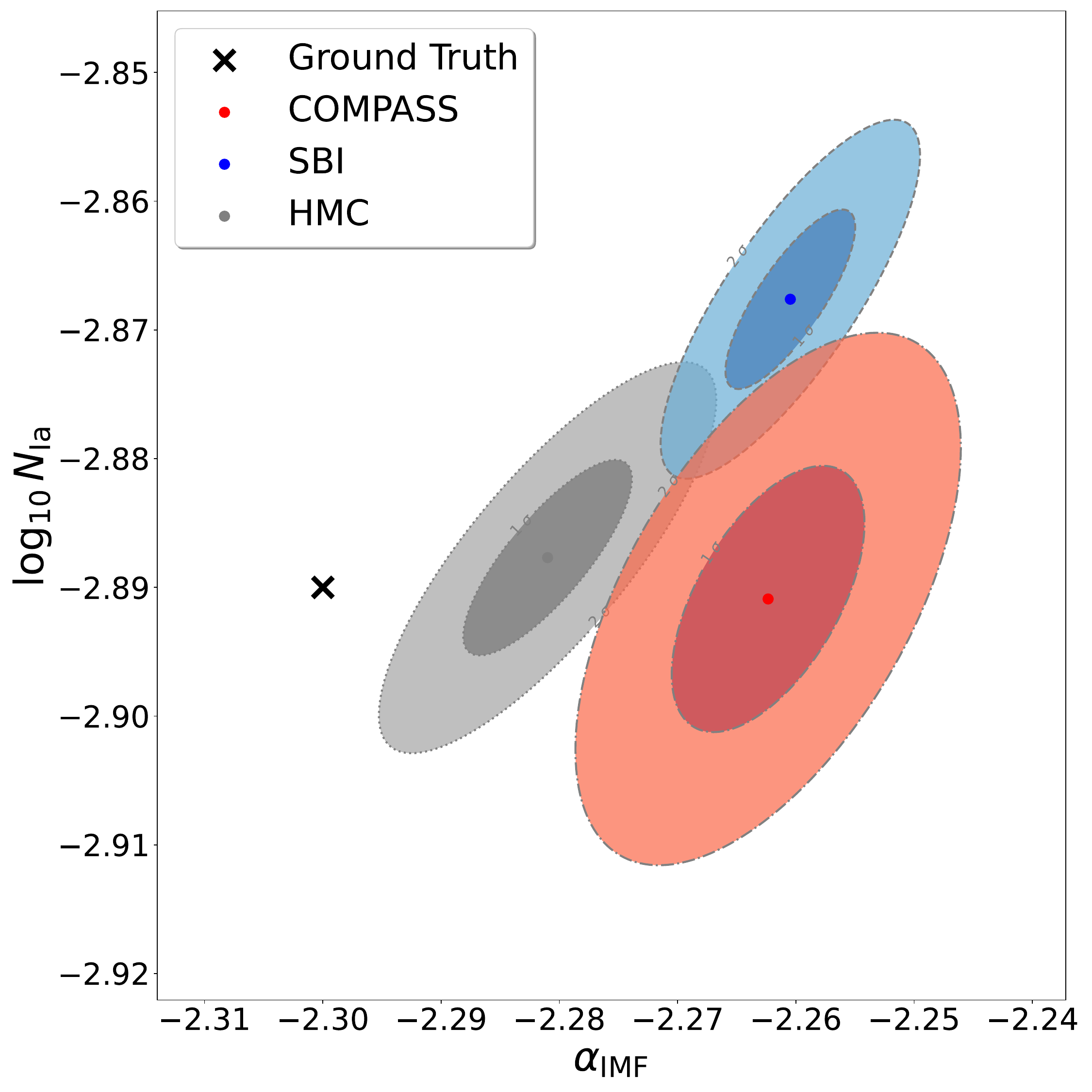}
    \includegraphics[width=0.49\linewidth]{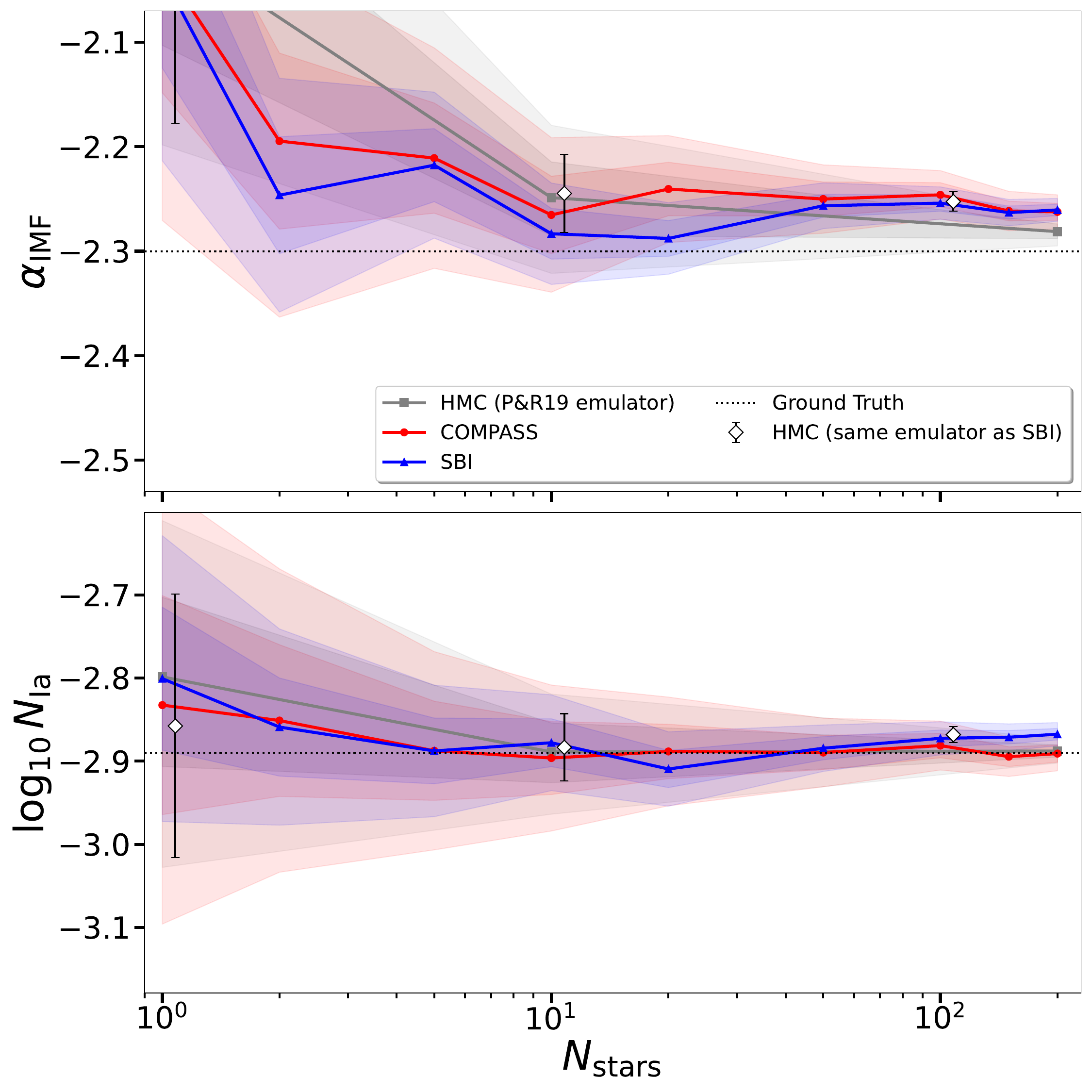}
\caption{Parameter inference from an IllustrisTNG Milky Way simulation.
Comparison of inference results for the global galactic parameters using \C (red), SBI (blue) and HMC (gray) on stellar abundances from a Milky Way like galaxies taken from the IllustrisTNG simulation.
(Left) Joint posterior distribution inferred from $200$ stars; methods and contours as in Fig.~\ref{fig:sbi-chempy-TNG}. The HMC runs use the \citet{Philcox_2019} noise realization, per-star errors and stellar ages of the same stars.
(Right) Convergence of the inferred parameters as a function of $N_{\rm stars}$; open diamonds as in Fig.~\ref{fig:sbi-chempy-TNG}.}
\label{fig:sbi-chempy-TNG-sim}
\end{figure}

For the IllustrisTNG galaxy (Fig.~\ref{fig:sbi-chempy-TNG-sim}), all methods recover $\log_{10}(N_{\rm Ia})$ reasonably well (\C $-0.1\sigma$, SBI $+3.2\sigma$, HMC $+0.3\sigma$), while $\alpha_{\rm IMF}$ is biased high: \C gives $\alpha_{\rm IMF} = -2.262 \pm 0.008$ ($+4.6\sigma$) and SBI is biased by $+7.2\sigma$. HMC with the \citet{Philcox_2019} emulator gives $+2.7\sigma$. We note that this HMC run uses the \citet{Philcox_2019} noise realization, per-star errors and stellar ages of these stars, whereas \C and SBI use $5\%$ noise and no ages. With the same emulator as SBI, HMC shows the same bias as \C and SBI ($N = 100$: $+5.1\sigma$, against $+4.7\sigma$ for \C and $+5.9\sigma$ for SBI at $N = 100$).
The bias is therefore common to the inference methods and suggests structural mismatch between the one-zone model and the hydrodynamical simulation potentially linked to differences in the detailed modeling approaches of one-zone models and full hydrodynamical models.
We emphasize however that \C reaches this accuracy at an order-of-magnitude lower computational cost than HMC.

\end{appendix}

\end{document}